\documentclass[preprint,superscriptaddress]{revtex4}
\usepackage[utf8]{inputenc}
\usepackage{array}
\usepackage{hhline}
\usepackage{graphicx}				
\usepackage{amssymb}
\usepackage{physics}
\usepackage{amsmath}
\usepackage[dvips]{epsfig}
\usepackage{subfigure}
\usepackage{caption}
\usepackage{amsfonts}
\usepackage{amsmath}
\numberwithin{equation}{section}
\usepackage{mathrsfs}
\usepackage{bm}
\usepackage[title]{appendix}
\makeatletter
\newcommand{\oset}[2]{%
  {\mathop{#2}\limits^{\vbox to -.5\ex@{\kern-\tw@\ex@
  \hbox{\scriptsize #1}\vss}}}}
\makeatother
\makeatletter
\newsavebox\myboxA
\newsavebox\myboxB
\newlength\mylenA
\newcommand*\xoverline[2][0.75]{%
    \sbox{\myboxA}{$\m@th#2$}%
    \setbox\myboxB\null% Phantom box
    \ht\myboxB=\ht\myboxA%
    \dp\myboxB=\dp\myboxA%
    \wd\myboxB=#1\wd\myboxA% Scale phantom
    \sbox\myboxB{$\m@th\overline{\copy\myboxB}$}%  Overlined phantom
    \setlength\mylenA{\the\wd\myboxA}%   calc width diff
    \addtolength\mylenA{-\the\wd\myboxB}%
    \ifdim\wd\myboxB<\wd\myboxA%
       \rlap{\hskip 0.5\mylenA\usebox\myboxB}{\usebox\myboxA}%
    \else
        \hskip -0.5\mylenA\rlap{\usebox\myboxA}{\hskip 0.5\mylenA\usebox\myboxB}%
    \fi}
\makeatother

\def\la{\lower4pt\hbox{$\buildrel\textstyle\ <\over\sim\ $}}
\def\ga{\buildrel\textstyle>\over\sim}
\def\simgreat{\buildrel \textstyle \lower3pt\hbox{$\sim$} \over >}
\def\simless{\buildrel \textstyle \lower3pt\hbox{$\sim$} \over <}
\newcommand\bsdot{\ensuremath{\boldsymbol{.}}}
\def\tensorK{\overset{\text{\tiny$\bm\leftrightarrow$}}{\mathbf{K}}}

\def\tensorD{\overset{\text{\tiny$\bm\leftrightarrow$}}{\mathbf{D}}}
\def\tensorI{\overset{\text{\tiny$\bm\leftrightarrow$}}{\mathbf{I}}}
\def\tensorT{\overset{\text{\tiny$\bm\leftrightarrow$}}{\mathbf{T}}}
\def\tensorR{\overset{\text{\tiny$\bm\leftrightarrow$}}{\mathbf{\mathcal{R}}}}
\renewcommand{\vec}[1]{\boldsymbol{\mathbf{#1}}}
\newcommand{\unv}[1]{\hat{\vec{#1}}}

\def\cchi{\raise2pt\hbox{$\chi$}} %slightly higher chi

\begin{document}
\title{Interaction of radio frequency waves with cylindrical density filaments -- scattering and radiation pressure}

%
%\date{\today}
%
\author{Spyridon I. Valvis}
\affiliation{National Technical University of Athens,
School of Electrical and Computer Engineering,
Association EURATOM-Hellenic Republic, GR-15773, Greece}
\author{Abhay K. Ram}
\affiliation{Plasma Science and Fusion Center, 
Massachusetts Institute of Technology, Cambridge, MA 02139.}
\author{Kyriakos Hizanidis}
\affiliation{National Technical University of Athens, 
School of Electrical and Computer Engineering, 
Association EURATOM-Hellenic Republic, GR-15773, Greece}

%
%\maketitle
%
\begin{abstract}
The propagation of radio frequency (RF) waves in tokamaks can be affected by filamentary structures, or blobs, that are
present in the edge plasma and the scrape-off layer. The difference in the permittivity between the surrounding
plasma and interior of a filament leads to reflection, refraction, and diffraction of the waves. This, in turn, can affect
the power flow into the core of the plasma and reduce the efficiency of heating and/or current generation. 
The scattering of RF waves -- lower hybrid, helicon, and ion cyclotron waves -- by a single cylindrical filament, 
embedded in a background plasma, is studied using a full-wave analytical theory developed previously [A. K. Ram
and K. Hizanidis, Phys. Plasmas \textbf{23}, 022504-1--022504-17 (2016)]. The theory assumes that the plasma in and around
a filament is homogeneous and cold. A detailed scattering analysis reveals a variety of common features
that exist among the three distinctly different RF waves. These common attributes can be inferred intuitively based on an examination
of the cold plasma dispersion relation. The physical intuition is a useful step to understanding experimental observations
on scattering, as well as results from simulations that include general forms of edge plasma turbulence. While a filament
can affect the propagation of RF waves, the radiation force exerted by the waves can influence the filament. 
The force on a filament is determined using the Maxwell stress tensor.
In 1905, Poynting was the first to evaluate and measure the radiation force on an interface separating  
two different dielectric media [J. H. Poynting, Phil. Mag. \textbf{9}, 393-406 (1905)]. For ordinary light propagating in
vacuum and incident on a glass surface, Poynting noted that the surface is ``pulled'' towards the vacuum. In a magnetized cold plasma,
there are two independent wave modes. Even if only one of these modes is excited by an RF antenna, a filament will couple power
to the other mode -- a consequence of electromagnetic boundary conditions. This facet of scattering has consequences on the
radiation force that go beyond Poynting's seminal contribution. The direction of the force depends on the polarization of the incident wave and
on the mode structure of the waves inside and in the vicinity of a filament. It can either pull the filament toward the
RF source or push it away. For slow lower hybrid waves, filaments are pulled in regardless of whether they are more or less
dense compared to the ambient plasma. For fast helicon and ion cyclotron waves, the direction of the force depends on 
the plasma and wave parameters; in particular, on the ambient density.
For all three waves, the radiation force is large enough to impact the motion of a filament and 
could be measured experimentally. This suggests a possibility of modifying the edge turbulence using RF waves.
\end{abstract}

%\pacs{52.20.Dq, 52.65.Cc}

\maketitle

\section{Introduction}
\label{sec:0}

The occurrence of blobs and filaments in the edge region of a tokamak plasma has been observed in experiments and discussed
in theories \cite{krash,grulke, maqueda,grulke1}. The propagation of radio frequency (RF) waves can be affected by the presence of these structures
\cite{rkk13,rk13,rk16}. Over the years, there have been a number of studies -- theoretical, computational, and experimental -- on the
propagation of RF waves through the turbulent edge region of a fusion device 
\cite{kohn,snicker,valvis,chel1,martin,arist,biswas,lau,chel2}. The premise of these studies is
to quantify the role of turbulence in scattering of RF power and in modifying wave properties such as their direction of propagation.
Various forms of RF waves play an important role in heating and in generating non-inductive current in fusion plasmas. Any modifications
to the waves in the edge region can reduce the RF power available in the core plasma for heating and current drive. 

In this paper we investigate the scattering of waves with frequencies below the electron cyclotron frequency -- 
in particular, the lower hybrid (LH), helicon, and ion cyclotron (IC) waves -- by a single cylindrical filament aligned
along the magnetic field. The wavelengths of these waves span a broad spatial scale ranging from
being comparable to the radial dimension of the filament to being many times longer. The theoretical analysis is
based on the analytical model discussed in \cite{rk16}. We assume that the plasma is homogeneous and cold inside the
filament and in the region surrounding it. There is a discontinuity in the plasma density and in its permittivity
across the interface of the filament. The Maxwell equations for a homogeneous plasma take on the form of a vector Helmholtz equation
which can be solved inside and outside the filament. The continuity of the electromagnetic
fields across the interface, consistent with Maxwell equations, gives a complete analytical solution for the scattering of an
incident plane wave. The magnitude of the discontinuity in plasma density at the interface is not restricted in this analysis.
The domain of validity of the analytical theory has been examined numerically \cite{zisis}.

The first part of the paper describes the influence of a filament on the propagation of LH, helicon, and IC waves.
We relate the physics of scattering with the dispersion characteristics of the cold plasma waves. Even a cursory 
examination of these characteristics can provide a good physical insight into the scattering process. An added 
advantage is that a detailed analysis of every scattering configuration is not necessary. The phase space of 
plasma parameters, both inside and outside the filament, that needs further examination can be constrained. 
Furthermore, a physical insight is useful for understanding simulation
results from more complicated scattering processes where multiple filaments are involved \cite{zisis}.

In the second part of the paper we analyze the impact of RF waves on a filament; in particular, the RF induced
radiation forces exerted on a filament. In 1905, Poynting reported on his theoretical and experimental research
on the radiation force exerted on a planar interface separating two dielectrics \cite{poynt}. By imposing conservation
of momentum, Poynting deduced that the force exerted on a vacuum-glass interface by ordinary light points towards the 
vacuum \cite{poynt,loudon}. In combination with the analytical model for scattering, we use the Maxwell stress 
tensor \cite{stratton,griffiths} to evaluate the radiation force
on a filament due to the different RF waves. Incidentally, Poynting's conclusion can also be derived using the
Maxwell stress tensor \cite{loudon,hirose}.
We find that the direction of the force induced by RF waves does not necessarily follow Poynting's conclusion.
The propagation of waves in a magnetized plasma,  
represented by a tensor permittivity, is different from wave propagation in a scalar dielectric. In particular, for a given frequency,
there are two waves with disparate wavelengths and polarizations that coexist in a cold plasma. For a particular choice of an incident
wave, the scattering off a filament couples power to the second wave. Thus, the contrast with Poynting's results 
is not surprising. Moreover, in the frequency domain of interest, evanescent waves are present in the edge region. 
These waves, whether excited by a RF source or generated through the scattering process, affect the radiation forces in
ways that are different from propagating electromagnetic waves. In ordinary dielectrics,
evanescent waves are of practical importance, e.g., in near-field scanning optical microscopy \cite{pawlink}. 

\section{Description of the geometry and the plasma}
\label{sec:1}

We consider a cylindrical plasma filament, with spatially homogeneous density, embedded in a uniform background plasma.
The filament has a circular cross-section with its axis aligned along the
ambient magnetic field line (Fig.~1). The axial extent of the filament is taken to be infinite which allows us
to neglect the effects of the end caps. Inherent in this assumption is that the RF fields are axially confined
to a spatial region that is smaller than the length of the filament. The magnetic field in the scattering region is
uniform and the plasma is presumed to be cold.

The relationship between the cylindrical coordinate system $(\unv{\rho},\; \unv{\phi},\; \unv{z})$, used as a basis in the theory, 
and the Cartesian coordinate system $\left(\unv{x},\; \unv{y},\; \unv{z}\right)$ can be expressed in the form taken by the position vector $\vec{r}$,
\begin{equation}
\vec{r} \ = \ x \; \unv{x} \; + \;  y \; \unv{y} \; + \;  z \; \unv{z} \ = \ \rho \; \unv{\rho} \; + \; z \; \unv{z},  \label{1.1}
\end{equation} 
where $\rho = \sqrt{x^2 +y^2}$, and 
$z$ is along the axial direction. The origin of the coordinate system is the center of the cylindrical filament.

\section{Propagation of electromagnetic waves in a plasma}
\label{sec:2}

The mathematical description of the propagation and scattering of RF waves in a cold plasma is
based on the linearized set of continuity and momentum fluid equations for electrons and ions.
These are combined with Faraday's and Ampere's equations \cite{rk16,stix} to obtain the spatial variation
of the RF electric field,
\begin{equation}
{\boldsymbol{ \nabla \times}} \Bigl\{ {\boldsymbol{ \nabla \times}} \; \mathbf{E} \left( \mathbf{r} \right) \Bigr\} \;  - \;
\frac{\omega^2}{c^2} \ \tensorK \left( \mathbf{r} \right) \boldsymbol{.} \; \mathbf{E} \left( \mathbf{r} \right) \ = \ 0, \label{fa}
\end{equation}
where $\omega$ is the angular frequency of the electromagnetic fields, $c$ is the speed of light, and
$\tensorK \left( \mathbf{r} \right)$ is the plasma permittivity tensor. We have assumed that the plasma equilibrium is time independent,
while the linearized perturbed 
electromagnetic fields
have a time dependence of the form $\displaystyle{{\rm e}^{\displaystyle{-i \omega t}}}$, where $t$ is the time. 
In the cylindrical coordinate system where the ambient magnetic
field is $\mathbf{B}_0 = B_0 \; \unv{z}$, and $\tensorK \left( \mathbf{r} \right)$ has the form \cite{stix},
\begin{equation}
\tensorK \ = \ 
\left( \begin{array}{ccccc} K_\rho  & \ \  & -i K_\phi & \ \  & 0 \\
\\
iK_\phi & \ \  & K_\rho  & \ \   & 0 \\
\\
0 & \ \   &  0 & \ \   & K_z \label{pt}
\end{array} \right),
\end{equation}
where,
\begin{eqnarray}
K_\rho \; & = &  \; 1 \ - \ \frac{\omega_{pe}^2}{\omega^2-\omega_{ce}^2} \ - \
\sum_{i} \ \frac{\omega_{pi}^2}{\omega^2-\omega_{ci}^2},  \nonumber \\
K_\phi \; & = &  \;  - \frac{\omega_{ce}}{\omega} \ \frac{\omega_{pe}^2}{\omega^2-\omega_{ce}^2} \ + \ 
\sum_{i} \ \frac{\omega_{ci}}{\omega} \ \frac{\omega_{pi}^2}{\omega^2-\omega_{ci}^2},  \\
K_z  \; & = & \; 1 \ - \ \frac{\omega_{pe}^2}{\omega^2} 
\ - \ 
\sum_{i} \ \frac{\omega_{pi}^2}{\omega^2}, \nonumber \label{pe}
\end{eqnarray}
$\omega_{pe}$ ($\omega_{pi}$) and $\omega_{ce}$ ($\omega_{ci}$) 
are the angular electron (ion) 
plasma frequency and cyclotron frequency, respectively, and the index $i$ represents
all the ion species in the plasma. The plasma and cyclotron frequencies can, in general, be functions
of space.
The permittivity tensor of the background plasma and of the filament 
are expressed in terms of their respective ion compositions and constant, but different, densities.
Subsequently, the elements of $\tensorK$ are constants in each region inside and outside the filament.

For a spatially independent $\tensorK$, Eq. \eqref{fa} has the form of a vector Helmholtz equation and
can be solved analytically in the cylindrical coordinate system using vector cylinder functions \cite{rk16}.
We assume that the incoming RF wave launched by an antenna is a plane electromagnetic wave of fixed frequency.
Additionally, we assume that the component of wave vector along the direction of the ambient magnetic field $k_z$ is
prescribed ab initio. Since the axis of the cylindrical filament is also aligned along the magnetic field, boundary
conditions imposed by Maxwell's equations require that $k_z$ be the same for all waves -- those inside the filament
and those that are scattered. This property leads to some simplification when solving Eq. \eqref{fa} \cite{rk16}.

In Section \ref{sec:3} below,
we derive, in cylindrical coordinates,
the necessary properties of the plane wave that are consistent with Eq. \eqref{fa}.
A description of the scattered waves and the waves inside the filament is outlined in Section \ref{sec:3}. 

\section{Dispersion characteristics and polarization of a  plane plasma wave}
\label{sec:3}

In cylindrical coordinates, the spatial variation of a plane wave is \cite{stratton},
\begin{equation}
\displaystyle{e^{{i \mathbf{k} \bsdot \mathbf{r} }}} \ = \ 
\displaystyle{e^{{i k_\rho \rho \cos \left( \phi - \phi_k \right) \; + \; i k_z z}}} \ = \
\sum_{m=-\infty}^{\infty} \ i^m \ {\text J}_m \left( k_\rho \rho \right) 
\displaystyle{e^{{ i m \left( \phi - \phi_k \right)}}} \; \displaystyle{e^{i k_z z}}, 
% \equiv \ \sum_{m=-\infty}^{\infty} \ i^m \ \psi_{1m} \ \displaystyle{e^{-im \phi_k}} \; \displaystyle{e^{ik_z z}},
\label{3.1}
\end{equation}
where ${\text J}_m$ is the Bessel function of the first kind of order $m$,
$\phi$ and $\phi_k$ are the azimuthal angles between the
$x$-axis and $\mathbf{r}$ and $\mathbf{k}$, respectively, and $k_\rho = \sqrt{k_x^2+k_y^2}$
with $\mathbf{k} \ = \ k_x \; \unv{x} \; + \;  k_y \; \unv{y} \; + \;  k_z \; \unv{z}$.

In the cylindrical coordinate system \cite{rk16},
\begin{equation}
\left(
\begin{array}{c}
\unv{\rho}\\
\unv{\phi}\\
\unv{z}
\end{array}
\right) \ \displaystyle{e^{i \mathbf{k} \bsdot \mathbf{r} }} \ = \ 
\sum_{m=-\infty}^{\infty} \ i^m \ \displaystyle{e^{-im \phi_k}} \ \left( \ \mathbf{a} \mathbf{l}_{m} \ + \
\mathbf{b} \mathbf{m}_{m} \ + \
\mathbf{c} \mathbf{n}_{m} \right). \label{3.2}
\end{equation} 
The right-hand side is a sum of three dyadics,  with,
\begin{equation}
\mathbf{a} \ = \ 
\frac{i}{k^2} \ \left(
\begin{array}{c}
- k_\rho \cos \left( \phi - \phi_k \right) \\ 
k_\rho \sin \left( \phi - \phi_k \right) \\
-k_z 
\end{array}
\right),
\mathbf{b} \ = \ 
\frac{i}{k_\rho} \ \left(
\begin{array}{c}
\sin \left( \phi - \phi_k \right) \\
\cos \left( \phi - \phi_k \right) \\ 
0 
\end{array}
\right),
\mathbf{c} \ = \ 
\frac{1}{k k_\rho} \ \left(
\begin{array}{c}
- k_z \cos \left( \phi - \phi_k \right) \\ 
k_z \sin \left( \phi - \phi_k \right) \\ 
k_\rho
\end{array}
\right). \label{3.3}
\end{equation} 
The vector cylinder functions $\mathbf{l}_{m}$, $\mathbf{m}_{m}$, and $\mathbf{n}_{m}$ are \cite{stratton},
\begin{eqnarray}
\mathbf{l}_m \left( \rho, \phi, z; k_\rho, k_z \right)& = & \left[ \left( \frac{\partial}{\partial \rho}
                  {\text J}_m \right) \unv{\rho} + \left\{
\frac{im}{\rho} \unv{\phi} + i k_z \ \unv{z} \right\} {\text J}_m \right] \displaystyle{e^{i k_z z + i m \phi}}, \nonumber \\
\mathbf{m}_m \left( \rho, \phi, z; k_\rho, k_z \right) & = & \left[ \frac{im}{\rho}
                   {\text J}_m \ \unv{\rho} - 
\left( \frac{\partial}{\partial \rho}  {\text J}_m \right) \unv{\phi} \right] \displaystyle{e^{i k_z z + i m \phi}},  \label{3.4} \\
\mathbf{n}_m \left( \rho, \phi, z; k_\rho, k_z \right)  & = & \left[ \frac{ik_z}{k} 
\left( \frac{\partial}{\partial \rho} {\text J}_m \right) \unv{\rho} -
\left\{ \frac{k_z}{k} \frac{m}{\rho} \unv{\phi} - \frac{k_\rho^2}{k} \unv{z} \right\} {\text J}_m \right] \displaystyle{e^{i k_z z + i m \phi}},
\nonumber 
\end{eqnarray}
where the argument of ${\text J}_m$ is $k_\rho \rho$.

For a plane wave the electric field is,
\begin{equation}
\vec{E}_P \left( \mathbf{r} \right) \ = \ \mathbf{E}_0 \left( \mathbf{k}, \omega \right) \
\displaystyle{e^{i \mathbf{k} \bsdot \mathbf{r}}}, \label{3.5}
\end{equation}
where $\vec{E}_0$ is the electric field vector which is independent
of space and time. Substituting this form into Eq. \eqref{fa} yields,
\begin{equation}
\tensorD  \left(\vec{k}, \omega \right) \; \bsdot \; \vec{E}_0 \left( \vec{k}, \omega \right) \ = \ 0, \label{3.6}
\end{equation}
where
\begin{equation}
\tensorD \left( \vec{k}, \omega \right) \ = \ \frac{c^2}{\omega^2} \left( \vec{k} \vec{k} - k^2 \tensorI \right) \; + \; \tensorK. \label{3.7}
\end{equation}
Here $\vec{k} \vec{k}$ is a dyadic and $\tensorI$ is the identity tensor.
For a non-zero electric field of the RF wave, we require that,
\begin{equation}
{\rm det} \left( \tensorD \left( \vec{k}, \omega \right) \right) \ = \ 0, \label{3.8}
\end{equation}
where det denotes the determinant of the tensor. Using \eqref{3.7} in \eqref{3.8} leads to the following algebraic equation,
\begin{equation}
n_{\rho}^4 K_{\rho} \ + \ n_{\rho}^2 \left[ K_{\phi}^2 - \left( K_{ \rho} + K_{z} \right) \left( K_{\rho} - n_{z}^2 \right) \right] \ + \
K_{z} \left[ \left( K_{\rho} - n_{z}^2 \right)^2 - K_{\phi}^2 \right] \ = \ 0,
\label{3.9}
\end{equation}
where the index of refraction $\vec{n}=c\vec{k}/\omega$.

For a prescribed $n_z$, the two solutions of the bi-quadratic equation are,
\begin{multline}
n_{\rho \pm}^2 \ = \ \frac{1}{2K_{\rho}}  \left( K_{\rho} + K_{z} \right) \left( K_{\rho} - n_{z}^2 \right) \ - \
\frac{1}{2K_{\rho}} K_{\phi}^2 \\ \pm \ 
\frac{1}{2K_{\rho}} \ 
\sqrt{ \left[ \left\{ \left(K_{\rho} - K_{z} \right) \left( K_{\rho} - n_{z}^2 \right) - K_{\phi}^2 \right\}^2 \ + \
4 n_{z}^2 K_{\phi}^2 K_{z} \right] }.
\label{3.10}
\end{multline}
We will associate one of the roots to a slow wave and the other to a fast wave depending on their
relative phase velocities. The association will become clear when we consider specific examples.

\subsection{Electric field polarizations}
\label{sec:4}

The electric field $\vec{E}_0$ in \eqref{3.5} can be written as,
\begin{equation}
  \vec{E}_0 \left( \vec{k}, \omega \right) \ = \ {\mathcal{E}_0}\left( E_{k \rho} \; \unv{\rho}_k \ +\ E_{k \phi} \; \unv{\phi}_k \ +
    \ E_{k z} \; \unv{z} \right),
\label{4.1}
\end{equation}
where $\mathcal{E}_0$  is the amplitude of the electric field, and $\left( E_{k \rho}, E_{k \phi}, E_{k z} \right)$ are the components of the
polarization vector along $\left( \unv{\rho}_k, \unv{\phi}_k, \unv{z} \right)$. The directional vector
$\left( \unv{\rho}_k, \unv{\phi}_k, \unv{z} \right)$ is in the cylindrical coordinate system defined in the wave vector space.

The polarization of the wave electric field follows from \eqref{3.6}. 
Depending on whether the wave is a slow wave or a fast wave, we will make use of one of the following two representations 
for the polarization,
\begin{align}
e_{k \rho} \ &= \ - \frac{n_{\rho } n_{z} \left( K_{\rho} - n_{\rho}^2 - n_{z}^2 \right)}{\left( K_{\rho} -n_{z}^2 \right)
\left( K_{\rho} - n_{\rho}^2 - n_{z}^2 \right) - K_{\phi}^2}, \nonumber \\ 
e_{k \phi} \ &= \ - \frac{i n_{\rho} n_{z} K_{\phi}}
{\left( K_{\rho} -n_{z}^2 \right)
\left( K_{\rho} - n_{\rho}^2 - n_{z}^2 \right) - K_{\phi}^2}, \label{4.2} \\ 
e_{k z} \ &= \ 1, \nonumber
\end{align}
or,
\begin{align} 
e_{k \rho} \ &= \ 1, \nonumber \\ 
e_{k \phi} \ &= \ - \frac{i K_{\phi}}{K_{\rho} - n_{\rho}^2 - n_{z}^2}, \label{4.3} \\ 
e_{k z} \ &= \ - \frac{n_{\rho} n_{z}}{K_{z} - n_{\rho}^2}, \nonumber  
\end{align}
where $n_\rho$ is one of the roots given in \eqref{3.10}. For the slow wave \eqref{4.2} is a useful form
for the polarization vectors, while \eqref{4.3} is appropriate for the fast wave.
The components $\left( E_{k \rho}, E_{k \phi}, E_{k z} \right)$ in \eqref{4.1} are defined as,
\begin{equation}
  \left( E_{k \rho}, \; E_{k \phi}, \; E_{k z} \right)\ = \  \dfrac{1}{\sqrt{ \left| e_{k \rho} \right| ^2 +
  \left| e_{k \phi} \right| ^2 + \left| e_{k z} \right| ^2 }} \
  \left( e_{k \rho},\; e_{k \phi},\; e_{k z} \right),
\label{4.4}
\end{equation}

\subsection{Electric field representation of a plane wave in cylindrical coordinates}
\label{sec:5}

In the wave vector space,
\begin{equation}
\left(
\begin{array}{c}
\unv{\rho}_k\\
\unv{\phi}_k\\
\unv{z}
\end{array}
\right) \ \displaystyle{e^{i \mathbf{k} \bsdot \mathbf{r} }} \ = \ 
\sum_{m=-\infty}^{\infty} \ i^m \ \displaystyle{e^{-im \phi_k}} \ \Bigl(\mathbf{a}_k \mathbf{l}_{m} \ + \
\mathbf{b}_k \mathbf{m}_{m} \ + \
\mathbf{c}_k \mathbf{n}_{m} \Bigr), \label{5.1}
\end{equation} 
where,
\begin{equation}
\mathbf{a}_k \ = \ 
\frac{i}{k^2} \ \left(
\begin{array}{c}
- k_\rho \\ 
0 \\
-k_z 
\end{array}
\right),\quad
\mathbf{b}_k \ = \ 
\frac{i}{k_\rho} \ \left(
\begin{array}{c}
0 \\
1 \\
0 
\end{array}
\right), \quad
\mathbf{c}_k \ = \ 
\frac{1}{k k_\rho} \ \left(
\begin{array}{c}
- k_z \\
0 \\
k_\rho
\end{array}
\right), \label{5.2}
\end{equation} 
and $k$ is the magnitude of $\mathbf{k}$.
The explicit form of the wave electric field is obtained by substituting the expressions in
\eqref{4.1}, \eqref{5.1}, and \eqref{5.2} into \eqref{3.5},
\begin{align}
 \vec{E}_P \left(\vec{r} \right) \ = \ \mathcal{E}_0\ \sum_{m=-\infty}^{\infty} \ i^m \bigg[ \bigg. & \left\{ -i E_{k \rho}
 {\rm J}_m' \left( k_{\rho} \rho \right)
- \frac{m}{\rho k_{\rho}} E_{k \phi} {\rm J}_m \left( k_{\rho} \rho \right) \right\} \unv{\rho}+  \nonumber \\
& \left\{ -i E_{k \phi} {\rm J}_m' \left( k_{\rho} \rho \right)
+ \frac{m}{\rho k_{\rho}} E_{k \rho} {\rm J}_m \left( k_{\rho} \rho \right)\right\} \unv{\phi} + \nonumber \\
&  \ \ \ \ \bigg. E_{kz} {\rm J}_m \left( k_{\rho} \rho \right) \unv{z} \bigg] \ e^{im \left( \phi - \phi_k \right)} \ e^{i k_{z} z},
\label{5.3}
\end{align} 
where $'$ denotes derivative with respect to the argument.

\section{Electromagnetic fields of the scattered waves and waves inside the filament}
\label{sec:6}

As $\tensorK$ is a function of $\omega$ only, the general solution of Eq. \eqref{fa} is obtained using the Fourier
representation of the electric field,
\begin{equation}
\vec{E} \left( \vec{r} \right) \ = \ \int d^3k \ \vec{E} \left( \vec{k} \right) e^{i \vec{k} \bsdot \vec{r}}\ = \ 
\int_{0}^{\infty} dk_\rho \ k_\rho \  \int_{0}^{2 \pi} d\phi_k \ \int_{-\infty}^{\infty} dk_z \ \vec{E}
\left( \vec{k} \right) e^{i k_\rho \rho \cos \left( \phi - \phi_k \right)} e^{i k_z z} .\label{6.1}
\end{equation}  
Substituting this form in \eqref{fa} yields,
\begin{equation}
  \int \ d^3k \ \tensorD \left( \vec{k}, \omega \right) \bsdot \; \vec{E}_k \left( \vec{k} \right) \
  e^{i k_\rho \rho \cos \left( \phi - \phi_k \right)} e^{i k_z z}  \ = \ 0, \label{6.2}
\end{equation}
where 
\begin{equation}
  \tensorD \left( \vec{k}, \omega \right) \ = \ \frac{c^2}{\omega^2} \left( \vec{k} \vec{k} -
    k^2 \tensorI \right) + \tensorK \left( \omega \right). \label{6.3}
\end{equation}
In general, Eq. \eqref{6.2} is satisfied if and only if, 
\begin{equation}
\tensorD \left( \vec{k}, \omega \right) \bsdot \; \vec{E}_k \left( \vec{k} \right) \ = \ 0. \label{6.4}
\end{equation}
A non-trivial solution for the $\vec{E}_k$ requires that ${\rm det} \tensorD \left( \vec{k}, \omega \right) = 0$. 
This requirement, as in Section \ref{sec:3}, leads to a dispersion relation connecting $k_\rho$, $k_z$, and $\omega$. 
The dispersion relation is of the same form as in \eqref{3.9}.

Since $k_z$ for the scattered waves and for waves inside the filament is the same as that of the incident plane wave,
we find that \cite{rk16},
\begin{align}
\vec{E} \left( \vec{r} \right) \ = \ 
\sum_{\ell=1}^2 \ \sum_{m=-\infty}^{\infty} \ i^m \ \mathcal{E}_{\ell m} 
\ \biggl[ \biggr. &
\left\{ -i E_{k \rho \ell} {\mathcal Z}_m' \left( k_{\rho \ell} \rho \right)- \frac{m}{\rho k_{\rho \ell}} E_{k \phi \ell} 
{\mathcal Z}_m \left( k_{\rho \ell} \rho \right) \right\} \unv{\rho} + \nonumber \\
& \left\{ -i E_{k \phi \ell} {\mathcal Z}_m' \left( k_{\rho \ell} \rho \right)+ \frac{m}{\rho k_{\rho \ell}} E_{k \rho \ell} 
{\mathcal Z}_m \left( k_{\rho \ell} \rho \right) \right\} \unv{\phi} + \nonumber \\
& \ \ \ \ E_{k z \ell} {\mathcal Z}_m \left( k_{\rho \ell} \rho \right) \unv{z} \biggl. \biggr] \ e^{im\phi} \ e^{i k_{0z} z}. \label{6.5}
\end{align} 
In this equation, $\mathcal Z$ is the Bessel function of the first
kind for waves inside the filament, and Hankel function of the first kind for the scattered
waves \cite{A-S}. The former ensures that the wave fields are non-singular inside the
filament, while the latter ensures that the scattered waves are propagating away from the
filament. The summation in $\ell$ is for the two roots of $n_\rho$ that are obtained
from the dispersion relation for the waves either inside the filament or in the background
plasma ($n_{\rho 1} = n_{\rho +}$, $n_{\rho 2} = n_{\rho -}$).
This indicates that the wave fields inside the filament and the scattered waves have to
include both natural modes of the cold plasma. The incoming plane wave, that is excited
by an antenna with its propagation characteristics described by one particular root of the
dispersion relation, can couple power to the other plasma wave in the presence of a
density filament. $\mathcal{E}_{\ell m}$ is the amplitude of the $m$-th Fourier mode of the ${\ell}$-th plasma wave.

\section{Boundary conditions}
\label{sec:7}

At the interface separating the filament from the background plasma,
Maxwell's equations lead to the following boundary conditions \cite{griffiths},
\begin{align}
\unv{\rho} \; \bsdot \; \Bigl( \vec{D}_I + \vec{D}_S \Bigr) \Big|_{\rho=a} \ &= \ \unv{\rho} \; \bsdot \; \vec{D}_F \Big|_{\rho=a}, \label{7.1} \\ 
\unv{\rho} \; \bsdot \; \Bigl( \vec{B}_I + \vec{B}_S \Bigr) \Big|_{\rho=a} \ &= \ \unv{\rho} \; \bsdot \; \vec{B}_F \Big|_{\rho=a},  \label{7.2} \\ 
\unv{\rho} \times \Bigl( \vec{E}_I + \vec{E}_S \Bigr) \Big|_{\rho=a} \ &= \ \unv{\rho} \times \vec{E}_F \Big|_{\rho=a},  \label{7.3} \\ 
\unv{\rho} \times \Bigl( \vec{B}_I + \vec{B}_S \Bigr) \Big|_{\rho=a} \ &= \ \unv{\rho} \times \vec{B}_F \Big|_{\rho=a}. \label{7.4} 
\end{align} 
The subscripts $I$, $S$, and $F$ refer to the incident, scattered, and filamentary wave fields, respectively,
$\vec{D}=\epsilon_0 \; \tensorK \bsdot \vec{E}$ is the wave electric displacement field, $\epsilon_0$ is the
free-space permeability, and $\vec{E}$ 
and $\vec{B}$ are the wave electric and magnetic fields, respectively.
The four sets of boundary conditions follow from Gauss' law, Gauss' magnetism law, Faraday's law, and Ampere's law, respectively.
The left and right sides of Eqs. \eqref{7.1}-\eqref{7.4} are evaluated at the boundary of the filament $\rho=a$. 
The magnetic fields associated with all the waves are obtained from Faraday's equation,
\begin{equation}
\vec{B} \left( \vec{r} \right) \ = \ - \frac{i}{\omega} \; \nabla \times \vec{E} \left( \vec{r} \right). \label{7.5}
\end{equation}
It can be shown that, for a cold plasma dielectric,
only four of the six boundary conditions \eqref{7.1} -- \eqref{7.4} are independent \cite{rk16}. These four boundary
conditions uniquely determine the scattered wave fields and the fields inside the filament for a prescribed incident
plane wave.

The boundary conditions \eqref{7.1} -- \eqref{7.4} have to be satisfied for all $\phi$ and $z$, and for all times $t$. 
Consequently, the $\left( \phi, z, t \right)$ variations of all the
fields must be the same at $\rho = a$. It follows that $k_z$ is preserved in the scattering process. This validates our
earlier assumption that all waves have the same component of the wave vector along the direction of the magnetic field.

\section{Maxwell's stress tensor and the force on a filament}
\label{sec:8}

Apart from the scattering of RF waves by the filament, the RF waves can themselves exert a radiation force
on the filament. In this section, we determine the force on a filament using the Maxwell stress tensor.

The dyadic form of the Maxwell stress tensor in a dielectric medium is \cite{stratton,griffiths},
\begin{equation}
\tensorT \ = \ {\mathbf{E}}_R{\mathbf{D}}_R + {\mathbf{H}}_R{\mathbf{H}}_R - \frac{1}{2} \tensorI \Bigl( 
{\mathbf{E}}_R{\mathbf{D}}_R + {\mathbf{H}}_R{\mathbf{B}}_R \Bigr), \label{8.1}
\end{equation}
where the subscript $R$ indicates the real component of the corresponding field, $\mathbf{B} = \mu_0 \mathbf{H}$, and
$\mu_0$ is the permeability of free space. In terms of our complex field representation,
\begin{equation}
\begin{aligned}
\tensorT \ = \ & \frac{\epsilon_0}{4} \left( \mathbf{E} \tensorK \bsdot \mathbf{E} e^{-2 i \omega t} +
\mathbf{E}^* \tensorK {}^* \bsdot \mathbf{E}^* e^{2 i \omega t} +
\mathbf{E}^* \tensorK \bsdot \mathbf{E} +
\mathbf{E} \tensorK {}^* \bsdot \mathbf{E}^* \right) + \\ 
& \frac{\mu_0}{4} \Bigl( \mathbf{H}\mathbf{H} e^{-2 i \omega t} + \mathbf{H}^* \mathbf{H}^* e^{2 i \omega t} + 
\mathbf{H}^* \mathbf{H} + \mathbf{H} \mathbf{H}^* \Bigr) - \\ 
& \frac{1}{8} \ \tensorI \ \Bigl[ \epsilon_0 \left( 
\mathbf{E} \bsdot \tensorK \bsdot \mathbf{E} e^{-2 i \omega t} +
\mathbf{E}^* \bsdot \tensorK {}^* \mathbf{E}^* e^{2 i \omega t} +
\mathbf{E}^* \bsdot \tensorK \bsdot \mathbf{E} + 
\mathbf{E} \bsdot \tensorK {}^* \bsdot \mathbf{E}^* \right) + \Bigr. \\ 
& \mkern45mu \Bigl. \mu_0 \Bigl( \mathbf{H} \bsdot \mathbf{H} e^{-2 i \omega t} + \mathbf{H}^* \bsdot \mathbf{H}^* e^{2 i \omega t} + 
\mathbf{H}^* \bsdot \mathbf{H} + \mathbf{H} \bsdot \mathbf{H}^* \Bigr) \Bigr],
\end{aligned}
\label{8.2}
\end{equation}
where $^*$ indicates complex conjugate. The time average of \eqref{8.2} over one period of the wave cycle leads to,
\begin{equation}
\left< \tensorT \right> \ = \ \frac{1}{2} \ {\rm Re} \left[ \epsilon_0 \mathbf{E} \tensorK {}^* \bsdot \mathbf{E}^* + 
\mu_0 \mathbf{H} \mathbf{H}^* - \frac{1}{2} \; \tensorI \; \left( \epsilon_0 {\mathbf E} \bsdot \tensorK {}^* \bsdot {\mathbf E}^*
+ \mu_0 \left| {\mathbf H} \right|^2 \right) \right],
\label{8.3}
\end{equation}
where $\rm Re$ indicates the real part of the bracketed quantity.
The time-averaged force on a filament of axial length $L_z$ is,
\begin{equation}
{\pmb{\mathcal F}} \ = \ \int \ \left< \tensorT \right> \bsdot \; d{\mathbf A} \ = \ a \; \int_{0}^{L_z} dz \int_{0}^{2\pi} \; d\phi 
\ \left< \tensorT \right> \bsdot \; \unv{\rho}. \label{8.4}
\end{equation}
where $\mathbf A$ is the surface surrounding the cylindrical filament. The normal to the surface of
the filament is along $\unv{\rho}$. 
The projection of the time-averaged stress tensor along $\unv{\rho}$ is,
\begin{equation}
\begin{aligned}
\left< \tensorT \right> \bsdot \; \unv{\rho} \ = \ & \frac{1}{2} \ {\rm Re} \Bigl[ \epsilon_0 {\mathbf E} \Bigl( K_\rho E_\rho^* +
i K_\phi E_\phi^* \Bigr) + \mu_0 {\mathbf H} H_\rho^* \Bigr] - \\
& \frac{\epsilon_0}{4} \ \Bigl[ K_\rho \left|E_\rho \right|^2 +
K_\rho \left|E_\phi \right|^2 + K_z \left| E_z \right|^2 + 2 K_\phi {\rm Im} \Bigl( E_\rho^* E_\phi \Bigr) \Bigr] \unv{\rho} \ - \ 
\frac{\mu_0}{4} \ \left|{\mathbf H} \right|^2 \unv{\rho}, \label{8.5}
\end{aligned}
\end{equation}
where $\rm Im$ is the imaginary part of the expression within the parentheses. This is the force, per unit area, exerted
on the surface of the filament by the RF waves inside and outside the filament.
On the surface of the filament,
\begin{equation}
\left< \tensorT \left( \rho = a \right) \right> \bsdot \; \unv{\rho} \ = \ \left< \tensorT \left( \rho = a \right) \right>_b  
\bsdot \unv{\rho} \ - \ \left< \tensorT \left( \rho = a \right) \right>_f \bsdot \; \unv{\rho}, \label{8.6}
\end{equation} 
where $\left< \tensorT \right>_b$ and $\left< \tensorT \right>_f$ are the stress tensors corresponding to the total RF
fields in the background plasma and inside the filament, respectively. The negative sign on the right hand side of
\eqref{8.6} follows from the convention that the outward pointing normal at the surface of the filament is positive.
Explicitly, for the background plasma,
\begin{equation}
\begin{aligned}
\left< \tensorT \left( \rho = a \right) \right>_b \bsdot \; \unv{\rho} \ = \ & \frac{1}{2} \ {\rm Re} 
\Bigl[ \epsilon_0  \Bigl( {\mathbf E_I} + {\mathbf E_S} \Bigr) \Bigl\{ K_\rho^B  \Bigl( E_{I\rho}^* + E_{S \rho}^* \Bigr) +
i K_\phi^B \Bigl(  E_{I \phi}^* + E_{S \phi}^* \Bigr) \Bigr\} \Bigr.+ \\
& \mkern45mu  \Bigl. \mu_0 \Bigl( {\mathbf H}_I + {\mathbf H}_S \Bigr)
\Bigl( H_{I \rho}^* + H_{S \rho}^* \Bigr) \Bigr]_{\rho = a}\ - \ \\
& \frac{\epsilon_0}{4} \ \Bigl[ K_\rho^B \; \left| \Bigl( E_{I\rho} + E_{S\rho} \Bigr) \right|^2 +
K_\rho^B \; \left| \Bigl( E_{I \phi} + E_{S \phi} \Bigr) \right|^2 + K_z^B \; \left| \Bigl( E_{I z} + E_{S z} \Bigr) \right|^2 + \Bigr. \\
& \mkern36mu
\Bigl. 2 K_\phi^B \ {\rm Im} \Bigl( E_{I \rho}^* E_{I \phi} +  E_{S \rho}^* E_{S \phi}\Bigr) \Bigr]_{\rho = a} \  \unv{\rho} \ - \ 
\frac{\mu_0}{4} \ \Bigl[ \left| \Bigl( {\mathbf H}_I + {\mathbf H}_S \Bigr) \right|^2 \Bigr]_{\rho = a} \ \unv{\rho}, \label{8.7}
\end{aligned}
\end{equation}
where the right hand side is to be evaluated at $\rho = a$, and $K_\rho^B$, $K_\phi^B$, and $K_z^B$ are components of the plasma
permittivity tensor evaluated for the parameters of the background plasma. Analogously, for the filament,
\begin{equation}
\begin{aligned}
\left< \tensorT \left( \rho = a \right) \right>_f \bsdot \; \unv{\rho} \ = \ & \frac{1}{2} \ {\rm Re} 
\Bigl[ \epsilon_0 {\mathbf E_F} \Bigl( K_\rho^F  E_{F\rho}^*  +
i K_\phi^F E_{F \phi}^* \Bigr) \ + \ \mu_0 {\mathbf H}_F  H_{F \rho}^* \Bigr]_{\rho = a} \ - \\
& \frac{\epsilon_0}{4} \ \Bigl[ K_\rho^F \; \left|E_{F\rho} \right|^2 +
K_\rho^F \; \left|E_{F \phi}  \right|^2 + K_z^F \; \left|E_{F z} \right|^2 + \Bigr. \\
& \mkern36mu
\Bigl. 2 K_\phi^F \ {\rm Im} \Bigl( E_{F \rho}^* E_{F \phi} \Bigr) \Bigr]_{\rho = a} \ \unv{\rho} \ - \ 
\frac{\mu_0}{4} \ \left[ \left|{\mathbf H}_F \right|^2 \right]_{\rho = a} \ \unv{\rho}. \label{8.8}
\end{aligned} 
\end{equation}
The four independent boundary conditions, which follow from \eqref{7.1}--\eqref{7.4}, lead to the following relations,
\begin{equation}
\begin{aligned}
\Bigl[ K_\rho^B \Bigl( E_{I \rho} + E_{S \rho} \Bigr) - i K_\phi^B \Bigl( E_{I \phi} + E_{S \phi} \Bigr) \Bigr]_{\rho = a} \ = \ &
\Bigl[ K_\rho^F E_{F \rho} - i K_\phi^F E_{F \phi} \Bigr]_{\rho = a}, \\
\Bigl[ {\mathbf H}_I + {\mathbf H}_S \Bigr]_{\rho = a} \ = \ &  {\mathbf H}_F \Big|_{\rho = a} , \\
\Bigl[ E_{I \phi} + E_{S \phi} \Bigr]_{\rho = a} \ = \ & E_{F \phi} \Big|_{\rho = a}, \\ 
\Bigl[ E_{I z} + E_{S z} \Bigr]_{\rho = a} \ = \ & E_{F z} \Big|_{\rho = a}. \label{8.9}
\end{aligned} 
\end{equation}
It follows that the three components of $\left< \tensorT \right> \bsdot \;\unv{\rho}$ are,
\begin{align}
\unv{\rho} \; \bsdot \left< \tensorT \left( \rho = a \right) \right> \bsdot \; \unv{\rho} \ = \ & \frac{\epsilon_0}{4} \; \Bigl[ 
K_\rho^B \left|E_{I \rho} + E_{S \rho} \right|^2 - K_\rho^F \left|E_{F \rho} \right|^2 + \Bigr. \nonumber \\
& \mkern30mu \Bigl. \left( K_\rho^F - K_\rho^B \right) \left| E_{F \phi} \right|^2 +
\left( K_z^F - K_z^B \right) \left| E_{F z} \right|^2 \Bigr]_{\rho = a}, \label{8.10} \\
\unv{\phi} \; \bsdot \left< \tensorT \left( \rho = a \right) \right> \bsdot \; \unv{\rho} \  = \ & 0, \label{8.11} \\
\unv{z} \; \bsdot \left< \tensorT \left( \rho = a \right) \right> \bsdot \; \unv{\rho} \  = \ & 0. \label{8.12}
\end{align}
Thus, the net force on the surface of the filament is only in the radial direction. There are no forces in the azimuthal and
axial directions.

Since all the wave fields have the same dependence $e^{ik_zz}$ on the $z$-coordinate, \eqref{8.10} is independent of the
axial length of the filament. From \eqref{8.4}, the force along the radial direction, per unit axial length, is,
\begin{equation}
{\mathcal F}_\rho \ = \ a \; \int_{0}^{2 \pi} \; d\phi \; \unv{\rho} \; \bsdot \left< \tensorT \left( \rho = a \right) \right> \bsdot \; \unv{\rho}. 
\label{8.13}
\end{equation}
The dimensions of ${\mathcal F}_\rho$ are N m$^{-1}$. The Cartesian $x$ and $y$ components of the force are, respectively,
\begin{equation}
\left( \begin{array}{c}{\mathcal F}_x \\ {\mathcal F}_y \end{array} \right) 
 \ = \ a \; \int_{0}^{2 \pi} \; d\phi \; \left( \begin{array}{c} \cos \phi \\ \sin \phi \end{array} \right) 
\unv{\rho} \; \bsdot \left< \tensorT \left( \rho = a \right) \right> \bsdot \; \unv{\rho}. 
\label{8.14}
\end{equation}

\section{Cartesian coordinate representation and normalizations}
\label{sec:9}

We will be displaying our numerical results in the Cartesian coordinate system. The relevant rotation matrix for the
transformation from cylindrical coordinates $\left( \unv{\rho},\; \unv{\phi},\; \unv{z} \right)$ to the Cartesian system
$\left( \unv{x},\; \unv{y},\; \unv{z} \right)$ is, 
\begin{equation}
\tensorR \left( \phi \right) \ = \ 
\left( \begin{array}{ccccc} \cos \phi  & \ \  &  - \sin \phi & \ \  & 0 \\
\sin \phi & \ \  & \cos \phi  & \ \   & 0 \\
0 & \ \   &  0 & \ \   & 1 
\end{array} \right).
\label{9.1}
\end{equation}
For the space spanned by the wave vector $\vec{k}$, the transformation tensor is $\tensorR \left( \phi_k \right)$ with $\phi$ replaced
by $\phi_k$ in \eqref{9.1}.
Thus, the electric field polarizations in \eqref{4.1} transform to the Cartesian system according to,
\begin{equation}
\left( 
\begin{array}{c} 
E_{k x} \\
E_{k y} \\
E_{k z} 
\end{array}
\right)
\ = \ \tensorR \left( \phi_k \right) \ \bsdot \ 
\left(
\begin{array}{c}
E_{k \rho} \\
E_{k \phi} \\
E_{k z }
\end{array}
\right)
\label{9.2}
\end{equation}

The time-averaged Poynting vector for the wave fields is,
\begin{equation}
\Bigl< {\mathbf{S}} \left( t \right) \Bigr> \ = \ \frac{1}{2} {\rm Re} \Bigl( {\mathbf{E}} \times {\mathbf{H}}^* \Bigr). \label{9.3}
\end{equation}
The normalized Poynting vector is defined as,
\begin{equation}
\mathbf{P} \ = \ \dfrac{ \Bigl< {\mathbf{S}} \left( t \right) \Bigr> }{\displaystyle \frac{1}{2} \sqrt{\frac{\epsilon_0}{\mu_0}} 
\left| \mathcal{E}_0^2 \right|}
\ = \ 
\frac{ \Bigl< {\mathbf{S}} \left( t \right) \Bigr> }{S_I},
 \label{9.4}
\end{equation}
where $\left| \mathcal{E}_0  \right|$ is the amplitude of the incident wave field given in \eqref{4.1}, and $S_I$ is the
magnitude of the Poynting vector for the incident field.

The normalized radial force on the surface of the filament by the wave fields is defined as, 
\begin{equation}
F_{\rho} \left( \phi \right) \ = \ c \ \frac{\unv{\rho} \; \bsdot \left< \tensorT \left( \rho = a \right) \right> \bsdot \; \unv{\rho}}
{S_I}. \label{9.5}
\end{equation}

\section{Scattering of lower hybrid waves}
\label{sec:10}

\subsection{Dispersion characteristics}
\label{sec:10A}
 
In Figs. \ref{fig:fig2}, \ref{fig:fig3}, and \ref{fig:fig4}, we illustrate various properties of the
dispersion relation \eqref{3.10} in the lower hybrid range of frequencies. These figures are useful
in limiting the parameter space for exploring the scattering of LH waves by a density filament. If we define
the complex ``wavelength'' with the following notation,
\begin{equation}
\Lambda_{\rho} \ = \ \left( \lambda_\rho, \ \tilde{\lambda}_{\rho} \right), \label{10.1}
\end{equation}
where the two terms in the parenthesis on the right side are the real and imaginary parts, respectively, with
\begin{equation}
\lambda_\rho \ = \ {\rm Re} \left( \frac{c}{\nu n_\rho}  \right), \quad \quad
\tilde{\lambda}_{\rho} \ = \  {\rm Im} \left( \frac{c}{\nu n_\rho} \right),
\label{10.2}
\end{equation} 
then the figures
show the variation in $1/\lambda_\rho$ as a function
of local density, for different $B_0$ (Fig. \ref{fig:fig2}), $n_z$ (Fig. \ref{fig:fig3}), and wave frequency
$\nu = \omega/2\pi$ (Fig. \ref{fig:fig4}). In each figure, the two roots of \eqref{3.10} are indicated by the letter $S$ for
the slow wave root and $F$ for the fast wave root. The paired dispersion curves $(S1, F1)$ in each figure correspond
to the same set of parameters: $B_0 = 4.5$ T, $\nu = 4.6$ GHz, and $n_z = 2$. The plasma is assumed to be
be composed of electrons and deutrons -- we will assume this to be the plasma composition for all our numerical
calculations. The paired dispersion curves $(S2, F2)$, $(S3, F3)$ differ from figure to
figure. 

For the density range shown in the figures, the fast wave is cutoff
below a certain density that depends on $B_0$ and wave parameters. Below the cutoff density, ${\rm Re} ( n_\rho ) = 0$ and
${\rm Im} ( n_\rho ) \ne 0$, indicating the wave is an evanescent mode. The slow wave is a propagating LH wave with
${\rm Im} ( n_\rho ) = 0$. The exception is for the pair $(S2, F2)$. At $n_e \approx 7 \times 10^{19}$ m$^{-3}$, the
roots merge and become complex conjugate pairs with $\big| {\rm Im} ( n_\rho ) \big| \ne 0$ and ${\rm Re} ( n_\rho ) > 0$. The
accessibility of LH waves to higher densities is limited due to the occurrence of this confluence point \cite{stix}. 

The figures \ref{fig:fig2}, \ref{fig:fig3}, and \ref{fig:fig4}
show that, over a significant range of electron densities $\left( \le 5\times 10^{19} \ {\rm m}^{-3} \right)$ expected in the scrape-off layer,
there is not much difference in the radial index of refraction for the LH waves as $n_z$, $B_0$, and $\nu$ are
varied. Consequently, in our numerical studies on the scattering of LH waves, we will use the parameters corresponding
to the dispersion branches $(S1, F1)$ unless stated otherwise; thus, $B_0 = 4.5$ T, $\nu = 4.6$ GHz, and $n_z = 2$.
A useful feature of plotting $1/\lambda_\rho$ is that it is easy
to compare the wavelength of the wave to the radial extent of the filament.

\subsection{Excitation of plasma waves by the filament}
\label{sec:10B}

The physics aspects of wave scattering by a filament can be illustrated by the following example.
We assume that the background plasma density is $2.25 \times 10^{19}$ m$^{-3}$ and the density inside the filament is
$2 \times 10^{19}$ m$^{-3}$, i.e., the filament has depleted density. 
From the curves for $(S1, F1)$ in Fig. \ref{fig:fig2}, the slow and fast waves are propagating normal modes in the background plasma.
However, inside the filament, the slow wave is a propagating normal mode and the fast mode is
evanescent. In Table \ref{table:t1}, we list $n_\rho$, $\Lambda_\rho$, electric
field polarizations, and the real Poynting vector for the two normal modes in the background plasma
and in the filament plasma. 

In the studies that follow, we will assume, without loss of generality, that the incident plane wave 
is propagating in the $x$-$z$ plane in the background plasma. Then, in \eqref{3.1}, 
$\phi_k = 0$, and the relationship between the Cartesian components and the cylindrical components of the wave vector, and of the 
polarization fields, are trivially connected through the transformation \eqref{9.2},
\begin{equation}
n_{0x} \ = \ n_{0 \rho}, \ \ E_{0x}\ = \ E_{0 \rho}, \ \ E_{0y} \ = \ E_{0 \phi}. \label{10.3}
\end{equation}
From \eqref{3.5} and \eqref{4.1}, the
incident plane wave has the following form in Cartesian coordinates,
\begin{align}
\vec{E}_I \ &= \ {\mathcal{E}_0} \ \left(  E_{0x} \; \unv{x} \ + \ E_{0y} \; \unv{y} \ + \
E_{0z} \; \unv{z} \right) \ e^{\displaystyle{i \left( k_{0x}x \; + \; k_{0z}z \right)}}  \nonumber \\
&\equiv \ {\mathcal{E}_0} \ \left( \mathbb{E}_{0x} \; \unv{x} \ + \ \mathbb{E}_{0y} \; \unv{y} \ + \ \mathbb{E}_{0z} \; \unv{z} \right)
 \label{10.4},
\end{align}
where, in the second expression, the exponential phase factor has been included in $\mathbb{E}$. Since the incoming wave is planar,
the physics of scattering is more transparent if the numerical results are displayed in the Cartesian coordinate system.

In the ensuing sections, we use the following notation. The subscripts $\left( x,\; y \right)$ will indicate components
in the Cartesian system, while $\left( \rho,\; \phi \right)$ will be components in the cylindrical system. The subscripts $S$ and
$F$ will be used for slow and fast waves, respectively, and superscripts $b$ and $f$ will indicate background and filament plasmas,
respectively. The subscript $0$ will be used for the incoming plane wave which is initially set up in the background plasma.

\subsubsection{Scattering of a slow lower hybrid plane wave}
\label{sec:10C}
 
Consider a slow LH wave with $n_{0z} = 2$ incident on a filament of radius $a = 1$ cm. From Table \ref{table:t1},
$n_{0x}^b = 14.612$ or $\lambda_{0x}^b = 0.45$ cm. The normalized electric field components are, 
\begin{equation}
\begin{split}
{\rm Re} \left( \mathbb{E}_{0x} \right) \ &= \  0.995\; \cos \left(k_{0x} x + k_{0z} z \right), \\
{\rm Re} \left( \mathbb{E}_{0y} \right) \ &= \ -0.014\; \sin \left(k_{0x} x + k_{0z} z \right), \\
{\rm Re} \left( \mathbb{E}_{0z} \right) \ &= \ 0.098\; \cos \left(k_{0x} x + k_{0z} z \right).
\end{split}
\label{10.5}
\end{equation}
The components of the Poynting vector are,
\begin{equation}
P_{0x} \ = \ -0.092,\ \ \  P_{0y} \ = \ 0, \ \ \  P_{0z} \ = \ 0.996.
\label{10.6}
\end{equation}
The negative sign in $P_{0x}$  affirms that the slow LH wave is a backward wave \cite{stix}.

From Table \ref{table:t1}, we note that the slow wave properties inside and outside the filament are approximately
the same. For this reason, it is to be expected that the incident plasma wave will couple effectively to the slow wave inside the filament.
Meanwhile, the fast wave inside the filament is evanescent with $\tilde{\lambda}_{\rho F}^f   = -12.5$, so that
in the expression for the electric fields inside the filament \eqref{6.5}, 
the argument of the Bessel functions $k_{\rho F}^f \rho = 2 \pi i \rho / 12.5$ is imaginary $\left(\rho \le a \right)$. 
The Bessel function ${\rm J}_m$ of
an imaginary argument is related to the modified Bessel function of the first kind ${\rm I}_m$ \cite{A-S}:
${\rm J}_m \left( k_{\rho F}^f \rho \right) = i^m {\rm I}_m \left( \left| k_{\rho F}^f \right| \rho \right)$. Since ${\rm I}_m$ increases monotonically
as a function of $\rho$, achieving its maximum value at $\rho = a$, we expect an enhancement of
the electric field near the interface if the incident slow wave couples power to the fast wave inside the filament. While the slow wave has
$P_{S \phi}^f = 0$ inside the filament, for the fast wave $P_{F \phi}^f \ne 0$. Thus, any coupling to the
fast wave should result in $P_{F \phi}^f \ne 0$ inside the filament and, as a consequence of the boundary conditions \eqref{7.1}--\eqref{7.4}, in the
surrounding plasma.
 
The numerical solutions resulting from the analytical theory support this simple reasoning. Figure \ref{fig:fig5} shows the real part of the Cartesian 
components of the total electric field $\vec{E}_T \; = \; \vec{E}_I + \vec{E}_S + \vec{E}_F$ normalized to 
$\ \left| \mathcal{E}_0  \right|$. Figures \ref{fig:fig5a} and \ref{fig:fig5c} show the planar wavefronts that 
are slightly distorted by the presence of the filament. As mentioned above, this is to be expected since the properties of the slow wave
inside and outside the filament are approximately the same. Furthermore, since the incoming slow wave has $P_{0y} = 0$ while
$\left( P_{0 x}, \; P_{0 y} \right) \ne 0$, the planar wave fronts exist for ${\rm Re} \left( E_{Tx} \right)$ and ${\rm Re} \left( E_{Tz} \right)$ only.
From \eqref{10.5}, for the incident wave, the maximum amplitudes of
${\rm Re} \left( \mathbb{E}_{0x} \right)$ and ${\rm Re} \left( \mathbb{E}_{0z} \right)$ are $0.995$ and $0.098$, respectively. 
The maximum amplitudes of ${\rm Re} \left( E_{Tx} \right)$ and ${\rm Re} \left( E_{Tz} \right)$
are approximately the same as shown in Figs. \ref{fig:fig5a} and \ref{fig:fig5c}.
The wavefronts in Fig. \ref{fig:fig5b} for the $y$-component of the electric field are, definitely, not planar.
This is an affect of the evanescent fast wave, generated inside the filament, which retains some spatial
aspects of the cylindrical geometry. The figure shows an enhancement of the electric
field near the boundary of the filament in agreement with the discussion in the previous paragraph. 
Additionally, the maximum value of ${\rm Re} \left( E_{Ty} \right)$ 
is much larger than the maximum value of ${\rm Re} \left( \mathbb{E}_{0y} \right)$ -- the larger value of ${\rm Re} \left( E_{k \phi F}^f \right)$
(Table \ref{table:t1}) being the contributing factor. The enhanced fields of the evanescent fast wave near the boundary generate a propagating fast
wave in the background plasma.

The coupling to the fast wave has consequences on the flow of wave energy. Figure \ref{fig:fig6} shows the three Cartesian components of the
Poynting vector $\vec{P}$ for the complete set of electromagnetic fields. 
The power flow in the $y$-direction inside and outside the filament is a direct result of coupling to the fast mode since only
the fast wave inside the filament has a non-zero power flow in the $y$-direction. Figures \ref{fig:fig6a} and \ref{fig:fig6c} show the
diffraction pattern due to scattering in the wake of the filament. 

\subsubsection{Scattering of a fast lower hybrid plane wave}
\label{sec:10D}
 
The scattering is interestingly different if, instead of the slow LH wave,
the incident plane wave is the fast LH wave. From Table \ref{table:t1}, we note that
$\lambda_{0x} / a = 9.25 \gg 1$, i.e., the wavelength of the incident wave is much longer than the radial dimension
of the filament. Consequently, the electric field of the incident plane wave will have a very small spatial variation across the
filament. Since the fast wave is evanescent inside the filament and its polarization, especially the $z$-component,
is quite different from the incident wave, we do not expect a coupling between the fast waves inside and outside the
filament. The boundary conditions \eqref{7.1}--\eqref{7.4} can only be satisfied if we account for the slow wave inside the filament. 
Given that $\lambda_{\rho S}^f / a \approx 0.5$, we expect
two radial wavelengths of the slow wave to fit inside the filament. 
The contribution to \eqref{6.5} from the propagating 
slow wave depends on ${\rm J}_m \left( k_{\rho S}^f \rho \right)$ and ${\rm J}_m' \left( k_{\rho S}^f \rho \right)$ with $k_{\rho S}^f$
being real and $0 \le \rho \le a$. We find that only for $m=0, 1, {\rm and}\ 2$, do the Bessel functions have two ``wavelengths'' inside
the filament (Fig. \ref{fig:fig7}). With this limitation on the azimuthal mode number, we expect the wave fields inside the filament
to be radially structured having a wavelength of $0.47$ cm, and an azimuthal variation corresponding to $m = 1 \; {\rm and} \; 2$. 

Figures \ref{fig:fig8} and \ref{fig:fig9} show the scattering of the incident fast wave by the filament. The generation of the
slow wave inside the filament is evident. The radial variation is as expected as are the $m=1\ {\rm and} \ 2$ azimuthal structures.
The physics of the generation of the slow wave inside the filament is simple. The electric field of
the incident plane wave induces dipole oscillations at the interface of the filament. These oscillations, in turn, generate a propagating
wave inside the filament that is consistent with the geometry and with the electromagnetic boundary conditions. 
The filament behaves like an antenna and excites cylindrical slow waves in the background plasma. 
While the incident plane wave has $P_{0y} = 0$, the scattering leads to Poynting flux in the
$y$-direction. Unlike the case of an incident slow wave where $P_{0y} \ne 0$ due to the presence of an evanescent fast wave inside
the filament, here it is the propagating slow wave inside the filament that leads to $P_{0y} \ne 0$. The incident plane wave
has $k_{0y} = 0$. The cylindrical wavefronts of the slow wave lead to $k_{yS}^f \ne 0$ which, in turn, leads to $P_{y} \ne 0$.
It is worth noting that, over the spatial scales shown in the figures, there is no obvious presence of the incident plane wave -- its wavelength
being over $9$ cm. In Fig. \ref{fig:fig10a}, we display results over an extended range. The planar phase front of the incident wave is
clearly discernable, as is the uniformity of the incident wave field over the filament cross-section. Figure \ref{fig:fig10b} shows that
the filament affects the spatial variation of power flow over a much wider region compared to its cross-section.

\subsubsection{Scattering of a fast lower hybrid plane wave by a filament with smaller radius}
\label{sec:10E}

If the radius of the filament is smaller than the radial wavelength of the slow wave, the electric field structure inside the filament
changes. Even so, the affect on the surrounding plasma is the same as in Fig. \ref{fig:fig8}. The real part of the $x$ and $z$ components of
the total electric field in the presence of a filament of radius $a = 0.4$ cm are shown in Fig. \ref{fig:fig11}. From the
radiation patterns we note that the filament behaves like a dipole antenna \cite{lai}. Inside the filament, the $m = 1$ pattern is
seen in \ref{fig:fig11b} for the ${\rm Re} \left( E_{Tz} \right)$; 
the radial structure follows from the Bessel function of order 1, ${\rm J}_1 \left( k_{\rho S}^f \rho \right)$, with
$0 \ge \rho \ge 0.4$ cm (see Fig. \ref{fig:fig7a}). The azimuthal variation of ${\rm Re} \left( E_{Tx} \right) $ in 
Fig. \ref{fig:fig11a} has the structure of a $m = 2$ mode -- the $x$-component of the field having an extra $\sin \left( \phi \right)$
multiplier when converting from a cylindrical coordinate system to the Cartesian system. 

If we reduce the radius of the filament to $0.25$ cm, the radiation patterns are similar to those shown in Fig. \ref{fig:fig11}. 
This leads to a compelling observation. Even if the wavelength of the incident RF wave is much longer than the radial extent of
the filament -- in this case the ratio is greater than 10 -- the scattered fields are significantly modified by the presence
of the filament. Consequently, it is not justified to ignore the effect of turbulence on RF waves even if the spatial scale length of
the turbulence is much shorter than the RF wavelength.

\subsection{Scattering of an evanescent lower hybrid wave}
\label{sec:10F}

So far we have studied the scattering of waves that propagate in the background plasma. However, evanescent waves can exist in the
low density plasma in the vicinity of an RF source. This is evident from the dispersion relation in Fig. \ref{fig:fig2}. For scattering
of an evanescent wave, we interchange the plasma parmeters for the background and the filament in Table \ref{table:t1}. In this
case, the filament density $2.25 \times 10^{19}$ m$^{-3}$ is larger than the background density $2 \times 10^{19}$ m$^{-3}$. 
An incident fast wave is an evanescent mode for which the electric field amplitude decays with distance. 
Inside the filament, both the slow and fast waves are propagating modes with $\lambda_{\rho S}^f / a < 1$
and $\lambda_{\rho F}^f / a \gg 1$ for $a = 1$ cm. 
From our discussions above, it is reasonable to expect that an incident (evanescent) fast wave
will excite the slow wave inside the filament; essentially no power being coupled to the fast mode as its wave characteristics 
do not match those of the incident wave. Indeed, as shown
in Figures \ref{fig:fig12} and \ref{fig:fig13}, that is exactly what we observe in the simulations. The excitation of the cylindrical slow 
wave inside the filament generates a cylindrical slow wave in the surrounding plasma. The filament, in effect, is instrumental
in coupling power from an evanescent wave to a propagating wave in the background plasma. The incident wave has $P_{0x} = 0$. The
propagating wave excited by the filament has a $P_x \ne 0$ which carries some of the incident power in towards the core plasma.

\subsection{Scattering when the density differential is large }
\label{sec:10G}
 
An analysis of experimental data shows that the density fluctuations in the edge region can be significantly larger than the background density
\cite{zweben1,zweben2}. We will consider such a case and show that the physics of RF scattering follows along the same path discussed above.
In the following, we assume that the electron density inside the filament is $4 \times 10^{19}$ m$^{-3}$, while
the background density is $2 \times 10^{19}$ m$^{-3}$. 
From Fig. \ref{fig:fig2}, we note that both the slow and fast LH waves are propagating modes inside the filament while only the
slow LH wave propagates in the background plasma. The dispersion equation \eqref{3.10} yields,
\begin{alignat}{3} 
&\Lambda_{\rho S}^b \ = \ \left( 0.47,\; 0 \right)\ {\rm cm}, \quad && \Lambda_{\rho F}^b \ = \ \left( 0,\; -12.5 \right)\ {\rm cm}, \label{10g.1}\\
&\Lambda_{\rho S}^f \ = \ \left( 0.358,\; 0 \right)\ {\rm cm}, \quad && \Lambda_{\rho F}^f \ = \ \left( 2.165,\; 0 \right)\ {\rm cm}. \label{10g.2}
\end{alignat} 
 
If the incoming wave is the slow LH wave, then $\lambda_{\rho S}^b$ is comparable to $\lambda_{\rho S}^f$
but much shorter than $\lambda_{\rho F}^f$. This situation is similar to the one discussed in section \ref{sec:10C}. 
Figures \ref{fig:fig14a} and \ref{fig:fig14b} show the contours of ${\rm Re} \left( E_{Tx} \right)$ and ${\rm Re}\left( E_{Ty} \right)$, respectively.
These results bear remarkable resemblance
to the corresponding plots in Fig. \ref{fig:fig5} even though the parameters are very different.

If the incoming LH wave is the evanescent fast wave, we expect both the fast and slow LH waves to be excited inside
the filament. However, from \eqref{10g.2}, for $a = 1$ cm, $\lambda_{\rho F}^f / a > 1$ while $\lambda_{\rho S}^f / a < 1$,
so that the field structure inside
the filament will be dominated by the slow wave. This, in turn, will couple power to the slow propagating wave in the background plasma with
a radial wavelength of $0.47$ cm. The contours of Re$\left( E_{Tx} \right)$ and Re$\left( E_{Ty} \right)$ in
Figs. \ref{fig:fig15a} and \ref{fig:fig15b}, respectively, clearly illustrate this scattering process. 
Whereas the incident wave has no power flow in the $x$-direction, the scattered wave carries power in this direction. The similarity
with Figs. \ref{fig:fig12a} and \ref{fig:fig12b} indicates that the physics of the scattering process is analogous to that in sub-section \ref{sec:10F}. 

\subsection{Scattering of lower hybrid waves with a different frequency}
\label{sec:10H}

For comparison, we assume that all the parameters are the same as in sub-section \ref{sec:10G} except for the wave frequency. For
$\nu = 2.45$ GHz, we find that,
\begin{alignat}{3} 
&\Lambda_{\rho S}^b \ = \ \left( 0.451,\; 0 \right)\ {\rm cm}, \quad \quad && \Lambda_{\rho F}^b \ = \ \left( 4.789,\; 0 \right)\ {\rm cm}, \label{10h.1}\\
&\Lambda_{\rho S}^f \ = \ \left( 0.333,\; 0 \right)\ {\rm cm}, \quad \quad && \Lambda_{\rho F}^f \ = \ \left( 1.989,\; 0 \right)\ {\rm cm}. \label{10h.2}
\end{alignat} 
The main difference, when comparing with Eqs. \eqref{10g.1} and \eqref{10g.2}, is that
the fast wave is a propagating wave at this lower frequency (Fig. \ref{fig:fig4}). Consequently, if the incident
wave is the slow LH wave, we do not expect much difference in the scattering due to the frequency change. We find that the results are similar
to those shown in Figs. \ref{fig:fig14a} and \ref{fig:fig14b}. When the incident wave is the fast LH wave, we can deduce the effect of the
filament from these two conditions: $\lambda_{\rho F}^b/ a  > 4$ and $\lambda_{\rho F}^f/ a  \approx 2$. The first condition implies that the
electric field of the incoming wave is essentially uniform over the cross-section of the filament. The second conditions implies that it
will be difficult to set up a fast wave eigenmode inside the filament. Since $\lambda_{\rho S}^f / a < 1$, the incident wave will 
excite the short wavelength slow wave inside the filament that couples power to the scattered slow wave outside the filament.
It follows that the total field should look essentially the same as in Figs. \ref{fig:fig15a} and \ref{fig:fig15b}. Not surprisingly,
numerical simulations support this intuitive argument. The details of the cylindrical wave patterns are different between
the two frequencies, but the global structure of the scattering is essentially the same.

\section{Scattering of helicon waves}
\label{sec:11}
 
The scattering of LH waves provides a useful base for studying the scattering of lower frequency waves.
Toward this end, we will first consider the scattering of helicon waves and, in the subsequent section, scattering of IC waves.
  
Helicon waves exist in a frequency range that is below the lower hybrid frequency, but well above the ion cyclotron frequency \cite{bers,stix}.
It is believed that helicon waves can efficiently induce plasma currents in toroidal fusion devices, and experiments in DIII-D are being
planned to test this premise \cite{pinsker}. The roots of the cold plasma dispersion relation \eqref{3.10} are
plotted in Fig. \ref{fig:fig16} for DIII-D type parameters that will be used in our study. 
The slow helicon wave has a resonance -- the lower hybrid resonance -- at
$n_e^{LHR} \approx 2.25 \times 10^{19}$ m$^{-3}$. For electron densities higher than $n_e^{LHR}$, the slow wave is evanescent. The fast helicon
wave -- experimentally favored as it can access high plasma densities -- has a cutoff which is a function of $n_z$ \cite{bers}. 
The density at which the cutoff occurs can be deduced from \eqref{3.10}. For our choice of parameters, the cutoff density is
$n_{e}^C \approx 3.5 \times 10^{18}$ m$^{-3}$ (Fig. \ref{fig:fig16}). 

The density range covered in Fig. \ref{fig:fig16} can be divided into four distinct regions. In each region the wave physics and
wave scattering are different. Even though we have carried out
detailed numerical simulations for parameters corresponding to a specific region, the narrative will follow the following rule. For situations that
bear similarities to LH scattering, we will discuss the physics without showing any figures. Otherwise, we will supplement
the discussion with graphs. In all the cases we will assume that the incident plane wave is the fast helicon wave.

\noindent
{\underline {\it Case 1: $n_e \le n_e^C$}} $\quad$
For densities below $n_e^C$, the incident fast wave is evanescent.
The propagating slow wave has $\lambda_{\rho S}^{b,f} \ga 0.424$ cm in the density range $\left[ 10^{17}, 3.5 \times 10^{18} \right]$ m$^{-3}$. 
In the filament and the background plasma, the slow wave will be excited by coupling of waves at the boundary.
The wave field structure inside the filament will depend on the argument 
$2 \pi \rho / \lambda_{\rho S}^f$ of the Bessel functions. If $a / \lambda_{\rho S}^f > 1$,
the field patterns will be similar to those in Figs. \ref{fig:fig8} and \ref{fig:fig12},
otherwise the fields will be of the form shown in Fig. \ref{fig:fig11}. In either case, there will be an outflow of power due to the scattered
slow cylindrical waves propagating in the background plasma.

\noindent
{\underline {\it Case 2: $n_e^C < n_e < n_e^{LHR}$}} $\quad$ For these densities the fast and the slow helicon waves are propagating modes with 
$\lambda_{\rho F}^{b,f} > 2.53$ cm while $0 < \lambda_{rho S}^{b,f} \la 0.424$ cm. It should be noted that, in the vicinity of the
lower hybrid resonance, the cold plasma model for the dispersion relation breaks down -- thermal effects need to be included in
the wave description \cite{bers}. For filaments with $a \la 1$ cm, $\lambda_{\rho F}^{b,f} > a$ while for the slow waves $\lambda_{\rho S}^{b,f} \la  a$.
The electric fields inside the filament are primarily those of the slow wave, and the total fields are similar to those
in Figs. \ref{fig:fig8} and \ref{fig:fig10}. The number of cylindrical wavelengths inside the filament depends on the density.
For example, for $n_e^f = 1.5 \times 10^{19}$ m$^{-3}$, $1/\lambda_{\rho S}^f \approx 7.74$ cm$^{-1}$, and the argument of the Bessel
functions is $\approx 48.63 a$. For $a = 1$ cm, the number of oscillations of the Bessel functions of low order is about $7$.
The results from full numerical simulations confirm this estimate.

\noindent  
{\underline {\it Case 3: densities in the vicinity of the lower hybrid resonance}} $\quad$  
If the background density is between $n_e^C$ and $n_e^{LHR}$ and
the density inside the filament is $> n_e^{LHR}$, then the fast and slow waves are propagating modes in the background plasma, while
only the fast wave is a propagating mode inside the filament, with the slow wave being an evanescent mode. For an incident fast wave, we do not
expect any slow cylindrical waves being generated inside the filament. However, the evanescent wave can still couple power to the propagating slow
wave outside the filament. Consider the following example where the background plasma density is $10^{19}$ m$^{-3}$ and the density inside
the filament is $3 \times 10^{19}$ m$^{-3}$. Then,
\begin{alignat}{3}
& \Lambda_{\rho F}^b \ = \ \left( 6.02, \; 0 \right) \ {\rm cm}, \quad \quad && \Lambda_{\rho S}^b \ = \ \left( 0.204, \; 0 \right) \ {\rm cm}, 
\label{11.1} \\
&\Lambda_{\rho F}^f \ = \ \left( 1.88, \; 0 \right) \ {\rm cm}, \quad \quad && \Lambda_{\rho S}^f \ = \ \left( 0, \; -0.09 \right) \ {\rm cm}. 
\label{11.2}
\end{alignat}
Inside the filament, the evanescent slow mode is localized close to the surface since the
real (imaginary) part of even order (odd order) Bessel functions for imaginary $\Lambda_{\rho S}^f$ in \eqref{11.2}
peak near $\rho \approx a$. 
For $a = 1$ cm, the wavelength of the fast wave, inside and outside the filament, is longer than the radius of the filament.
Thus, the fast wave fields inside the filament will have very weak spatial variation. These expectations, based on the previous results for LH
waves, are borne out by numerical simulations.
The contours of Re$\left( E_{Tx} \right)$ in Fig. \ref{fig:fig17} show the uniformity of fields inside the filament, the narrow region 
of enhanced fields near the boundary, and the cylindrical slow wave propagating away from the filament.
The scattered slow wave leads to side-scattering of the incoming fast wave and affects the power  
flow into the core.

\noindent 
{\underline {\it Case 4: $n_e > n_e^{LHR}$}} $\quad$ The only propagating mode is the fast wave  with $\lambda_{\rho F}^{b,f} < 2.53$ cm. 
The electric field due to the evanescent slow wave will be localized near the surface of the filament given by the maxima of the Bessel 
functions and the Hankel functions of the first kind for complex argument. We expect that there will be some resemblance
to the results shown in Fig. \ref{fig:fig5}. For illustrative purposes, we assume the background plasma density and the density inside
the filament to be $3 \times 10^{19}$ m$^{-3}$ and $5 \times 10^{19}$ m$^{-3}$, respectively. The corresponding (complex) wavelengths are,
\begin{alignat}{3}
& \Lambda_{\rho F}^b \ = \ \left( 1.88, \; 0 \right) \ {\rm cm}, \quad \quad && \Lambda_{\rho S}^b \ = \ \left( 0, \; -0.09 \right) \ {\rm cm}, 
\label{11.3} \\
&\Lambda_{\rho F}^f \ = \ \left( 1.12, \; 0 \right) \ {\rm cm}, \quad \quad && \Lambda_{\rho S}^f \ = \ \left( 0, \; -0.14 \right) \ {\rm cm}. 
\label{11.4}
\end{alignat} 
Figure \ref{fig:fig18} shows contours of Re$\left( E_{Tx} \right)$ with $a = 1$ cm -- comparable to the 
wavelength of the fast wave. The planar phase front of the incoming
wave gets distorted by the filament, with a shadow in its wake. The presence of large amplitude slow wave fields near the boundary is evident
even though we have suppressed the maximum amplitude in order to display the rest of the field pattern. 

\section{Scattering of ion cyclotron waves}
\label{sec:12}

A preferred means of heating toroidal plasma is by RF waves in the ion cyclotron range of frequencies; in particular, the fast Alfv\'en wave (FAW).
The cold plasma dispersion characteristics of these waves are plotted in Fig. \ref{fig:fig19} in the density range appropriate for the edge
plasma in a SPARC-like, high magnetic field tokamak \cite{lin}. Just like helicon waves, the slow IC wave propagates for densities less than 
$n_{e}^{LHR} \approx 4.27 \times 10^{17}$ m$^{-3}$ (Fig. \ref{fig:fig19}). For larger densities, the slow wave is evanescent.
The fast wave is evanescent until its cutoff density $n_{e}^C \approx 2.16 \times 10^{19}$ m$^{-3}$, after which it becomes a propagating wave.
Unlike helicon waves, the slow and fast waves propagate in distinctly different density regimes. In the discussion that follows,
we will assume that the incoming plane wave is the fast Alfv\'en wave. 

From the result for helicon waves, we can deduce that, for densities below $n_e^{LHR}$, the FAW wave will couple to the slow wave 
inside the filament; thereby exciting slow cylindrical waves in the background plasma. In the density range $\left[ n_e^{LHR}, \; n_e^C \right]$
both wave modes are evanescent. The Bessel functions
of imaginary argument, for the two modes inside the filament, peak near $\rho \approx a$. The Hankel functions of the first kind of an
imaginary argument ensure that the scattered wave fields decay away for $\rho > a$.
Consequently, the wave fields will peak in the vicinity of the boundary of the filament.

For densities greater than $n_e^C$, the FAW is a long wavelength mode with $a/ \lambda_{\rho F}^{b,f} \ll 1$ for any reasonable value of $a$.
Accordingly, we do not expect any wave like features, associated with the slow wave, inside the filament. The electromagnetic fields
will essentially be uniform over the cross-section except in the vicinity of the boundary where the fields have to match on to the incident and scattered fields.
In this density regime, there is no analogous situation for LH and helicon waves.
As an example, consider a filament with density $7 \times 10^{19}$ m$^{-3}$ surrounded by a plasma with density $4 \times 10^{19}$ m$^{-3}$. The
corresponding wavelengths are,
\begin{alignat}{3}
& \Lambda_{\rho F}^b \ = \ \left( 30.6, \; 0 \right) \ {\rm cm}, \quad \quad && \Lambda_{\rho S}^b \ = \ \left( 0, \; -0.45 \right) \ {\rm cm}, 
\label{12.1} \\
&\Lambda_{\rho F}^f \ = \ \left( 18.2, \; 0 \right) \ {\rm cm}, \quad \quad && \Lambda_{\rho S}^f \ = \ \left( 0, \; -0.36 \right) \ {\rm cm}. 
\label{12.2}
\end{alignat}
For a FAW incident on a filament with $a = 1$ cm, $a/\lambda_{\rho F}^{b,f} \ll 1$ and the fields inside the filament can be expected to be
uniform over the cross-section. Since the FAW is a propagating mode inside the filament, it cannot screen out the fields of 
the incident FAW. Thus, in the vicinity of $\rho \approx 0$ we expect the field amplitudes to be non-zero.
In Eq. \eqref{6.5} for wave fields inside the filament, 
only the $m = \pm 1$ azimuthal modes, corresponding to ${\rm J}'_{\pm 1} \left( k_{\rho F}^f a \right)$, are non-zero.
The $m = \pm 1$ feature is evident in the result from numerical simulations shown in Fig. \ref{fig:fig20a}.
For helicon waves, the presence of a similar feature is discernable in Fig. \ref{fig:fig17}.
The large amplitude electric fields inside the filament are responsible for an enhanced
flow of Poynting flux in the $z$-direction as shown in Fig \ref{fig:fig20b}. For comparison, the Poynting flux in $z$-direction
for the incoming FAW is $0.88$. Intuitively, one would expect the filament to play an insignificant role in the scattering process
since the wavelength of the incoming wave is much larger than the radial extent of the filament. However, the results show otherwise. A fraction
of the incident wave power will be converted to flow down the axis of the filament along the magnetic field line.

\section{Maxwell stress tensor -- radial force on a filament}
\label{sec:13}

In 1905, Poynting studied the radiation pressure of light at an interface separating two different dielectric media described
by scalar permittivities \cite{poynt,loudon}. 
He came to an interesting conclusion, which he followed up with supporting experiments. Poynting stated the following: ``In any real refraction
with ordinary light, there will be reflexion as well as refraction. The reflexion always produces a normal pressure, and the refraction a normal
pull. But with unpolarized light, a calculation shows that the refraction pull, for glass at any rate, is always greater than the
reflexion push, even at grazing incidence.'' The ``pull'' and ``push'' are defined with respect to the direction of propagation
of the incident light -- the pull being opposite to this direction while the push being along this direction. Thus, according to Poynting,
the incident light will pull the glass along the outward pointing normal to the interface. Poynting's research was a contributing factor
to the Minkowski-Abraham controversy regarding the definition of electromagnetic momentum within dielectric media \cite{loudon}.
The permittivity of a magnetized plasma is a tensor rather than a scalar. Consequently, as we have already noted,
not only is there a wider variety of waves that can exist in a plasma, compared to a scalar dielectric, but also the filament induces
coupling between these different waves. Following the Maxwell stress tensor formulation in section \ref{sec:8}, 
we will examine the radiation force on a filament by different plasma waves,
discussed in sections \ref{sec:10}, \ref{sec:11}, and \ref{sec:12}, and compare with
Poynting's observations. 

In order to have a meaningful quantitive measure, we will calculate the acceleration of a representative filament
due to the radiation force. From \eqref{8.14} and \eqref{9.5}, the force on a filament per unit incident power flux is defined as,
\begin{equation}
\left( \begin{array}{c}{\widetilde{\mathcal F}}_x \\ {\widetilde{\mathcal F}}_y \end{array} \right) 
 \ = \  \frac{1}{S_I} \; \left( \begin{array}{c}{\mathcal F}_x \\ {\mathcal F}_y \end{array} \right) \ = \ 
\frac{a}{c} \; \int_{0}^{2 \pi} \; d\phi \; \left( \begin{array}{c} \cos \phi \\ \sin \phi \end{array} \right) 
F_{\rho} \left( \phi \right) \ \ \ \ \frac{{\rm N \; m}^{-1}}{{\rm W \; m}^{-2}},
\label{13.1}
\end{equation}
where we have assumed that the filament has an axial length of $1$ m.
It turns out that, for all the scattering events we have studied, ${\mathcal F}_y \; = \; 0$. Various attempts at proving this result analytically
have not been successful; so we leave this as an exercise for the future. For ${\mathcal F}_x < 0$ the force is towards the RF source (``pull''),
while for ${\mathcal F}_x > 0$ the force is away from the source (``push''). The acceleration of the filament in the $x$-direction is,
\begin{equation}
{\mathfrak a}_x \ = \ \frac{\widetilde{\mathcal F}_x}{\mathfrak m}, \label{13.2}
\end{equation}
where ${\mathfrak m}$ is the mass of the filament. As in the scattering studies, a filament will have deutrons and electrons only,
so that,
${\mathfrak m} \ = \ 1.05 \times 10^{-26} \; a^2 \; n_e^f$ kg. Unless stated otherwise, in all the subsequent calculations we assume that
$a = 1$ cm, and the input RF power flux is $1$ kW\; m$^{-2}$. The primary plasma parameters --
magnetic field, wave frequency, and $n_z$ -- for the LH, helicon, and IC waves are as in figures \ref{fig:fig2} 
$\left( S_1,\, F_1 \right)$, \ref{fig:fig16}, and \ref{fig:fig19}, respectively. 

\subsection{Radiation force due to lower hybrid waves -- variation with  filament radius}
\label{sec:13A}
 
In sections \ref{sec:10C} and \ref{sec:10D} we discussed the scattering of slow and fast LH waves, respectively, by a filament.
The radial force on the filament $F_\rho$, given in \eqref{9.5}, is plotted as a function of the azimuthal angle $\phi$ for the slow wave
in Fig. \ref{fig:fig21a} and for the fast wave in Fig. \ref{fig:fig21b}. The difference
between the two results can be traced to the corresponding field plots in Figs. \ref{fig:fig5} and \ref{fig:fig8}. For the slow
wave, the fields at the surface of the filament have a wide range of variations in the azimuthal direction. The incoming
wave is distorted around the surface of the filament due to geometrical mismatch between the planar wave and the cylindrical
filament. For the fast wave the variation is almost sinusoidal as cylindrical waves generated within the filament dominate the
field pattern. The integration of these profiles over $\phi$ yields the force on the filament per unit power flow of
the incident wave,
\begin{equation}
\left. \widetilde{\mathcal F}_x \right|_{S,LH} \ = \ 4.23 \ \ \frac{{\rm nN \; m}^{-1}}{{\rm kW \; m}^{-2}}, \quad \quad 
\left. \widetilde{\mathcal F}_x \right|_{F,LH} \ = \ -56.13 \ \ \frac{{\rm nN \; m}^{-1}}{{\rm kW \; m}^{-2}}, 
\label{13.3}
\end{equation} 
for the slow and fast waves, respectively. The slow wave is ``pushing'' away the filament while the fast wave is ``pulling'' 
it in towards the RF source. The filament density assumed in the simulations $n_e^f = 2 \times 10^{19}$ m$^{-3}$ 
yields ${\mathfrak m} = 2.1 \times 10^{-11}$ kg. The corresponding accelerations are,
\begin{equation}
{\mathfrak a}_x \Big|_{S,LH} \ = \ 2.0 \times 10^2 \ \ {\rm m\; sec}^{-2}, \quad \quad 
{\mathfrak a}_x \Big|_{F,LH} \ = \ -2.7 \times 10^3 \ \ {\rm m\; sec}^{-2}.  
\label{13.4}
\end{equation} 
Even though the dynamics of a filament through a background plasma requires more physics, the acceleration gives a
measure of the effect of RF waves on the filament. Clearly, the direction of the force will play a role in the motion
of the filament. 

The scattering by a filament of smaller radius, $a < 1$ cm, was discussed in \ref{sec:10E}. For two different radii, the
acceleration due to the slow and fast LH waves is,
\begin{alignat}{3}
&{\rm for} \ a = 0.4 \ {\rm cm}: \quad & {\mathfrak a}_x \Big|_{S,LH} \ = \ 4.7 \times 10^2 \ \ {\rm m\; sec}^{-2}, \quad &
{\mathfrak a}_x \Big|_{F,LH} \ = \ -5.5 \times 10^3 \ \ {\rm m\; sec}^{-2}, \nonumber \\
&{\rm for} \ a = 0.25 \ {\rm cm}: \quad & {\mathfrak a}_x \Big|_{S,LH} \ = \ 7.5 \times 10^2 \ \ {\rm m\; sec}^{-2}, \quad & 
{\mathfrak a}_x \Big|_{F,LH} \ = \ -1.3 \times 10^4 \ \ {\rm m\; sec}^{-2}. 
\label{13.5}
\end{alignat} 
Comparing with \eqref{13.4}, we note that the acceleration increases as the radial dimension of the filament decreases while
preserving the direction of the force. 

\subsection{Radiation force due to propagating slow waves -- the push-pull effect}
\label{sec:13B}
 
In this subsection, we compare the radiation force due to the slow wave on a filament with density $n_{e}^f \, < \, n_{e}^b$ to 
that when $n_{e}^f \, > \, n_{e}^b$, for the LH, helicon, and IC waves. For LH waves the slow
wave is preferred in experiments, while for helicon and IC waves it is the fast wave. Nonetheless, we consider
the slow wave in all three frequency ranges to determine if there is any dependence of the radiation force on frequency.

\subsubsection{Lower hybrid waves}
\label{sec13BA} 
In section \ref{sec:13A}, the densities were such that $n_{e}^f < n_{e}^b$. 
If we only change the filament density to $n_{e}^f \, = \, 2.5 \times 10^{19} {\rm m}^{-3}$, then $n_{e}^f > \, n_{e}^b$ and
the acceleration due to the slow wave is,
\begin{equation}
{\mathfrak a}_x \Big|_{S,LH} \ = \ -1.6 \times 10^2 \ \ {\rm m\; sec}^{-2}.  
\label{13.6}
\end{equation}  
Comparing with the results in \eqref{13.4}, there is a change in sign of the force. This reversal of the ``push--pull'' effect 
was initially recognized by Poynting \cite{loudon}. In his formulation, for normal incidence on a planar surface
separating two different dielectric media, the direction of the force depended on the relative refractive indices of the media.
For light incident from a region of lower refractive index, the force was towards the region of lower refractive index. 
For light incident from a region of higher refractive index, the force was still towards the region of lower refractive index.
While the scattering from a filament is different from a planar interface, the reversal in sign is intriguing.
For the results in \eqref{13.4}, the wave is propagating from a region of higher refractive index (higher plasma density)
to a filament with lower refractive index (lower plasma density). For \eqref{13.6} the incident wave is in a region with
lower refractive index. The direction of the radiation force due to an incident slow wave follows Poynting's observations.
The force due to the slow wave is such that higher density (relative to the background density) filaments are pulled in
towards the RF source, while the lower density filaments are pushed away. The effect of the radiation force is to
create a density inversion in the vicinity of the source.

\subsubsection{Helicon waves}
\label{sec13BB}
 
For helicon waves, the slow mode is a propagating wave for $n_e^{b,f} < n_e^{LHR} \approx 2.25 \times 10^{19}$ m$^{-3}$ 
(Fig. \ref{fig:fig16}). For an ambient
density $n_e^{b} = 5 \times 10^{17}$ m$^{-3}$, assume two different filaments with densities $n_e^{f1} = 7 \times 10^{17}$ m$^{-3}$
and $n_e^{f2} = 3.75 \times 10^{17}$ m$^{-3}$, so that $n_e^{f1}  > n_e^{b}$ and $n_e^{f2} < n_e^b$. These densities are also below
the cutoff density of the fast wave. The respective accelerations
are,
\begin{equation}
\begin{aligned}
&{\rm for} \ n_e^{f1} > n_e^{b} : \quad & {\mathfrak a}_x \Big|_{S,H} \ &= \ -1.2 \times 10^4 \ \ {\rm m\; sec}^{-2}, \\
&{\rm for} \ n_e^{f2} < n_e^{b} : \quad & {\mathfrak a}_x \Big|_{S,H} \ &= \ 1.4 \times 10^4 \ \ {\rm m\; sec}^{-2}.  
\label{13.7}
\end{aligned}
\end{equation} 
The radiation force due to slow helicon waves has a push-pull behavior similar to that due to slow LH waves.

\subsubsection{Ion cyclotron waves}
\label{sec13BC}

The slow wave in the IC frequency range propagates for densities $n_e < n_e^{LHR} \approx 4.27 \times 10^{17}$ m$^{-3}$ 
(Fig. \ref{fig:fig19}). For $n_e^{b} = 2 \times 10^{17}$ m$^{-3}$, $n_e^{f1} = 3 \times 10^{17}$ m$^{-3}$,
and $n_e^{f2} = 1.5 \times 10^{17}$ m$^{-3}$, we have that $n_e^{f1}  > n_e^{b}$ and $n_e^{f2} < n_e^b$. The acceleration in each case is,
\begin{equation}
\begin{aligned}
&{\rm for} \ n_e^{f1} > n_e^{b} : \quad & {\mathfrak a}_x \Big|_{S,IC} \ &= \ -4.1 \times 10^4 \ \ {\rm m\; sec}^{-2}, \\
&{\rm for} \ n_e^{f2} < n_e^{b} : \quad & {\mathfrak a}_x \Big|_{S,IC} \ &= \ 1.2 \times 10^4 \ \ {\rm m\; sec}^{-2}. 
\label{13.8}
\end{aligned}
\end{equation}  
The acceleration due to slow IC waves is comparable to that due to the slow helicon waves. 
These results indicate that the tendency of the slow wave, for all three frequencies, is to pull in the 
higher density filaments towards the RF source
and push away the lower density filaments. 

\subsection{Radiation force due to fast waves}
\label{sec:13C}

For helicon and ion cyclotron waves, the preferred mode of propagation is the fast wave as it can access higher
plasma densities. In the low density region in the vicinity of an antenna, the fast wave is evanescent as
seen in Figs. \ref{fig:fig16} and \ref{fig:fig19}. The scattering studies have shown that the evanescent and the propagating
fast waves are affected by the presence of the shorter wavelength slow wave excited inside the filament. In this
subsection, we will examine, and compare, the radiation forces induced by the fast helicon wave and the FAW  in different
density regimes defined by $n_e^{LHR}$ and $n_e^{c}$.

\subsubsection{Low density region: \  $n_e^{b, f} \;  < \; n_e^{LHR},\; n_e^C$}
\label{sec:13CA}
 
The fast helicon wave (Fig. \ref{fig:fig16}) and the FAW (Fig. \ref{fig:fig19}) are
evanescent waves in this density regime. Only the slow waves are propagating modes.
For helicon waves, we assume $n_e^{b} = 5 \times 10^{17}$ m$^{-3}$ and two different filament densities, 
$n_e^{f1} = 7 \times 10^{17}$ m$^{-3}$ and $n_e^{f2} = 3.75 \times 10^{17}$ m$^{-3}$. 
The acceleration resulting from the radiation force of the fast helicon wave is, 
\begin{equation}
\begin{aligned}
&{\rm for} \ n_e^{f1} > n_e^{b} : \quad & {\mathfrak a}_x \Big|_{F,H} \ &= \ 8.6 \times 10^4 \ \ {\rm m\; sec}^{-2}, \\
&{\rm for} \ n_e^{f2} < n_e^{b} : \quad & {\mathfrak a}_x \Big|_{F,H} \ &= \ -1.6 \times 10^5 \ \ {\rm m\; sec}^{-2}. 
\label{13.8}
\end{aligned}
\end{equation} 
The higher density filament is being pushed away from the RF source by the evanescent fast helicon wave while the 
lower density filament is being pulled in. The radiation force is acting in such a way as to reduce the density
in front of the RF source. It is noteworthy that, in this case, the affect of the radiation force is opposite to Poynting's observations.
The contrast reinforces the differences between electromagnetic wave propagation in scalar dielectrics and in plasmas.

For IC waves, we assume $n_e^{b} = 2 \times 10^{17}$ m$^{-3}$ and two different filament densities,
$n_e^{f1} = 3 \times 10^{17}$ m$^{-3}$ and $n_e^{f2} = 1.5 \times 10^{17}$ m$^{-3}$. 
The acceleration induced by the FAW is,
\begin{equation}
\begin{aligned}
&{\rm for} \ n_e^{f1} > n_e^{b} : \quad & {\mathfrak a}_x \Big|_{F,IC} \ &= \ -4.3 \times 10^5 \ \ {\rm m\; sec}^{-2}, \\
&{\rm for} \ n_e^{f2} < n_e^{b} : \quad & {\mathfrak a}_x \Big|_{F,IC} \ &= \ 6.9 \times 10^5 \ \ {\rm m\; sec}^{-2}. 
\label{13.9}
\end{aligned}
\end{equation}  
Comparing with \eqref{13.8}, the radiation force due to the FAW has the opposite effect. The waves tend to
increase the density in the vicinity of the RF source which, in turn, could affect the coupling of the fast Alfv\'en wave
to the core plasma.

\subsubsection{Medium density region: \  $n_e^C \; < \; n_e^{b, f} \;  <  \; n_e^{LHR}$}
\label{sec:13CB}
 
This density regime, in which both the slow and fast waves are propagating modes, is applicable to helicon waves only.
For $n_e^{b} = 9 \times 10^{18}$ m$^{-3}$,  we consider the filament densities, 
$n_e^{f1} = 1.2 \times 10^{19}$ m$^{-3}$ and $n_e^{f2} = 6.0 \times 10^{18}$ m$^{-3}$.  
The resulting acceleration due to the fast helicon wave is, 
\begin{equation}
\begin{aligned}
&{\rm for} \ n_e^{f1} > n_e^{b} : \quad & {\mathfrak a}_x \Big|_{F,H} \ &= \ -3.9 \times 10^6 \ \ {\rm m\; sec}^{-2}, \\
&{\rm for} \ n_e^{f2} < n_e^{b} : \quad & {\mathfrak a}_x \Big|_{F,H} \ &= \ -1.0 \times 10^6 \ \ {\rm m\; sec}^{-2}. 
\label{13.10}
\end{aligned}
\end{equation}  
This result is quite different from all the previous cases we have considered; the radiation pressure, regardless of the
density of the filament relative to background, pulls in the filament towards the RF source. The force on either filament 
is comparably large. The presence
of a propagating slow wave has a definite role in the direction of the force as well as its magnitude.

\subsubsection{Medium density region: \  $n_e^{LHR} \; < \; n_e^{b, f} \;  <  \; n_e^C$}
\label{sec:13CC}

In this density regime, which applies to IC waves only, the slow and the fast waves are evanescent. 
Choosing $n_e^{b} = 3 \times 10^{18}$ m$^{-3}$,  we consider two different filament densities, 
$n_e^{f1} = 3.9 \times 10^{19}$ m$^{-3}$ and $n_e^{f2} = 2.1 \times 10^{18}$ m$^{-3}$. The radiation force due to
the FAW is,
\begin{equation}
\begin{aligned}
&{\rm for} \ n_e^{f1} > n_e^{b} : \quad & {\mathfrak a}_x \Big|_{F,IC} \ &= \ -1.6 \times 10^4 \ \ {\rm m\; sec}^{-2}, \\
&{\rm for} \ n_e^{f2} < n_e^{b} : \quad & {\mathfrak a}_x \Big|_{F,IC} \ &= \ -1.3 \times 10^5 \ \ {\rm m\; sec}^{-2}. 
\label{13.11}
\end{aligned}
\end{equation}  
The radiation force has the same characteristics as for the helicon wave in \eqref{13.10}, although the magnitude
of the acceleration is smaller.

\subsubsection{High density region: \  $ n_e^{b, f} \; >  \; n_e^C, \; n_e^{LHR}$}
\label{sec:13CD}

For these densities, the fast wave is a propagating mode for the helicon and IC waves; the slow wave is evanescent.
For  $n_e^{b} = 3 \times 10^{19}$ m$^{-3}$,  and two different filament densities, 
$n_e^{f1} = 3.6 \times 10^{19}$ m$^{-3}$ and $n_e^{f2} = 2.4 \times 10^{18}$ m$^{-3}$, the acceleration due to the helicon
wave is,
\begin{equation}
\begin{aligned}
&{\rm for} \ n_e^{f1} > n_e^{b} : \quad & {\mathfrak a}_x \Big|_{F,H} \ &= \ -2.3 \times 10^5 \ \ {\rm m\; sec}^{-2}, \\
&{\rm for} \ n_e^{f2} < n_e^{b} : \quad & {\mathfrak a}_x \Big|_{F,H} \ &= \ -9.9 \times 10^5 \ \ {\rm m\; sec}^{-2},  
\label{13.12}
\end{aligned}
\end{equation}  
and, the acceleration induced by the FAW is,
\begin{equation}
\begin{aligned}
&{\rm for} \ n_e^{f1} > n_e^{b} : \quad & {\mathfrak a}_x \Big|_{F,IC} \ &= \ 1.9 \times 10^5 \ \ {\rm m\; sec}^{-2}, \\
&{\rm for} \ n_e^{f2} < n_e^{b} : \quad & {\mathfrak a}_x \Big|_{F,IC} \ &= \ -2.4 \times 10^5 \ \ {\rm m\; sec}^{-2}.
\label{13.13}
\end{aligned}
\end{equation} 
The fast helicon wave pulls in lower and higher density filaments similar to the way it did in the medium density regime \eqref{13.10}.
However, the fast Alfv\'en wave pushes out the higher density filament while pulling in the lower density filament. Again, this
will affect the coupling of FAWs to a fusion plasma.

\section{Conclusions}
\label{conc}

There are two major parts of this paper. The first part is on the effect of a filament, present in an ambient
plasma, on the propagation of RF waves. The second part is on the effect of RF waves on the filament.

In the first part of this paper, we have developed a physical intuition for 
scattering of RF waves by a filament present in the edge plasma of a fusion device. In the frequency regime below the electron cyclotron
frequency, the characteristics of RF waves do not vary significantly as a function of density and magnetic field. A rudimentary analysis
of the dispersion relation provides ample information about the scattering process. For cold plasma, there are two modes of propagation -- a
slow wave and a fast wave. In densities typical of edge plasmas, either one or both of these waves can be evanescent over a range of
densities. The evanescent waves lead to an enhancement of the electric field amplitude in the vicinity of the surface of the
filament. For propagating waves, a relevant parameter to consider is the ratio of the radius of the filament to the wavelength of the
electromagnetic wave. Let this parameter be $\Gamma$.  A wave with $\Gamma > 1$ inside and outside the filament, is slightly modified by
the scattering process. There is some side scattering along with spatial fragmentation of the scattered power in the forward direction.
If an incident wave has $\Gamma << 1$, but the filament allows for a mode with $\Gamma > 1$, then the filament behaves like a dipole antenna
and excites this mode within and outside the filament. This occurs, for example, when a long wavelength fast wave is incident on a filament 
in which the short wavelength slow wave can propagate. However, if inside the filament the slow wave is evanescent then the fields amplify
near the surface and direct some of the incident power along the axis. Thus, the effect of scattering can be inferred by two elements; whether
the cold plasma modes are propagating or evanescent inside and outside the filament, and whether $\Gamma > 1$ or $\Gamma < 1$ inside and outside
the filament.

The second part of the paper is on the radiation force due to RF waves. We have evaluated the
acceleration induced by the radiation force on a filament using the Maxwell stress tensor. The evaluation of the stress tensor makes use
of the same theory developed in the first section on scattering. In the Cartesian coordinate system where the wave vector of
the incident wave is in the $x-z$ plane, the net radiation force on the filament in the $y$-direction is zero for all the cases we have
considered. The radiation force is in the $x$-direction only -- a positive
force, or acceleration, pushing the filament towards the core, while a negative force pulling the filament towards the RF source. 
We find that the slow LH wave pulls in higher density filaments and pushes away lower density filaments; higher and lower densities
being relative to the background density. Thus, the density tends to decrease as a function of the radial distance from the antenna.
The radiation force due to the fast helicon and the fast Alfv\'en waves varies as a function of the plasma density in the edge region.
For low densities, below the densities for lower hybrid resonance and fast wave cutoff, the helicon wave pushes away higher density
filaments and pulls in lower density filaments. The tendency of the helicon wave is to lower the density in front of the source. The
radiation force of the fast Alfv\'en wave has the opposite behavior; it increases the density in front of the source. For high
densities, above the densities for the lower hybrid resonance and fast wave cutoff, the helicon wave pulls in the lower and higher
density filaments; thereby, increasing the density in front of the source. The fast Alfv\'en wave pulls in lower density filaments and pushes
away higher density filament leading to a decrease in density in front of the source. For intermediate densities, the helicon and the fast
Alfv\'en waves tend to pull in the lower and higher density filaments leading to an increase in density in front of the RF source.

The stress tensor calculations have covered a broad range of plasma densities for all the three RF waves under consideration.
They illustrate the complicated effect of the radiation force in each frequency range. An analytical understanding is beyond the
present scope and is part of future research. The theoretical task is difficult as the radiation force 
includes all five waves -- the incident wave, the two scattered waves, and 
the two waves inside the filament -- regardless of whether they are propagating or evanescent. Furthermore, the radiation force
has a quadratic dependence on the electromagnetic fields. 
Nonetheless, the calculations reveal intriguing properties about the radiation force; in particular,  modifications 
to the plasma density in front of the RF source which could affect the coupling of waves to the plasma.

\section{Acknowledgements}
\label{ack}

AKR is supported by the US Department of Energy Grant numbers
DE-FG02-91ER-54109 and DE-FC02-01ER54648.

\vfill\eject

%figure1 
\begin{figure}[ht]
    \begin{center}
       {\scalebox{0.3} {\includegraphics{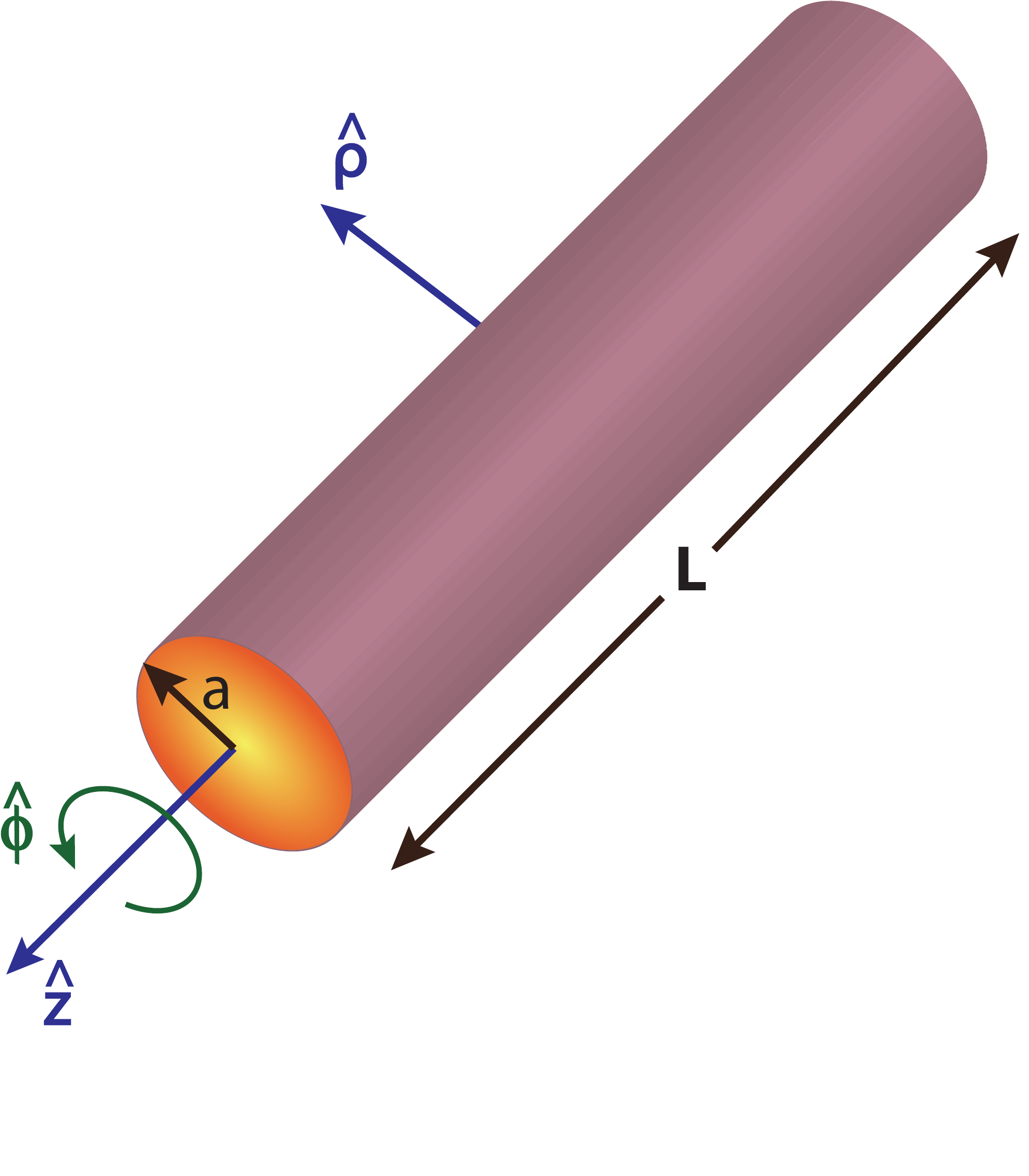}}}
        \captionsetup{justification=raggedright,singlelinecheck=false} %this is for putting the caption justified left 
        \caption{The cylindrical coordinate system: the 
                 ambient magnetic field is along the axial direction $\hat{z}$; $\hat{\rho}$ and $\hat{\phi}$ are 
                 the unit vectors along the radial and azimuthal directions, respectively.}
    \end{center}
\end{figure}

%figure2 
\begin{figure}[ht]
    \begin{center}
       {\scalebox{0.3} {\includegraphics{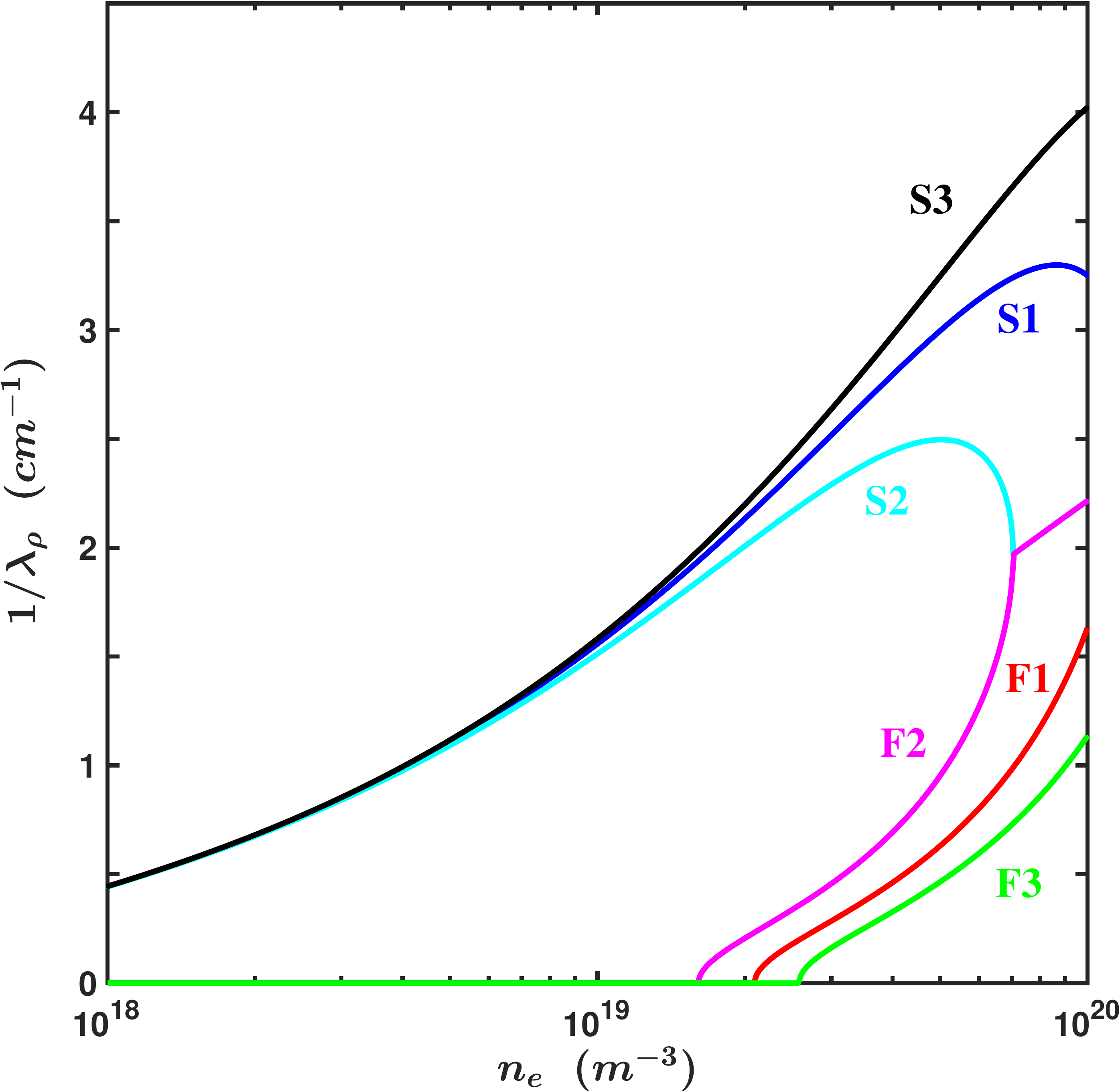}}}
        \captionsetup{justification=raggedright,singlelinecheck=false} %this is for putting the caption justified left 
        \caption{Variation of the cold plasma dispersion roots, obtained from \eqref{3.10}, as a function of 
density, for different strengths of the magnetic field. In this semi-log plot, the abscissa is 
electron density (per cubic meter) and the ordinate is $1/\lambda_\rho = {\rm Re} \left( n_\rho \nu/c \right)$ (cm$^{-1}$).
These results are for a plasma composed of deutrons and electrons, with $\nu = 4.6$ GHz and 
$n_z = 2$. The labels {\it S} and {\it F} indicate the slow wave and fast wave roots of \eqref{3.10}, respectively; 
$\left( S1, F1 \right)$ are the roots for $B_0 = 4.5$ T, 
$\left( S2, F2 \right)$ for $B_0 = 3.5$ T, and 
$\left( S3, F3 \right)$ for $B_0 = 5.5$ T.}  \label{fig:fig2}
    \end{center} 
\end{figure}

%figure3 
\begin{figure}[ht]
    \begin{center}
       {\scalebox{0.3} {\includegraphics{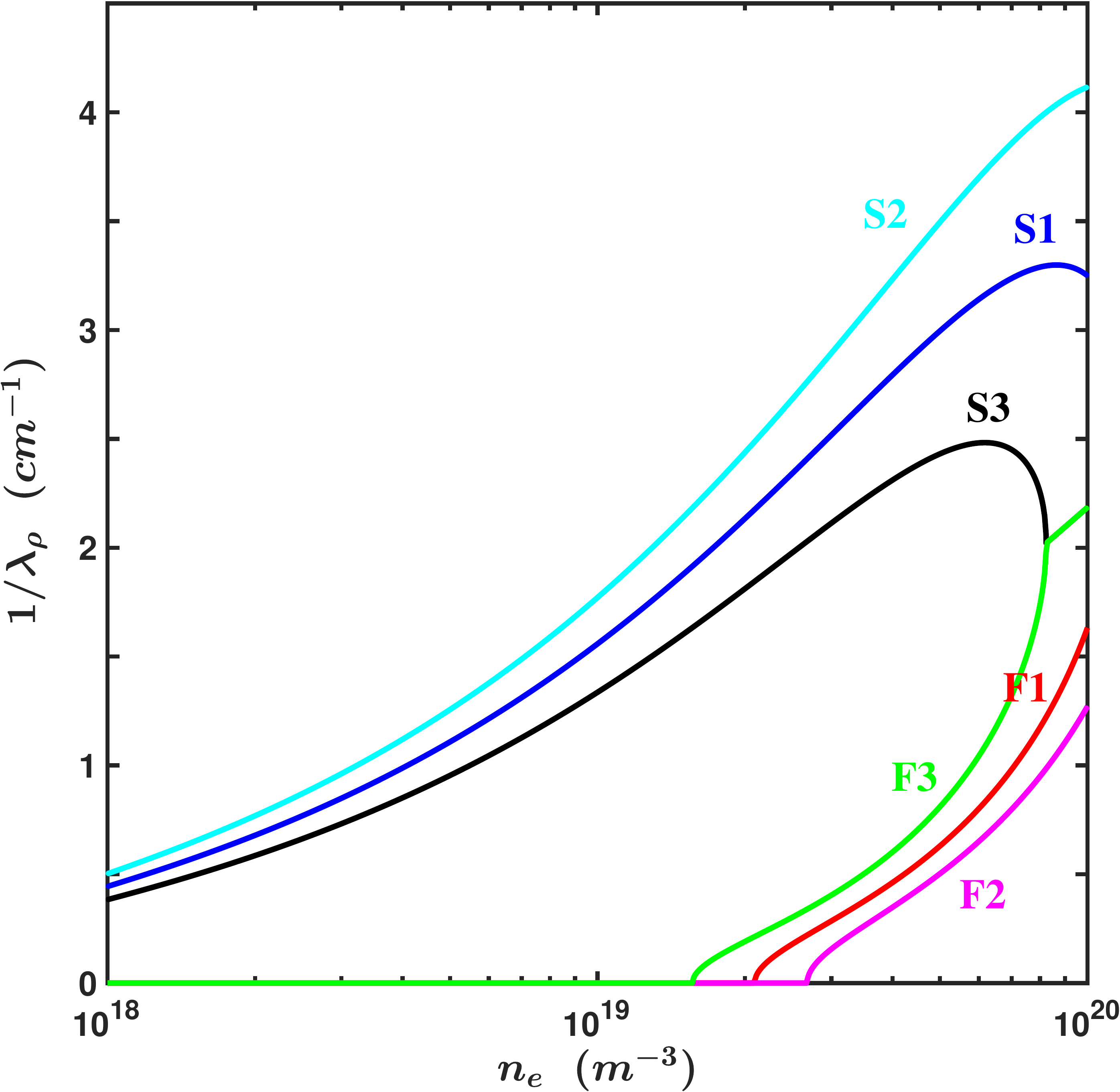}}}
        \captionsetup{justification=raggedright,singlelinecheck=false} %this is for putting the caption justified left 
        \caption{Variation of the cold plasma dispersion roots as a function of electron density, for different  $n_z$. 
Using the same convention as in Fig. \ref{fig:fig2}, these roots are for $\nu = 4.6$ GHz, $B_0 = 4.5$ T; 
$\left( S1, F1 \right)$ are the roots for $n_z = 2$ , 
$\left( S2, F2 \right)$ for $n_z=2.2$, and 
$\left( S3, F3 \right)$ for $n_z = 1.8$.} \label{fig:fig3}
    \end{center} 
\end{figure}

%figure4 
\begin{figure}[ht]
    \begin{center}
       {\scalebox{0.3} {\includegraphics{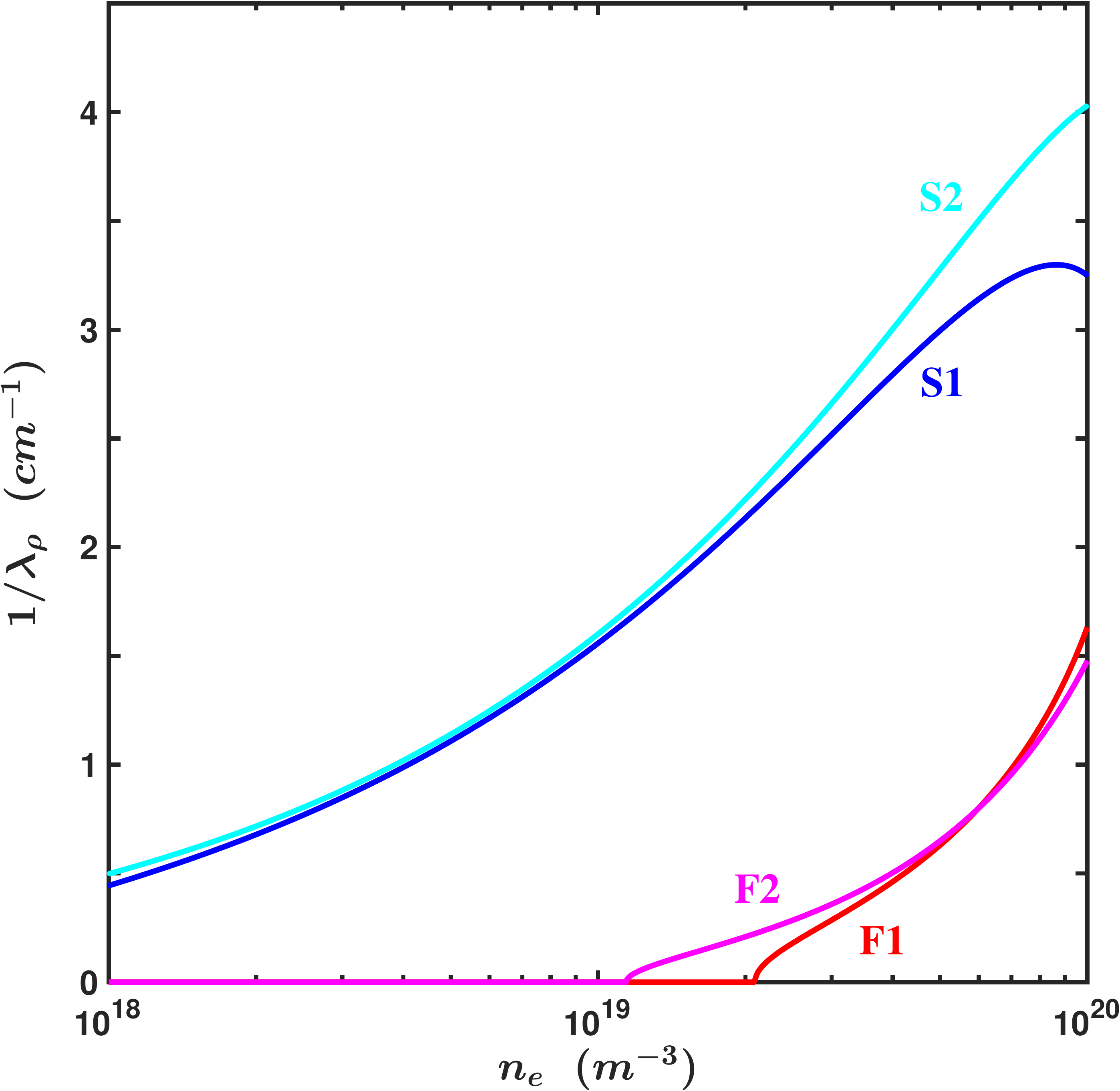}}}
        \captionsetup{justification=raggedright,singlelinecheck=false} %this is for putting the caption justified left 
        \caption{Variation of the cold plasma dispersion roots as a function of electron density, for two
different wave frequencies. 
Using the same convention as in Fig. \ref{fig:fig2}, these roots are for $n_z = 2$ and $B_0 = 4.5$ T ; 
$\left( S1, F1 \right)$ are the roots for $\nu = 4.6$ GHz and 
$\left( S2, F2 \right)$ for $\nu = 2.45$ GHz.} \label{fig:fig4}
    \end{center} 
\end{figure}
\begin{table}[h!]
%\caption{}
\label{}
\begin{tabular}{ |>{\centering}p{2cm}||>{\centering}p{3.5cm}|>{\centering}p{3.5cm}||>{\centering}p{3.5cm}|>{\centering\arraybackslash}p{3.5cm}|  }
\hline 
\multicolumn{5}{|c|}{Wave characteristics in the plasma} \\
\hline 
\multicolumn{1}{|c||}{} &
\multicolumn{2}{c||}{Background plasma} & 
\multicolumn{2}{c|}{Filament plasma} \\
\hline  
\multicolumn{1}{|c||}{} &
\multicolumn{2}{c||}{$n_e \ = \ 2.25 \times 10^{19} \ {\rm m}^{-3}$} & 
\multicolumn{2}{c|}{$n_e \ = \ 2 \times 10^{19} \ {\rm m}^{-3}$} \\
\hline  
& slow wave & fast wave & slow wave & fast wave \\
\hhline{|=#=|=#=|=|}
$n_{\rho \alpha}^{\beta}$ & $\left( 14.612, \ 0 \right)$ & $\left(0.705, \ 0 \right)$ & $\left( 13.902, \ 0 \right)$ & $\left( 0, \ 0.522 \right)$ \\
$\Lambda_{\rho \alpha}^\beta$ (cm) & $\left( 0.45, \ 0 \right)$ & $\left(9.25, \ 0 \right)$ & $\left( 0.47, \ 0 \right)$ & 
$\left( 0, \ -12.5 \right)$ \\
\hline 
$E_{k \rho \alpha}^{\beta}$ & $ \left( 0.995, \ 0 \right) $ & $\left( 0, \ -0.736 \right)$ & $\left( 0.995, \ 0 \right)$ & $\left( 0, \ -0.689 \right)$ \\
$E_{k \phi \alpha}^{\beta}$ & $ \left( 0, \ 0.014 \right) $ & $\left( 0.677, \ 0 \right)$ & $\left( 0, \ 0.014 \right)$ & $\left( 0.725, \ 0 \right)$ \\
$E_{k z \alpha}^{\beta}$ & $ \left( 0.098, \ 0 \right) $ & $\left( 0, \ -0.012 \right)$ & $\left( 0.103, \ 0 \right)$ & $\left( 0.01, \ 0 \right)$ \\
\hline 
$\vec{P}_\alpha^\beta$ & $\left( -0.092, \; 0, \; 0.996 \right)$& $\left( 0.151,\; 0, \; 0.998 \right)$ & $\left( -0.097,\; 0, \; 0.995 \right)$ & 
$\left( 0,\; 0.122, \; 0.992 \right)$ \\
\hline 
\end{tabular}
\caption{This table is for LH waves with $B_0 = 4.5$ T, $\nu = 4.6$ GHz, and $n_z = 2$.
The indices of refraction $n_{\rho \alpha}^{\beta}$ are obtained from \eqref{3.10}. The subscript $\alpha$ can be either $S$ or $F$ for 
the slow and fast LH waves, respectively; the superscript $\beta$ is either $b$ or $f$ representing the background and the filament plasmas, 
respectively. The complex $\Lambda$'s are defined in \eqref{10.1} and \eqref{10.2}. 
The $E$'s are components of the polarization vector defined in Section \ref{sec:4}. Inside each set of parenthesis, the first 
number is the real part and the second number is the imaginary part. The components of the Poynting vector $\vec{P}$, defined in \eqref{9.4}, 
are listed in the last row.} \label{table:t1}
\end{table}

%figure5 
\begin{figure}[htp]
\begin{center}
\subfigure[\label{fig:fig5a} \ Contours of ${\rm Re} \left( E_{T x} \right) $ ]{
       {\includegraphics[width=0.48\textwidth]{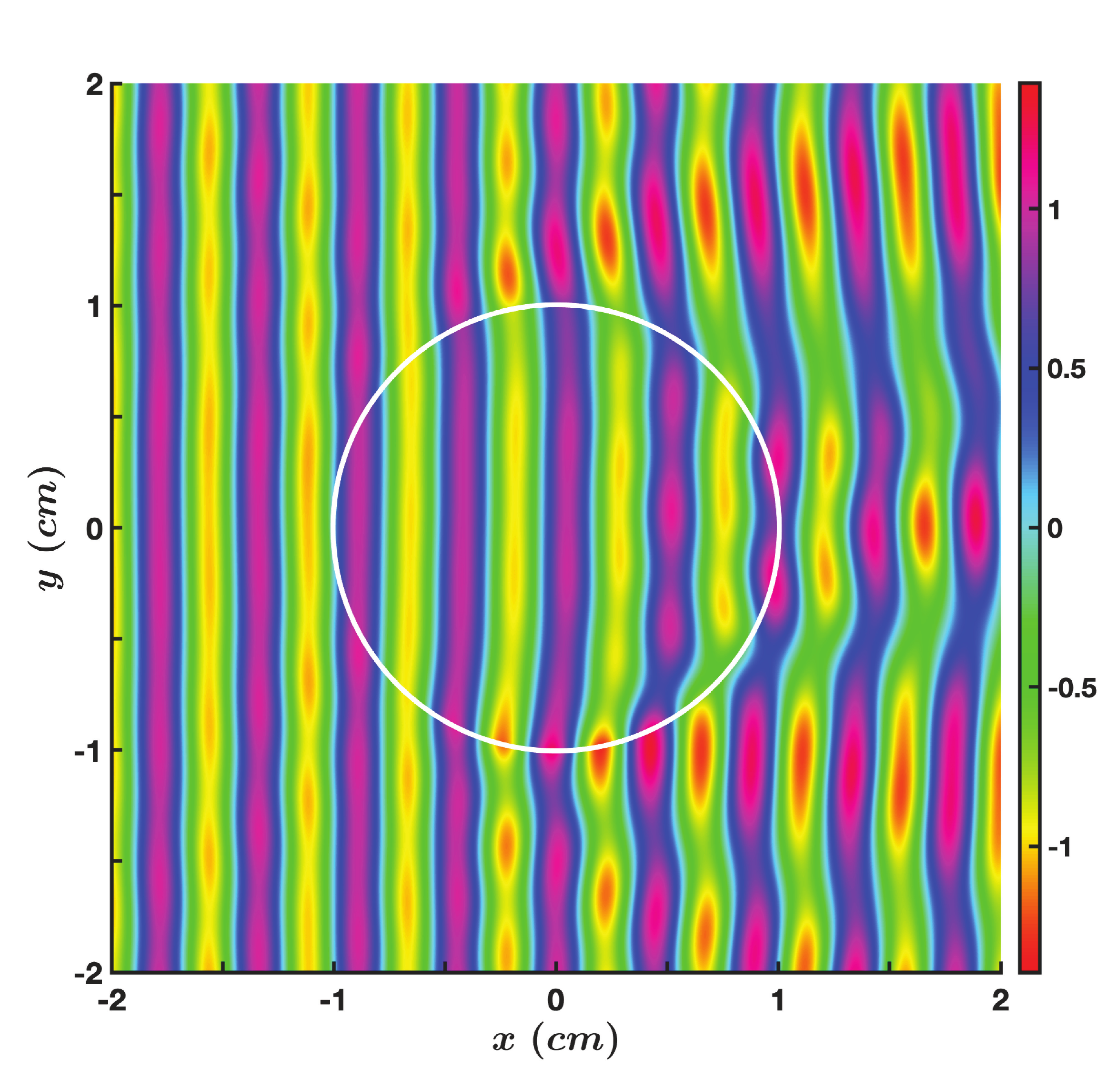}} 
} 
\subfigure[\label{fig:fig5b} \ Contours of ${\rm Re} \left( E_{T y} \right) $ ]{
       {\includegraphics[width=0.48\textwidth]{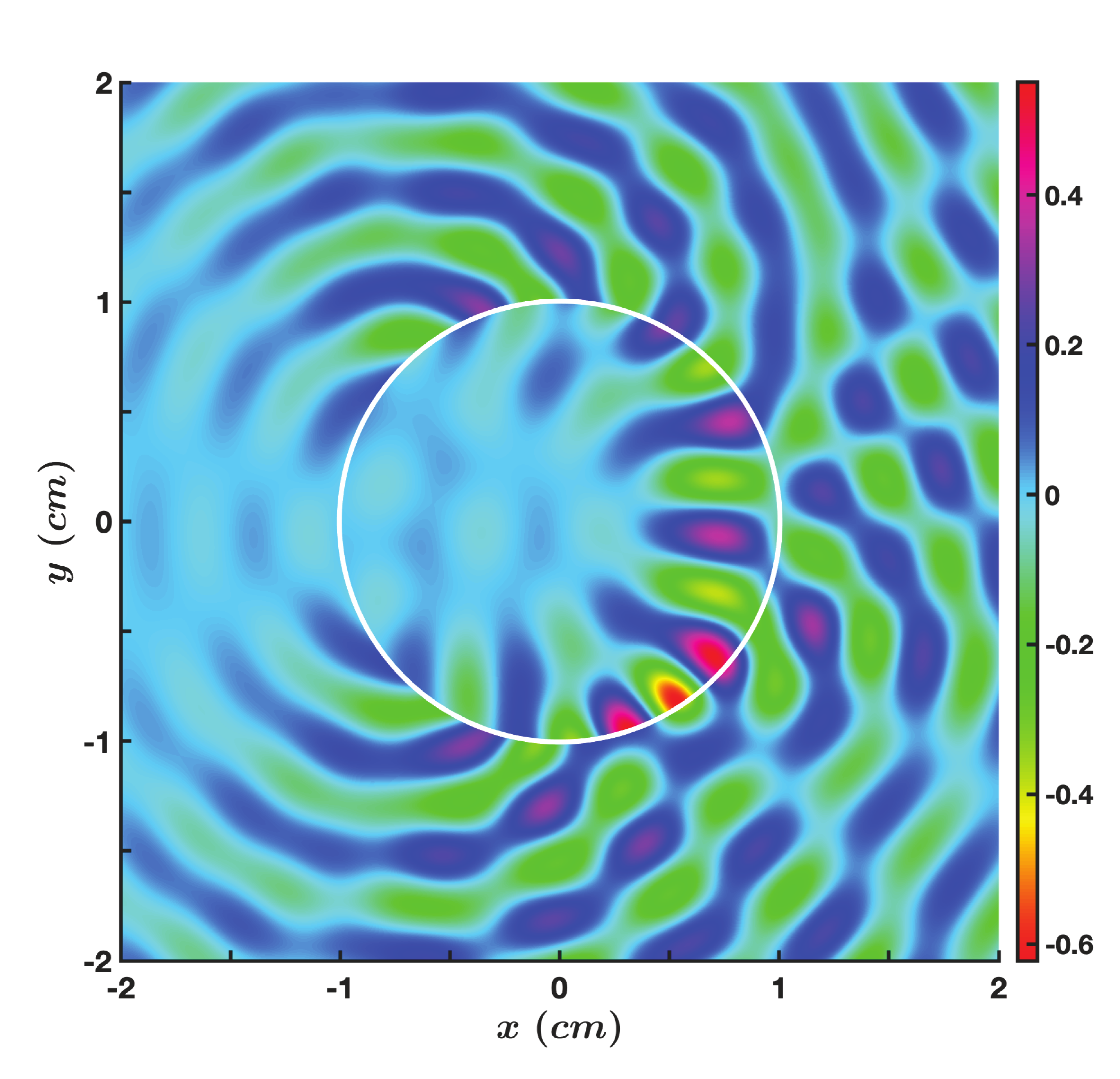}} 
} 
\subfigure[\label{fig:fig5c} \ Contours of ${\rm Re} \left( E_{T z} \right) $ ]{
       {\includegraphics[width=0.48\textwidth]{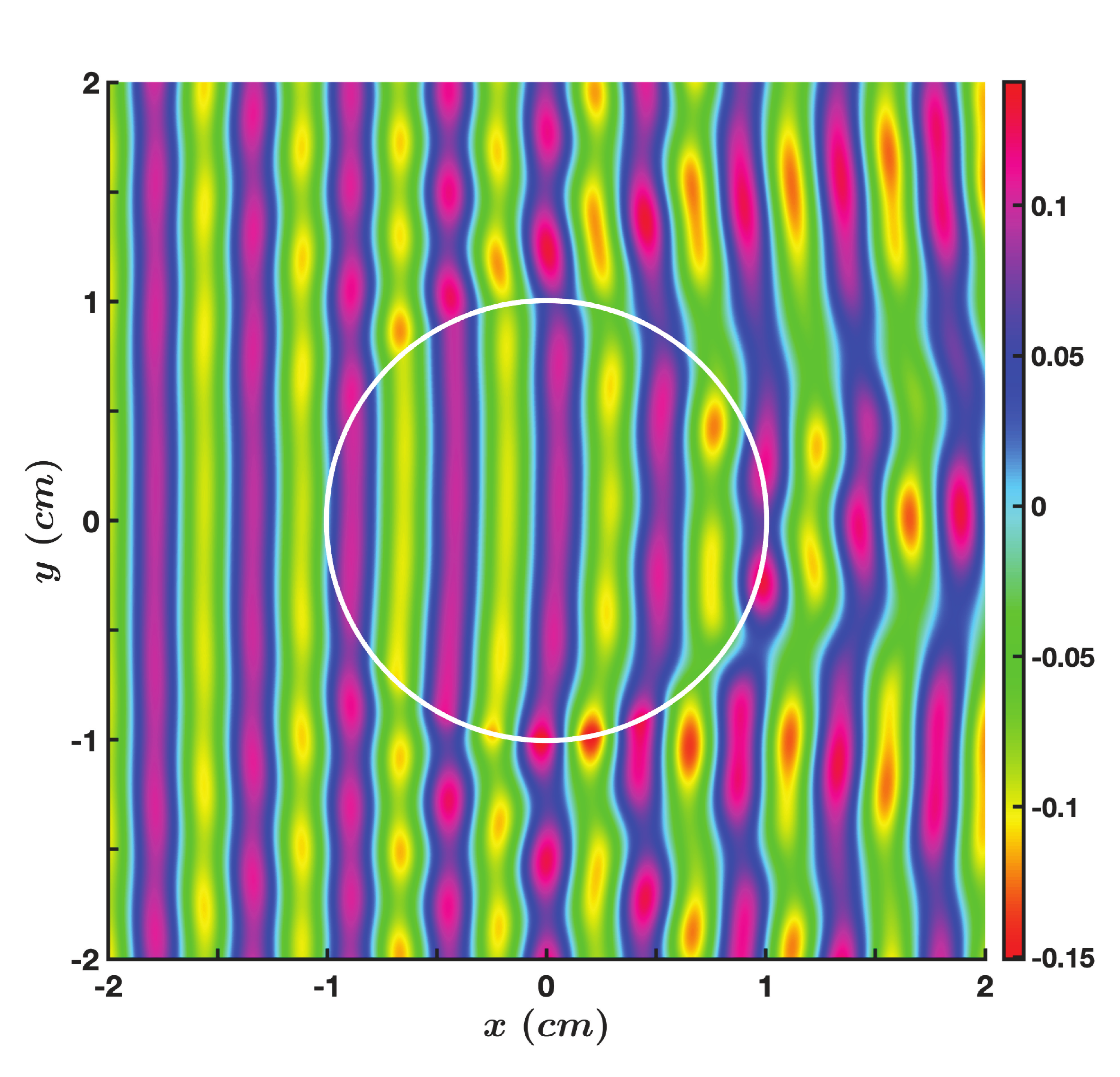}} 
} 
        \captionsetup{justification=raggedright,singlelinecheck=false} %this is for putting the caption justified left 
        \caption{ \label{fig:fig5} Contours of the Cartesian components of the total electric field $\vec{E}_T$
in the $x$-$y$plane, when the incident plane wave is a slow LH wave; $\vec{E}_T = \vec{E}_I + \vec{E}_S + \vec{E}_F$ is the vector sum
of the incident and scattered fields and fields inside the filament.
The wave is incident from the left, and the cross-sectional outline of the cylindrical filament (radius $a = 1$ cm) 
is shown in white. The plasma and wave parameters are given in Table \ref{table:t1}.}
\end{center}
\end{figure}

%figure6 
\begin{figure}[htp]
\begin{center}
\subfigure[\label{fig:fig6a} \ Contours of $P_x$ ]{
       {\includegraphics[width=0.48\textwidth]{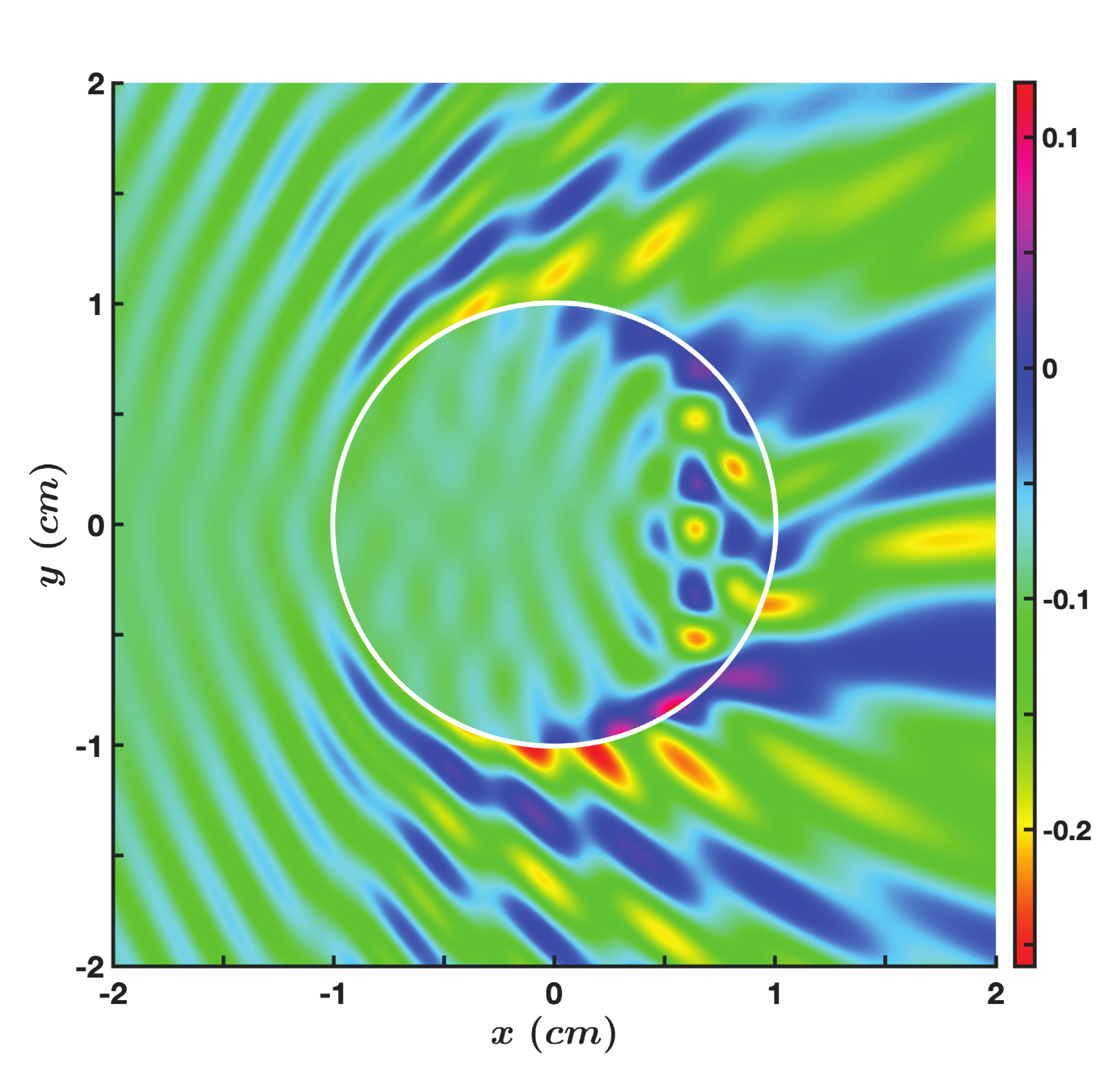}} 
} 
\subfigure[\label{fig:fig6b} \ Contours of $P_y$ ]{
       {\includegraphics[width=0.48\textwidth]{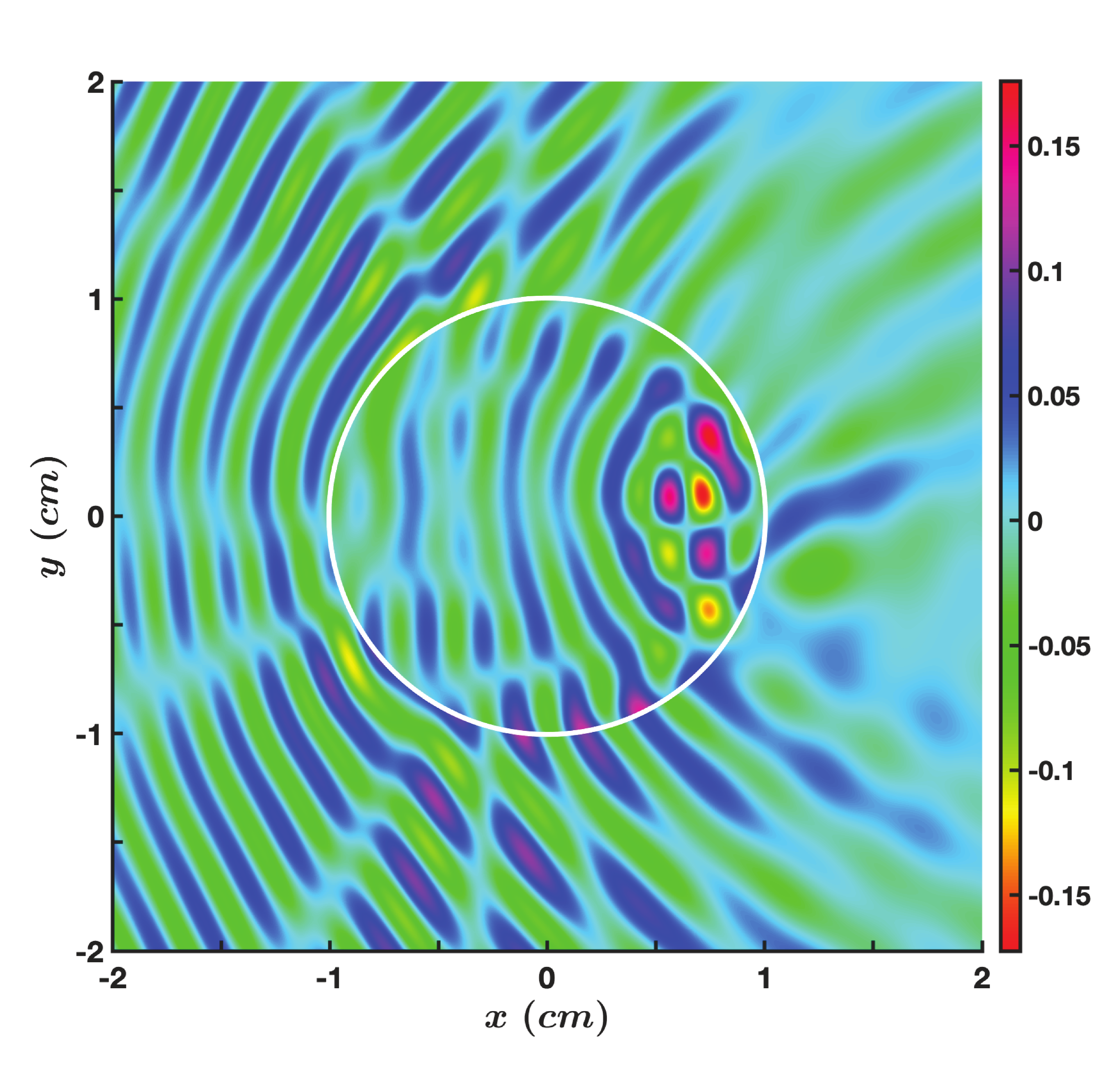}} 
} 
\subfigure[\label{fig:fig6c} \ Contours of $P_z$ ]{
       {\includegraphics[width=0.48\textwidth]{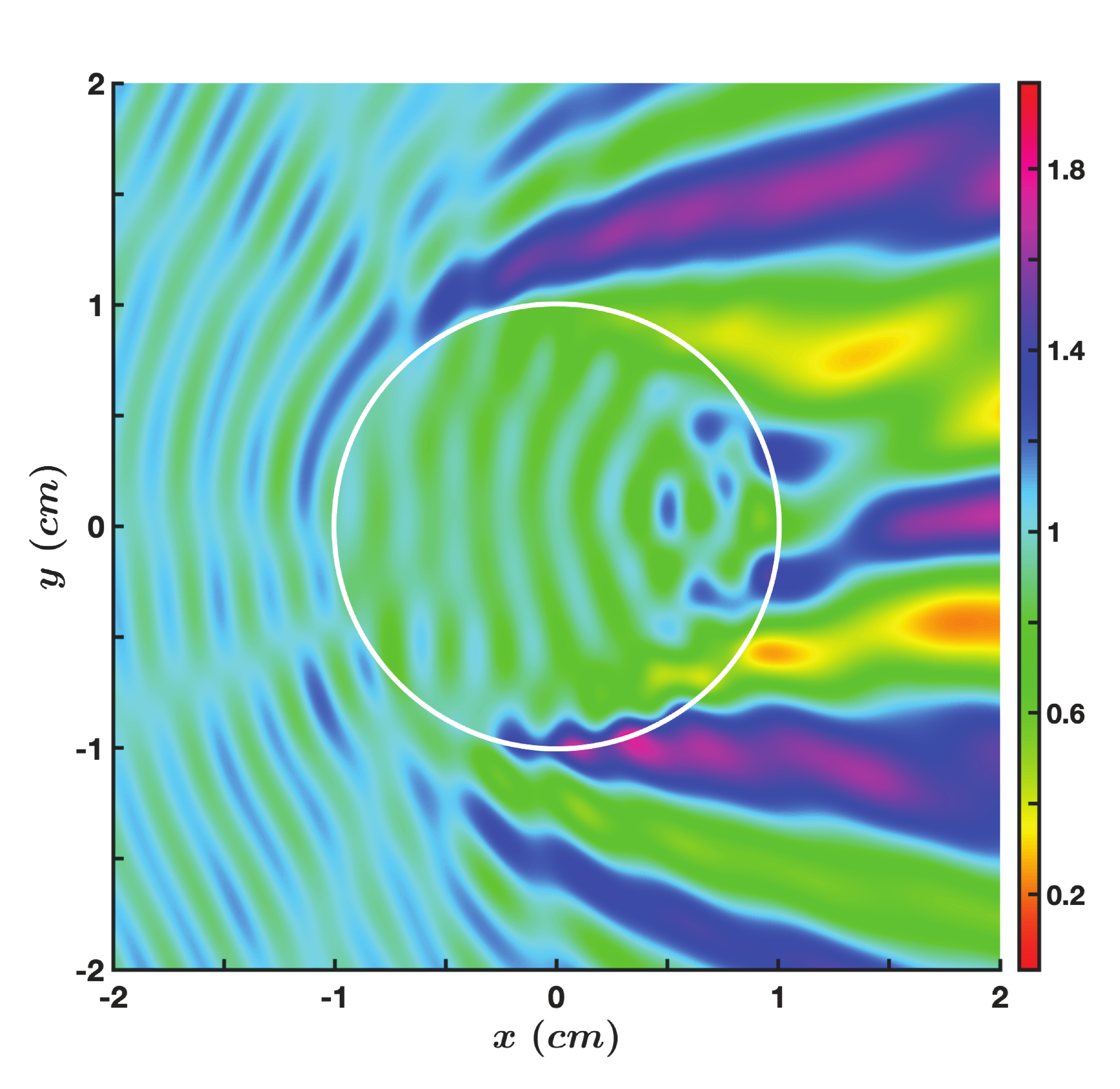}} 
} 
        \captionsetup{justification=raggedright,singlelinecheck=false} %this is for putting the caption justified left 
        \caption{ \label{fig:fig6} Contours of the three Cartesian components of the Poynting vector associated with 
the fields in Fig. \ref{fig:fig5}.}
\end{center}
\end{figure}

%figure7
\begin{figure}[htp]
\begin{center}
\subfigure[\label{fig:fig7a} \ ${\rm J}_m$ as a function of $\rho$ for $m = 0, 1, {\rm and} \ 2.$]{
       {\includegraphics[width=0.48\textwidth]{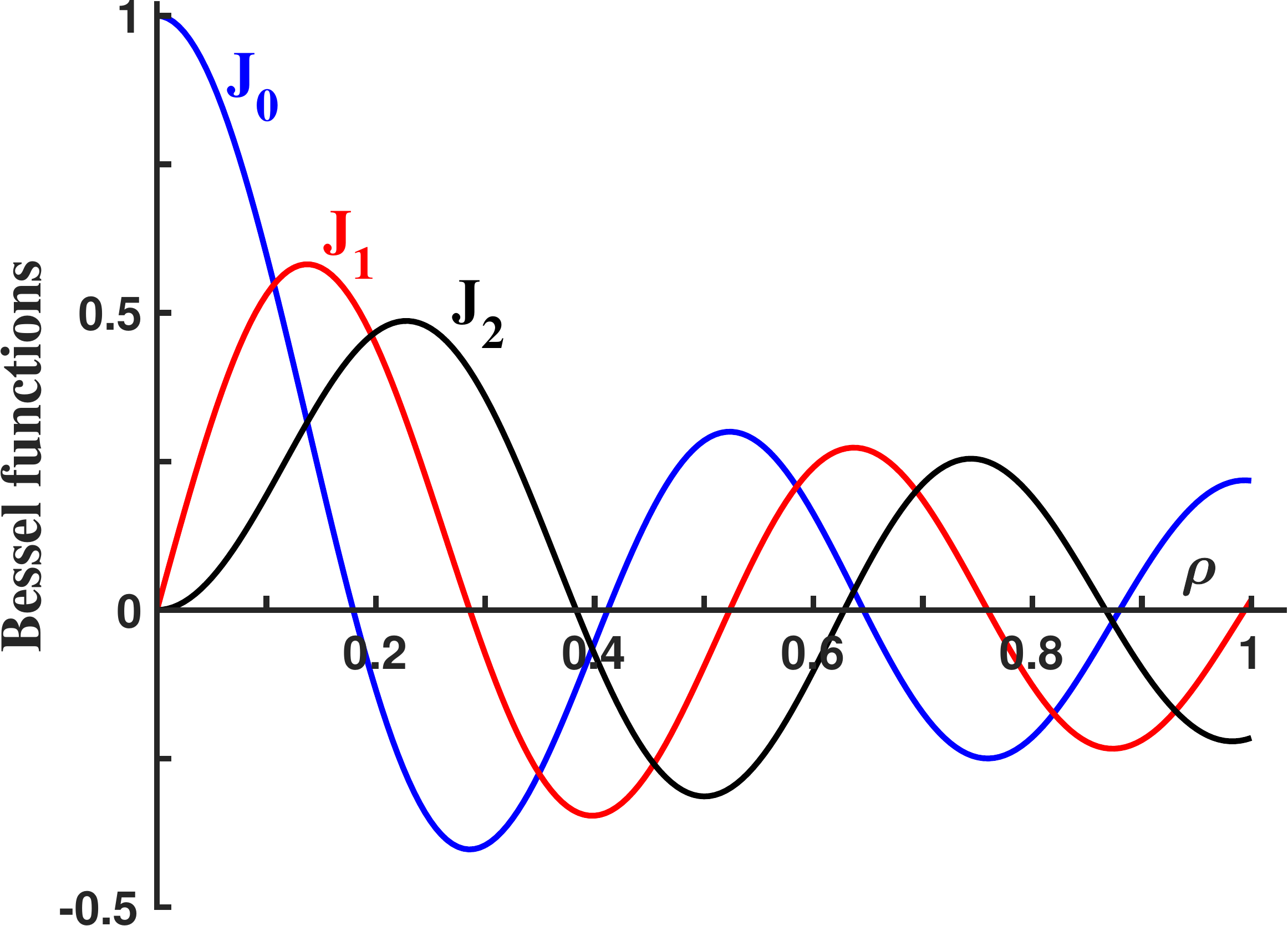}} 
} 
\subfigure[\label{fig:fig7b} \ ${\rm J}_m'$ as a function of $\rho$ for $m = 0, 1, {\rm and} \ 2.$]{
       {\includegraphics[width=0.48\textwidth]{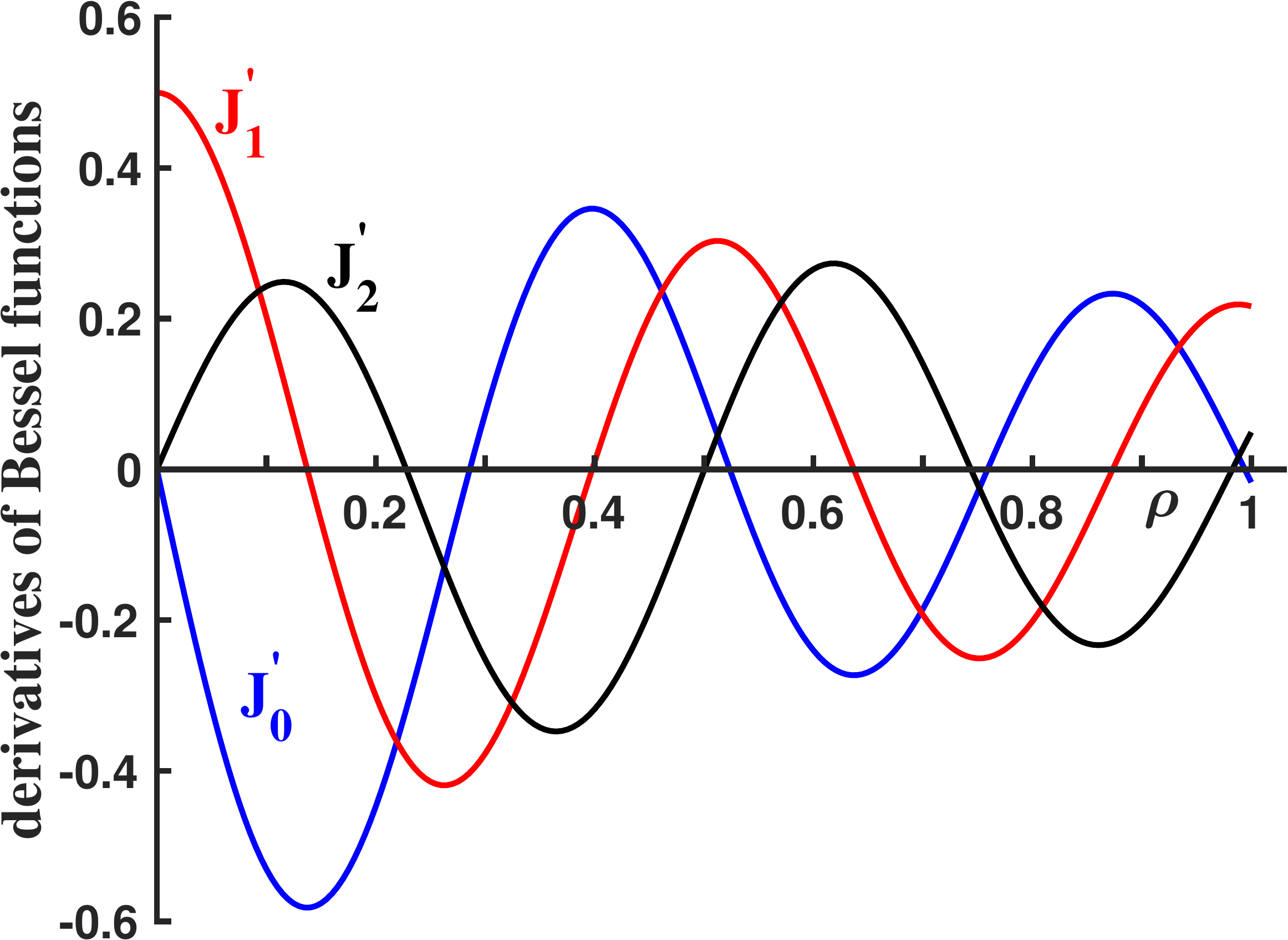}} 
} 
        \captionsetup{justification=raggedright,singlelinecheck=false} %this is for putting the caption justified left 
        \caption{ \label{fig:fig7} Variation of the Bessel functions (and derivatives) as a function of $\rho$ inside the filament.
The argument of the Bessel function is $k_{\rho S}^f \rho$, with $k_{\rho S}^f = \omega n_{\rho S}^f / c \approx 13.403$ cm$^{-1}$ being
the $\unv{\rho}$ component of the LH fast wave vector inside the filament. The parameters are as in Table \ref{table:t1}.}
\end{center}
\end{figure}

%figure8 
\begin{figure}[htp]
\begin{center}
\subfigure[\label{fig:fig8a} \ Contours of ${\rm Re} \left( E_{T x} \right) $ ]{
       {\includegraphics[width=0.48\textwidth]{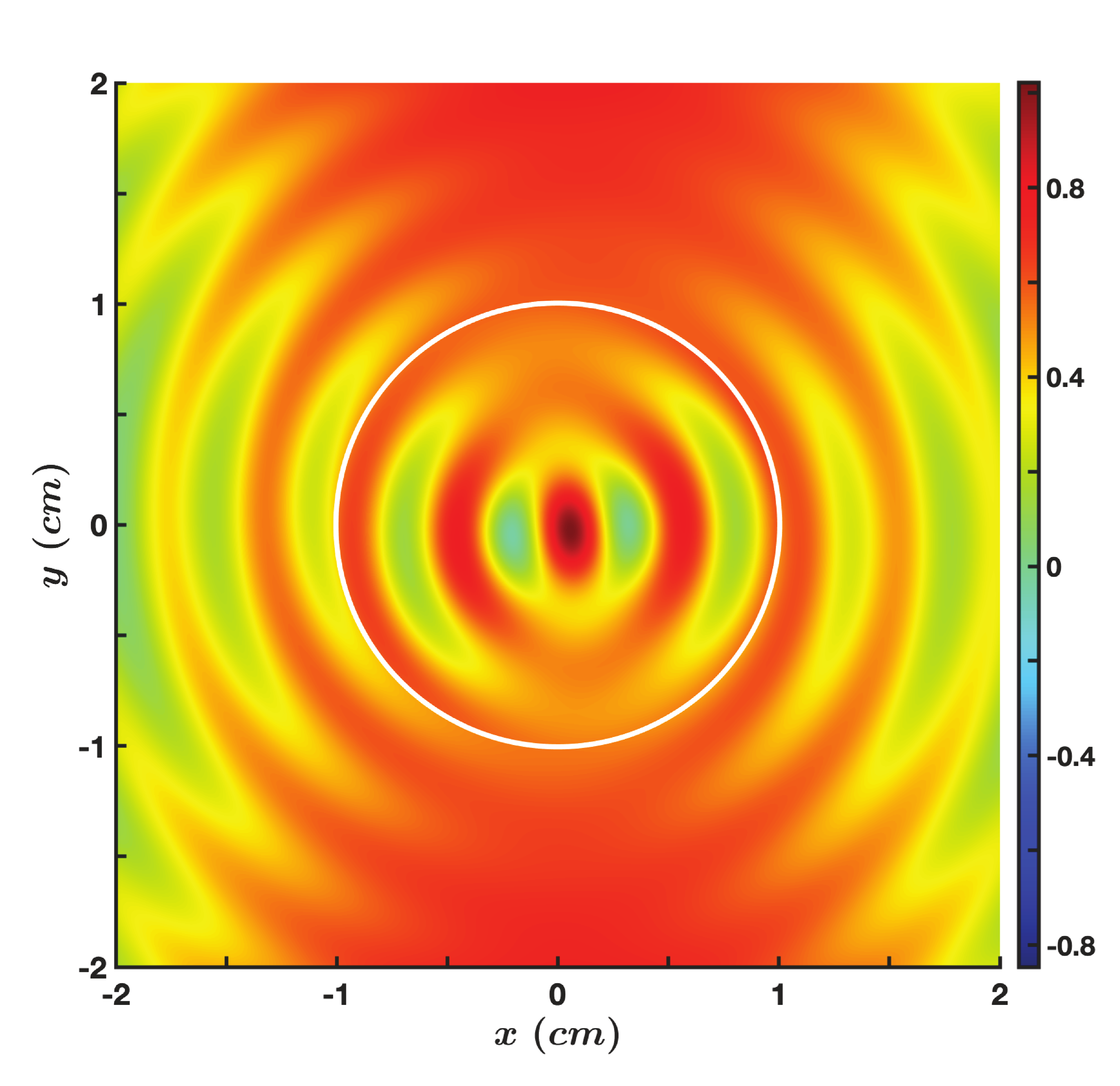}} 
} 
\subfigure[\label{fig:fig8b} \ Contours of ${\rm Re} \left( E_{T y} \right) $ ]{
       {\includegraphics[width=0.48\textwidth]{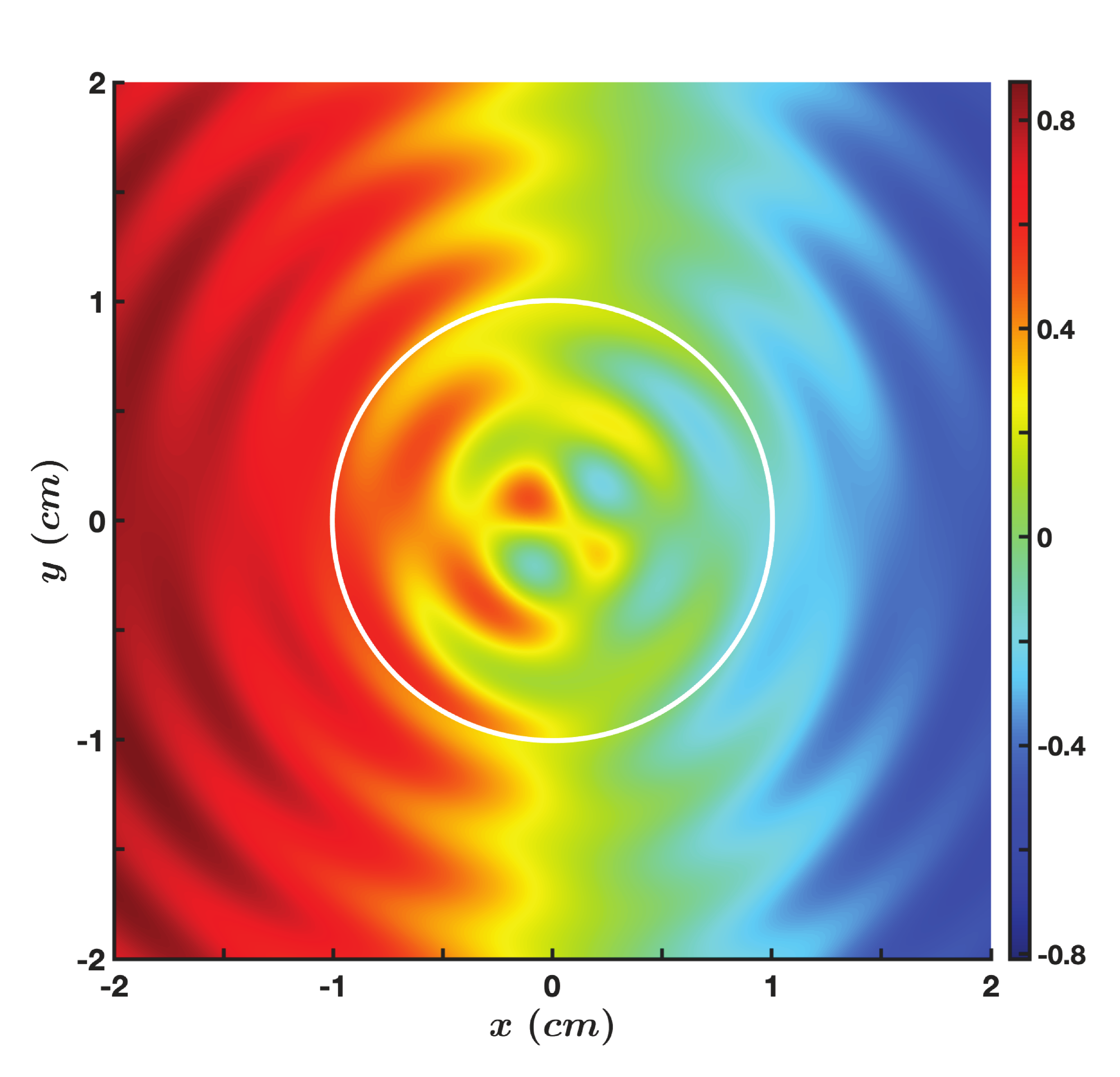}} 
} 
\subfigure[\label{fig:fig8c} \ Contours of ${\rm Re} \left( E_{T z} \right) $ ]{
       {\includegraphics[width=0.48\textwidth]{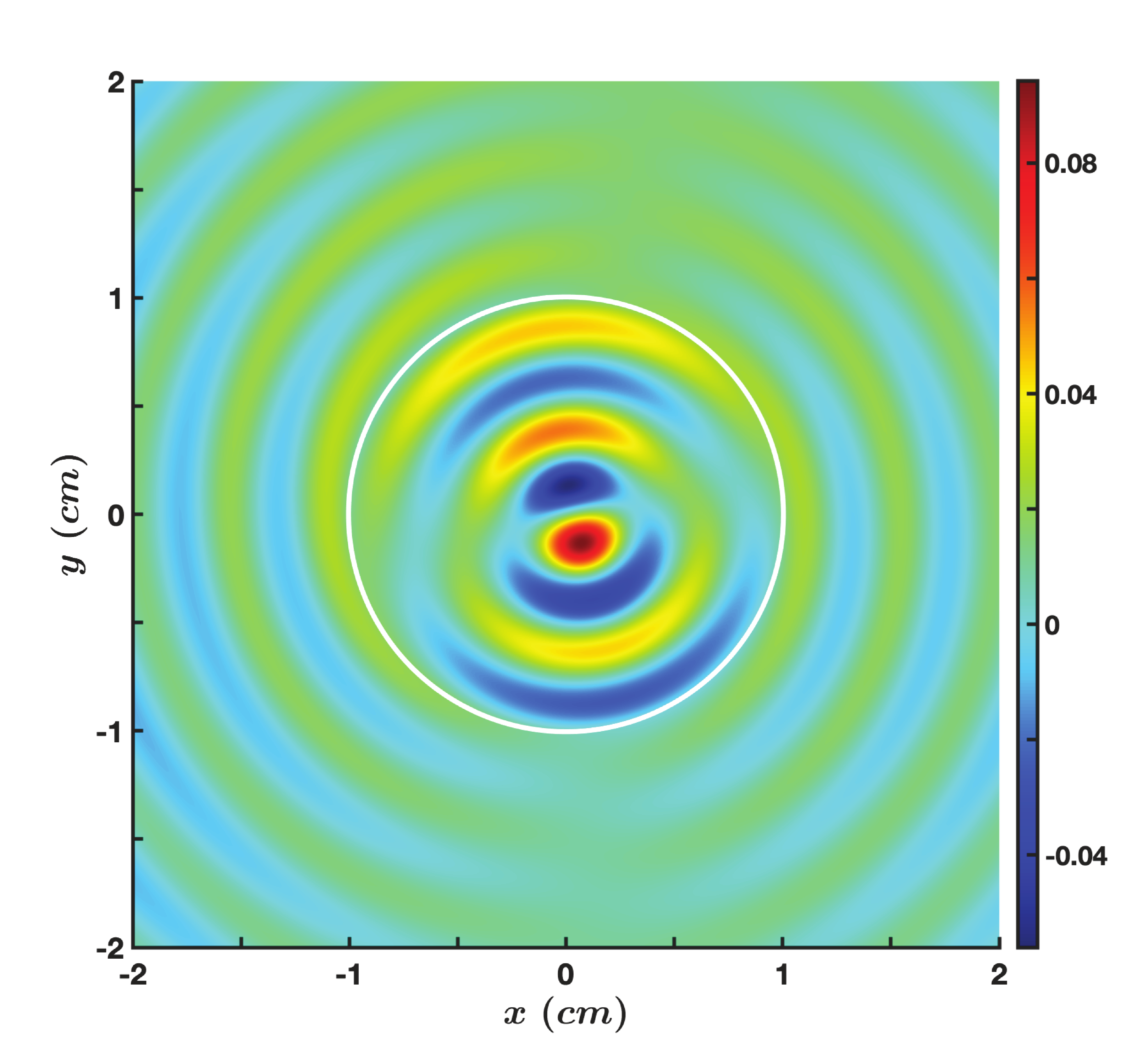}} 
} 
        \captionsetup{justification=raggedright,singlelinecheck=false} %this is for putting the caption justified left 
        \caption{ \label{fig:fig8} Contours of the Cartesian components of $\vec{E}_T$ when 
a fast LH wave is incident from the left. The parameters are the same as for Fig. \ref{fig:fig5}.}
\end{center}
\end{figure}

%figure9 
\begin{figure}[htp]
\begin{center}
\subfigure[\label{fig:fig9a} \ Contours of $P_x$ ]{
       {\includegraphics[width=0.48\textwidth]{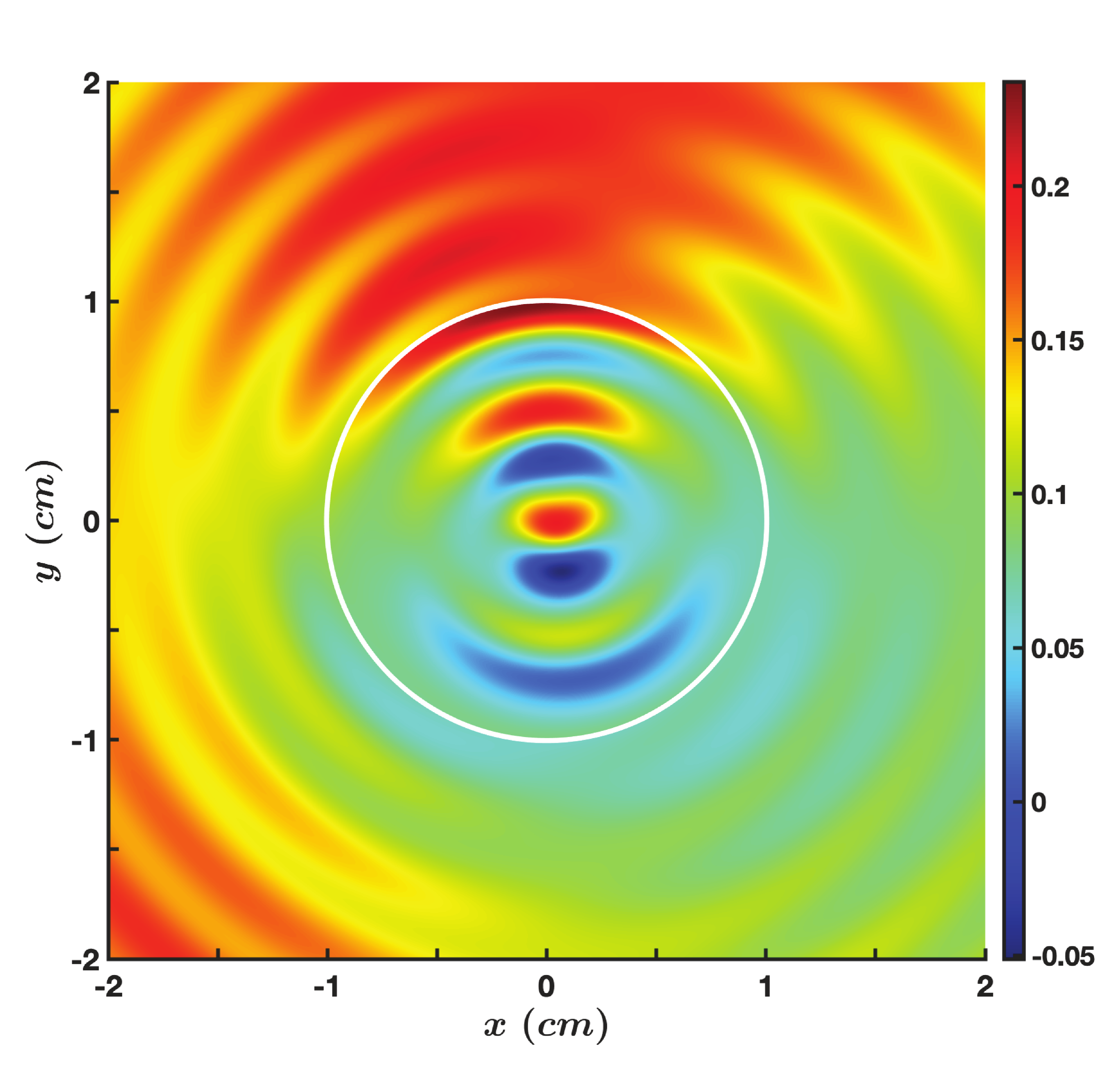}} 
} 
\subfigure[\label{fig:fig9b} \ Contours of $P_y$ ]{
       {\includegraphics[width=0.48\textwidth]{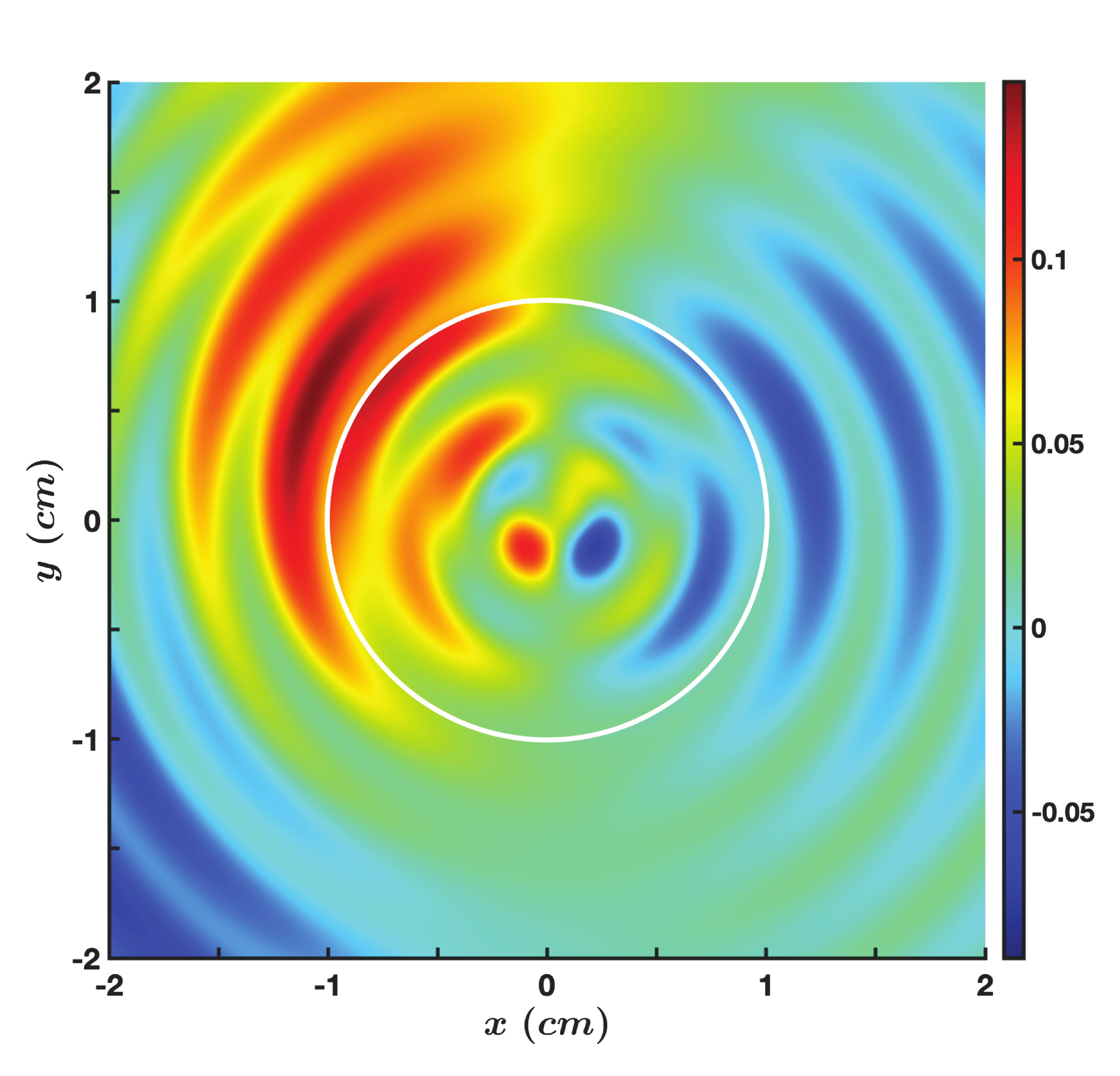}} 
} 
\subfigure[\label{fig:fig9c} \ Contours of $P_z$ ]{
       {\includegraphics[width=0.48\textwidth]{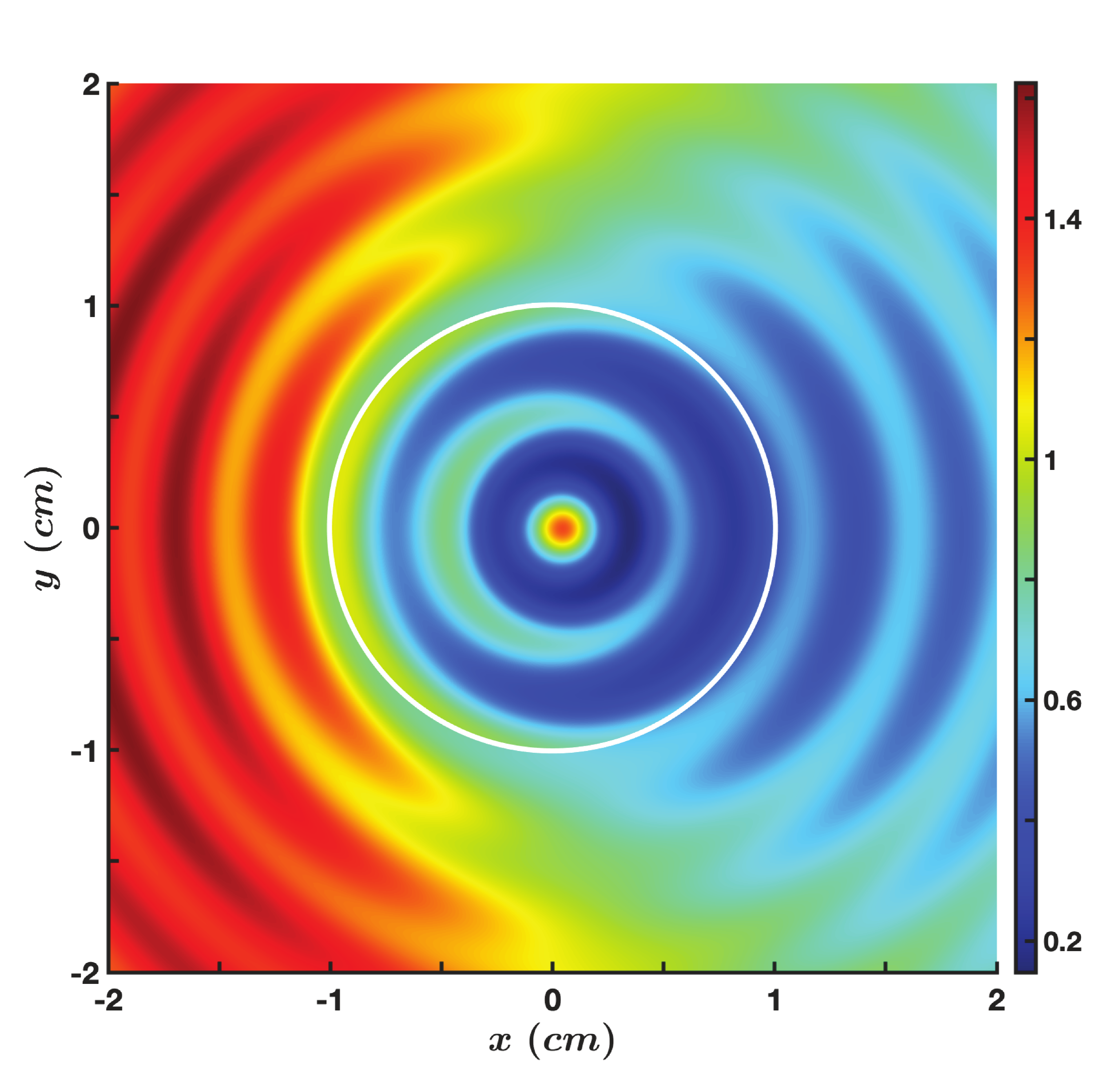}} 
} 
        \captionsetup{justification=raggedright,singlelinecheck=false} %this is for putting the caption justified left 
        \caption{ \label{fig:fig9} Contours of the three Cartesian components of the Poynting vector associated with 
the fields in Fig. \ref{fig:fig8}.}
\end{center}
\end{figure}

%figure10
\begin{figure}[htp]
\begin{center}
\subfigure[\label{fig:fig10a} \ Contours of ${\rm Re} \left( E_{T x} \right) $ ]{
       {\includegraphics[width=0.48\textwidth]{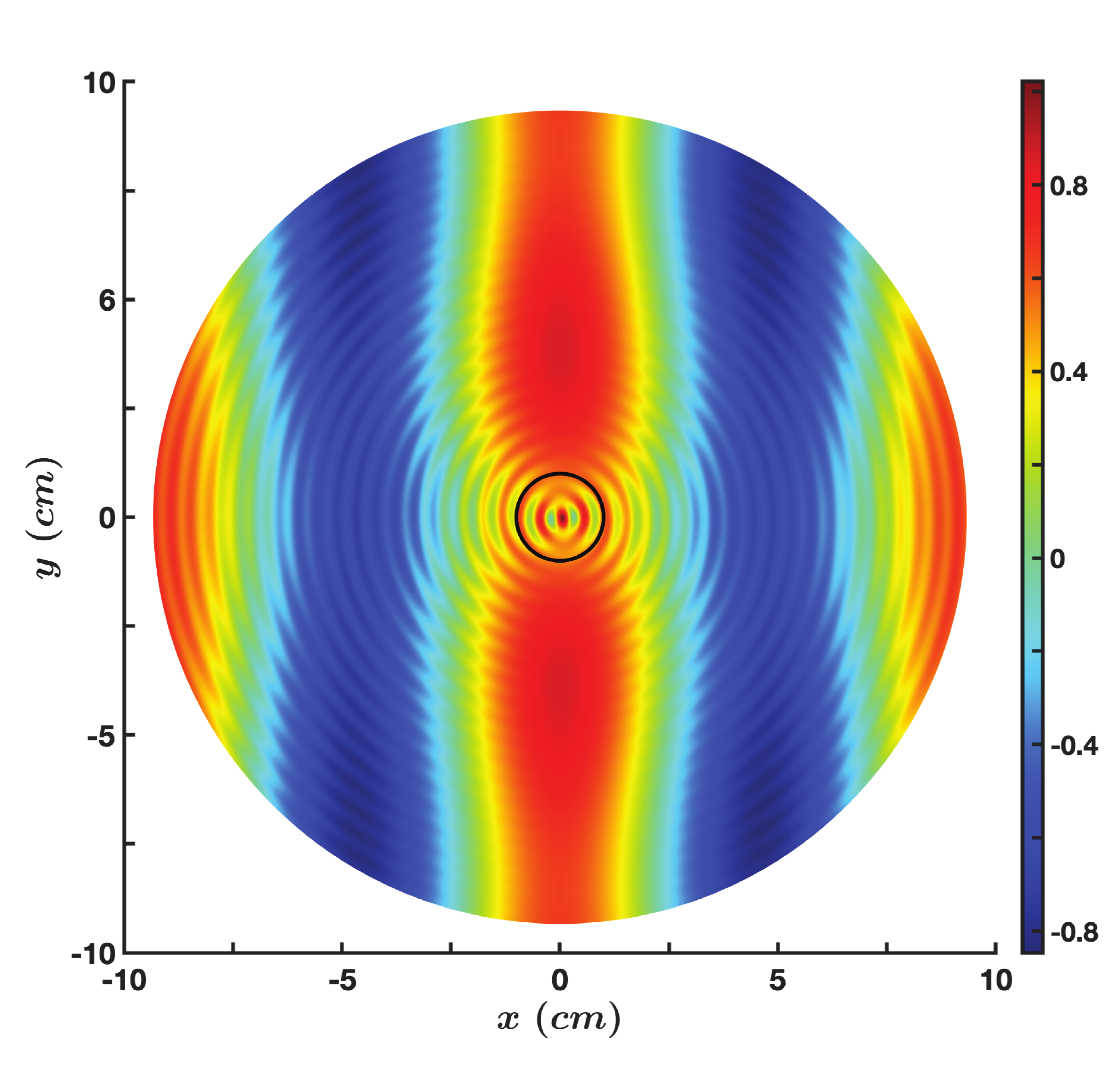}} 
} 
\subfigure[\label{fig:fig10b} \ Contours of $P_x$ ]{
       {\includegraphics[width=0.48\textwidth]{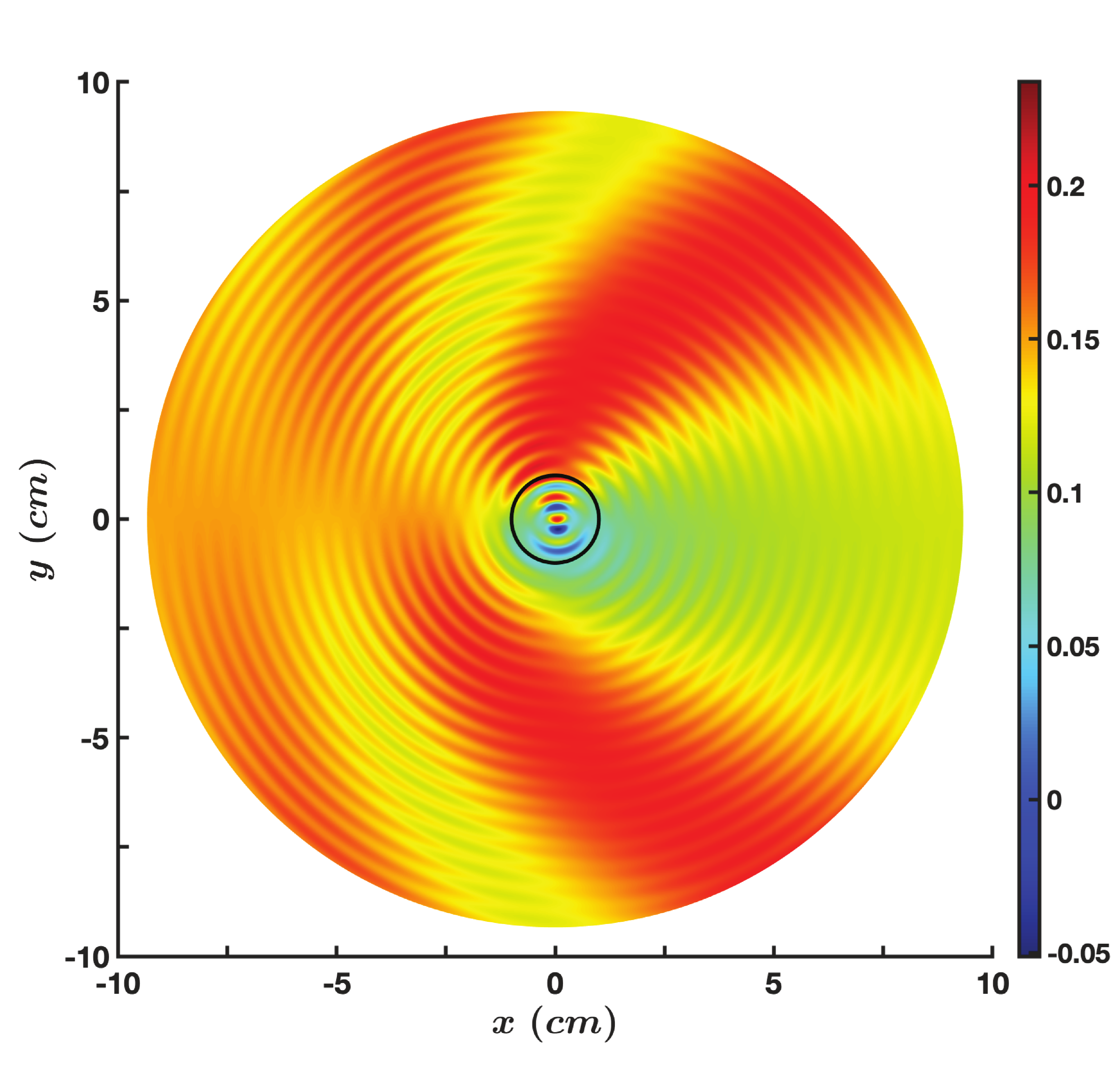}} 
} 
        \captionsetup{justification=raggedright,singlelinecheck=false} %this is for putting the caption justified left 
        \caption{ \label{fig:fig10} These figures are the same as Figs. \ref{fig:fig8a} and \ref{fig:fig9a}, respectively, except that
the display is over a wider region of space. The fast wave with its wavelength of $9.25$ cm is clearly discernable, as is the
effect of the slow wave on the power flow in the $x$-direction. The cross-section of the filament is in black.}
\end{center}
\end{figure}

%figure11
\begin{figure}[htp]
\begin{center}
\subfigure[\label{fig:fig11a} \ Contours of ${\rm Re} \left( E_{T x} \right) $ ]{
       {\includegraphics[width=0.48\textwidth]{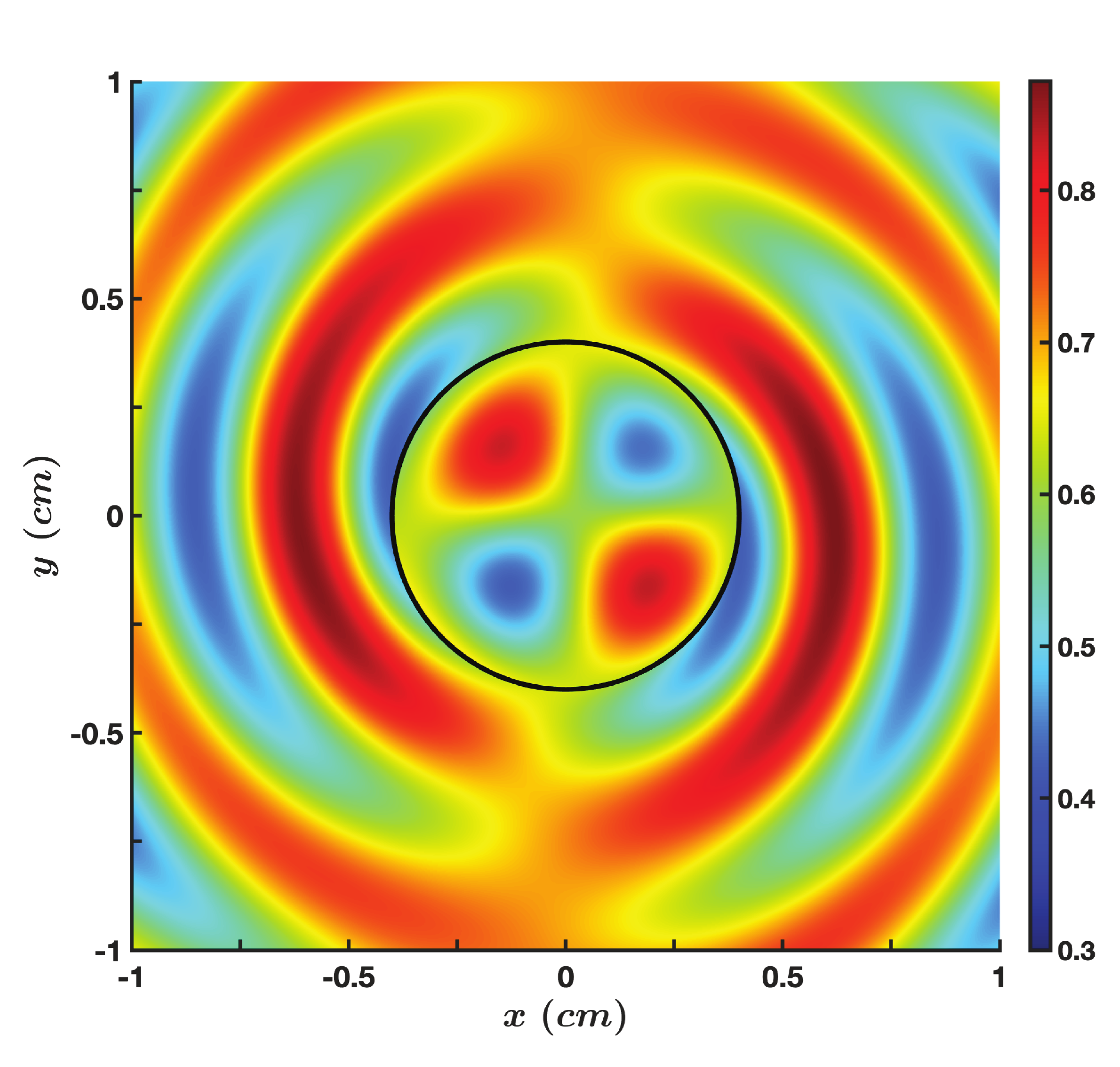}} 
} 
\subfigure[\label{fig:fig11b} \ Contours of ${\rm Re} \left( E_{T z} \right) $ ]{
       {\includegraphics[width=0.48\textwidth]{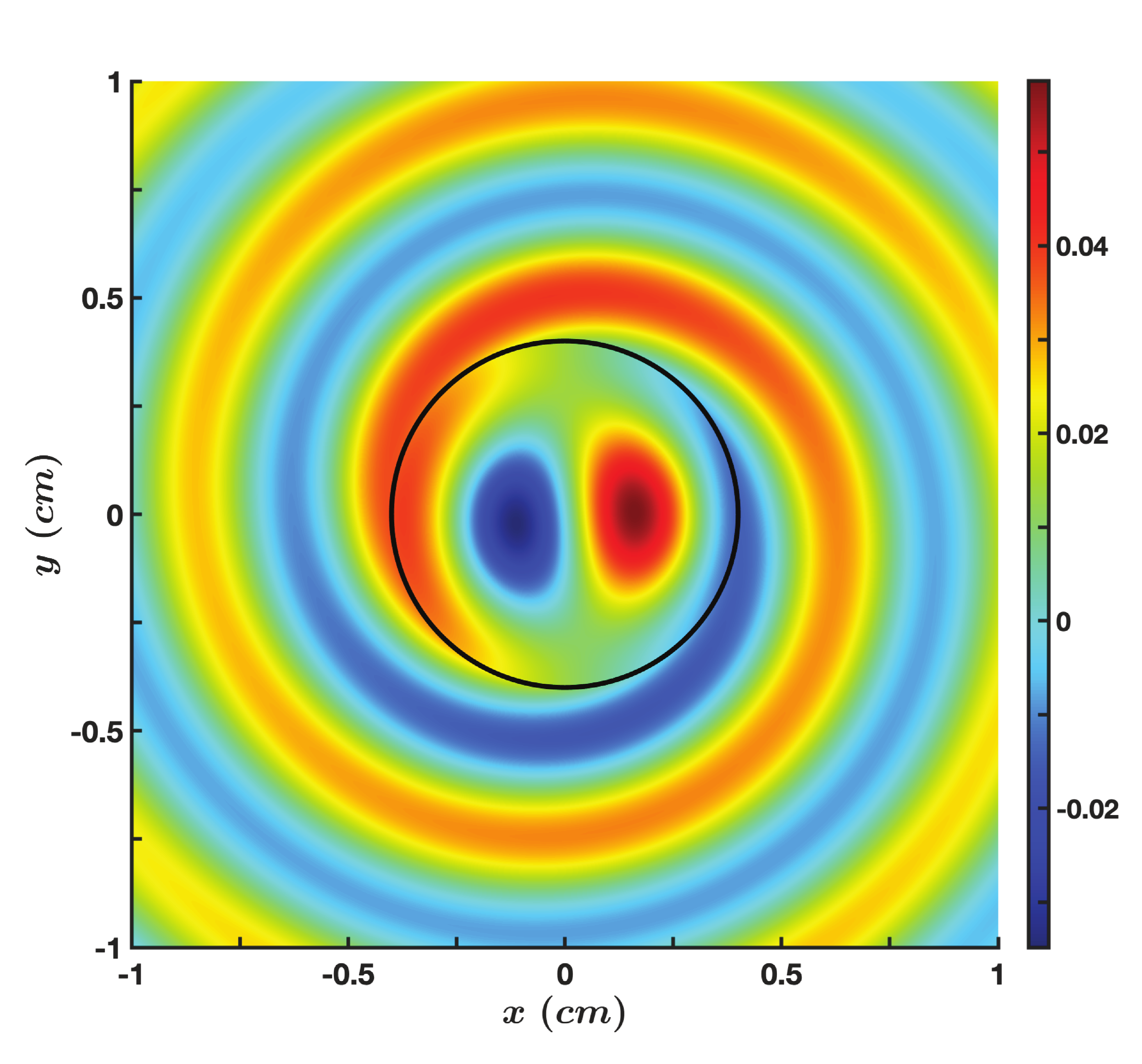}} 
} 
        \captionsetup{justification=raggedright,singlelinecheck=false} %this is for putting the caption justified left 
        \caption{ \label{fig:fig11} These results are for the same conditions and parameters as in Fig. \ref{fig:fig8} except that
$a = 0.4$ cm. The wavelength of the incoming fast LH wave is $9.25$ cm (see Fig. \ref{fig:fig10a}).}
\end{center}
\end{figure}

%figure12 
\begin{figure}[htp]
\begin{center}
\subfigure[\label{fig:fig12a} \ Contours of ${\rm Re} \left( E_{T x} \right) $ ]{
       {\includegraphics[width=0.48\textwidth]{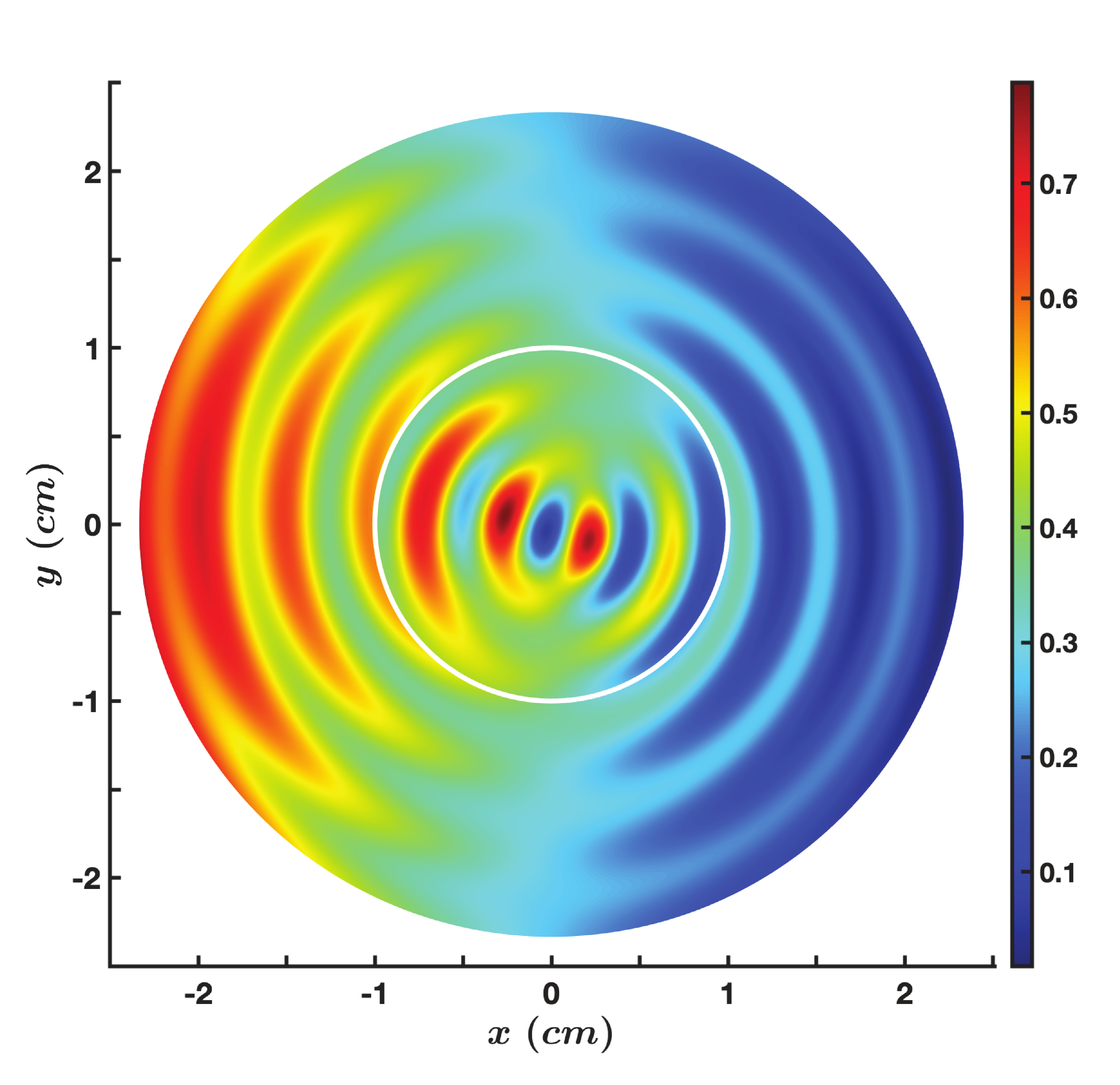}} 
} 
\subfigure[\label{fig:fig12b} \ Contours of ${\rm Re} \left( E_{T y} \right) $ ]{
       {\includegraphics[width=0.48\textwidth]{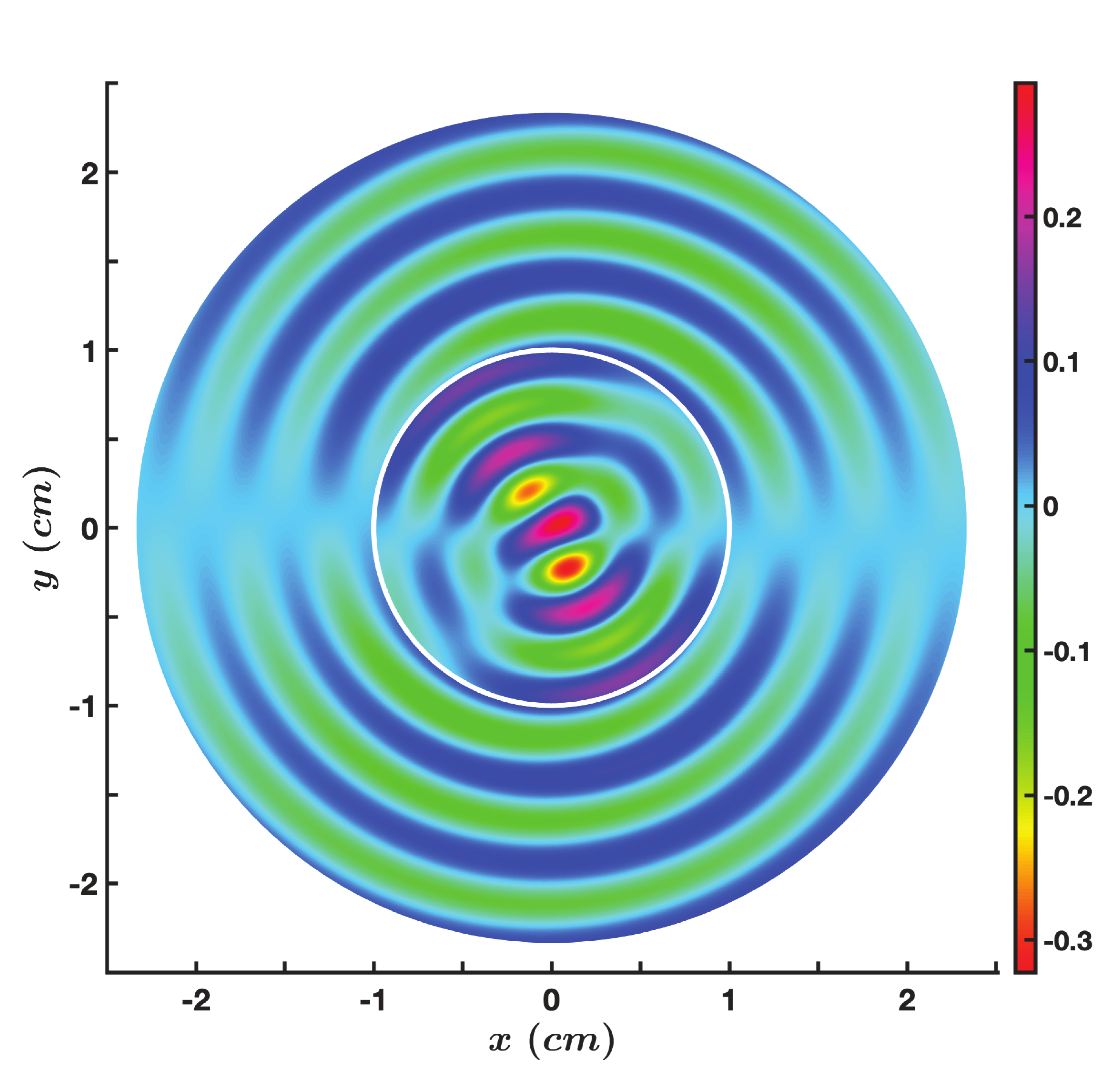}} 
} 
\subfigure[\label{fig:fig12c} \ Contours of ${\rm Re} \left( E_{T z} \right) $ ]{
       {\includegraphics[width=0.48\textwidth]{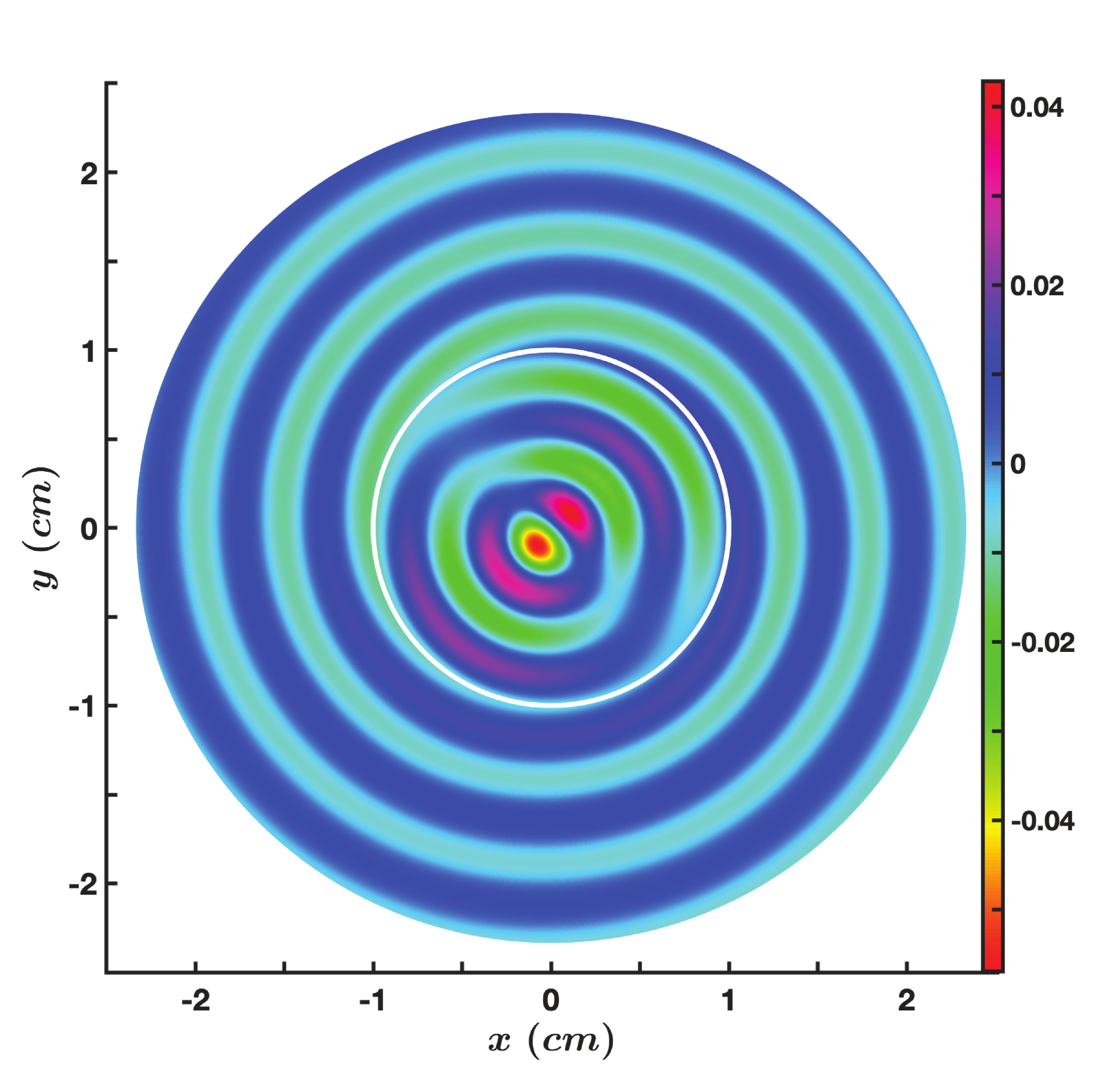}} 
} 
        \captionsetup{justification=raggedright,singlelinecheck=false} %this is for putting the caption justified left 
        \caption{ \label{fig:fig12} Contours of the Cartesian components of $\vec{E}_T$ 
when an evanescent LH fast wave, incident from the left, is scattered by a filament with $a = 1$ cm. The parameters are
the same as in Table \ref{table:t1} except that the background
and filament densities are interchanged, i.e., $n_e^b = 2 \times 10^{19}$ m$^{-3}$ and $n_e^f =  2.25 \times 10^{19}$ m$^{-3}$.}
\end{center}
\end{figure}

%figure13 
\begin{figure}[htp]
\begin{center}
\subfigure[\label{fig:fig13a} \ Contours of $P_x$ ]{
       {\includegraphics[width=0.48\textwidth]{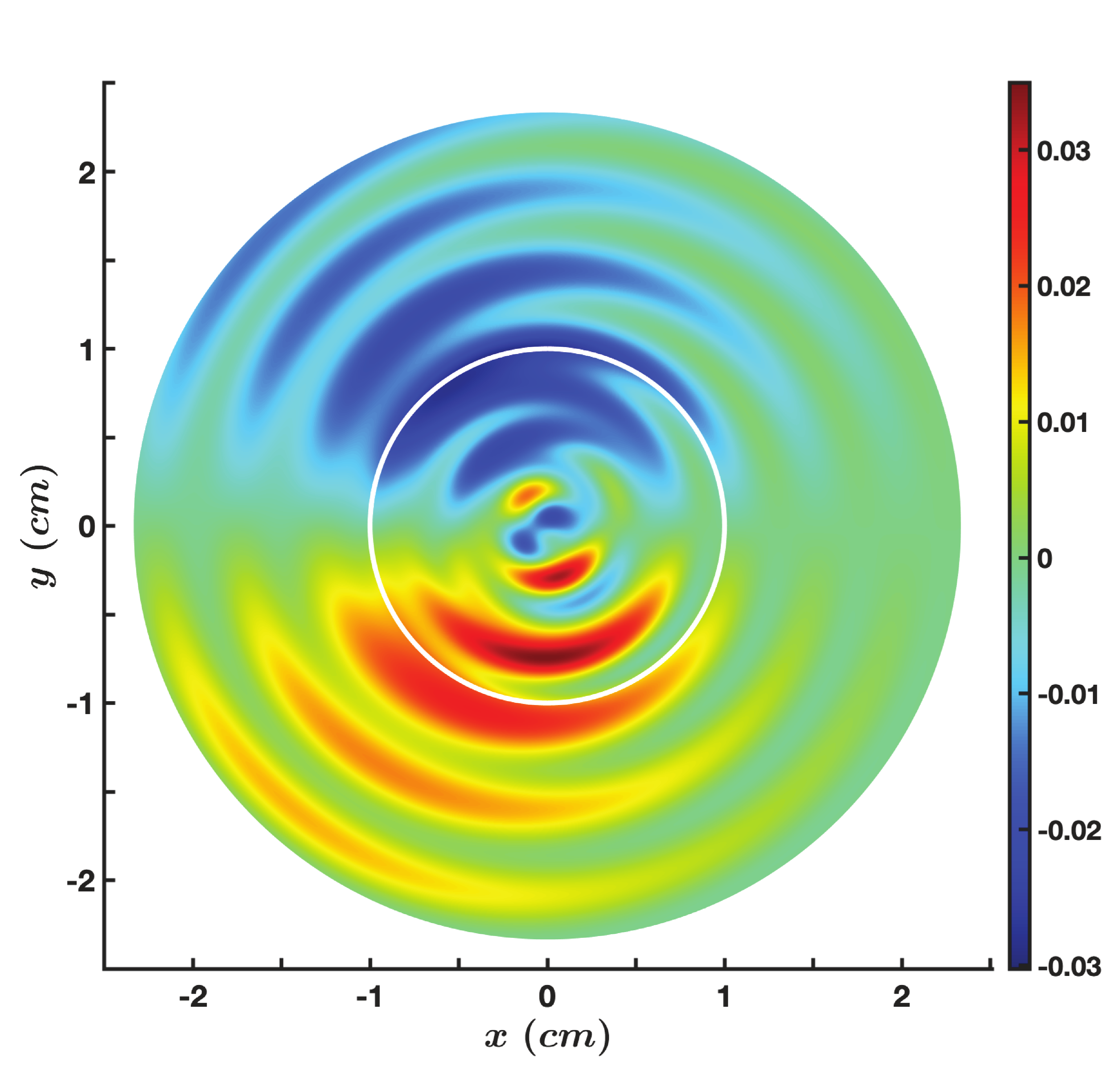}} 
} 
\subfigure[\label{fig:fig13b} \ Contours of $P_y$ ]{
       {\includegraphics[width=0.48\textwidth]{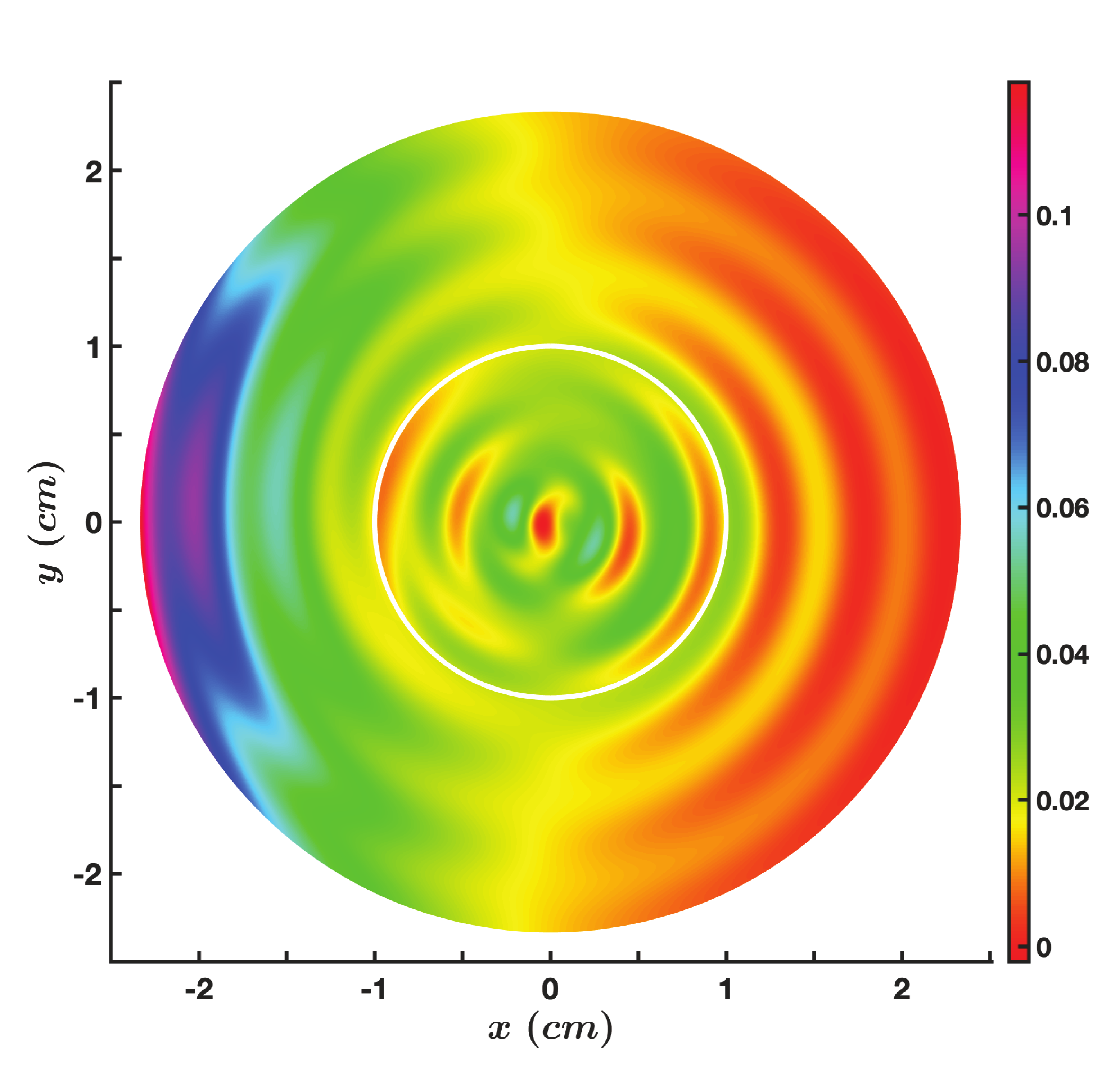}} 
} 
\subfigure[\label{fig:fig13c} \ Contours of $P_z$ ]{
       {\includegraphics[width=0.48\textwidth]{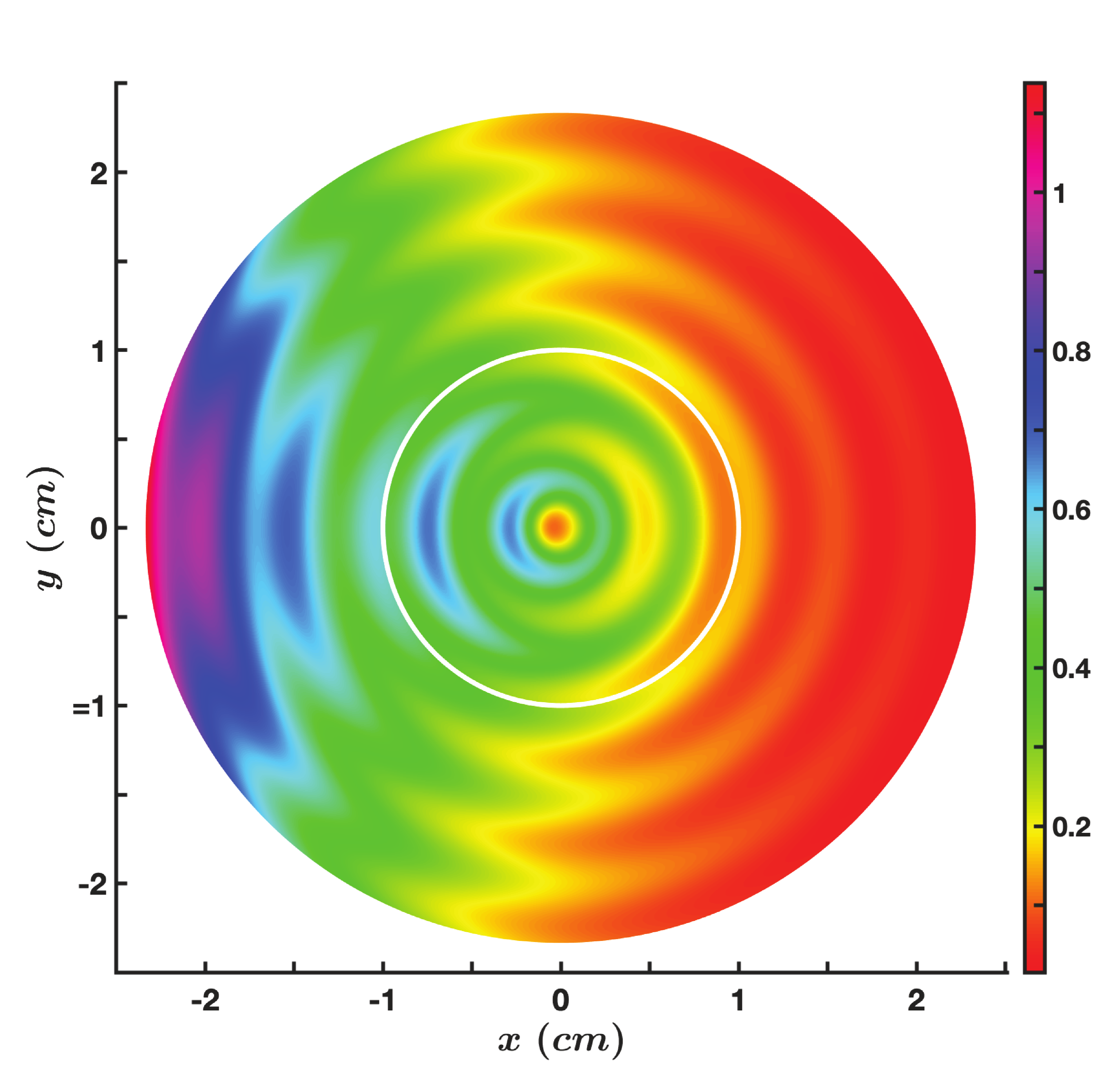}} 
} 
        \captionsetup{justification=raggedright,singlelinecheck=false} %this is for putting the caption justified left 
        \caption{ \label{fig:fig13} Contours of the three Cartesian components of the Poynting vector associated with 
the fields in Fig. \ref{fig:fig12}.}
\end{center}
\end{figure}

%figure14 
\begin{figure}[htp]
\begin{center}
\subfigure[\label{fig:fig14a} \ Contours of ${\rm Re} \left( E_{T x} \right) $ ]{
       {\includegraphics[width=0.48\textwidth]{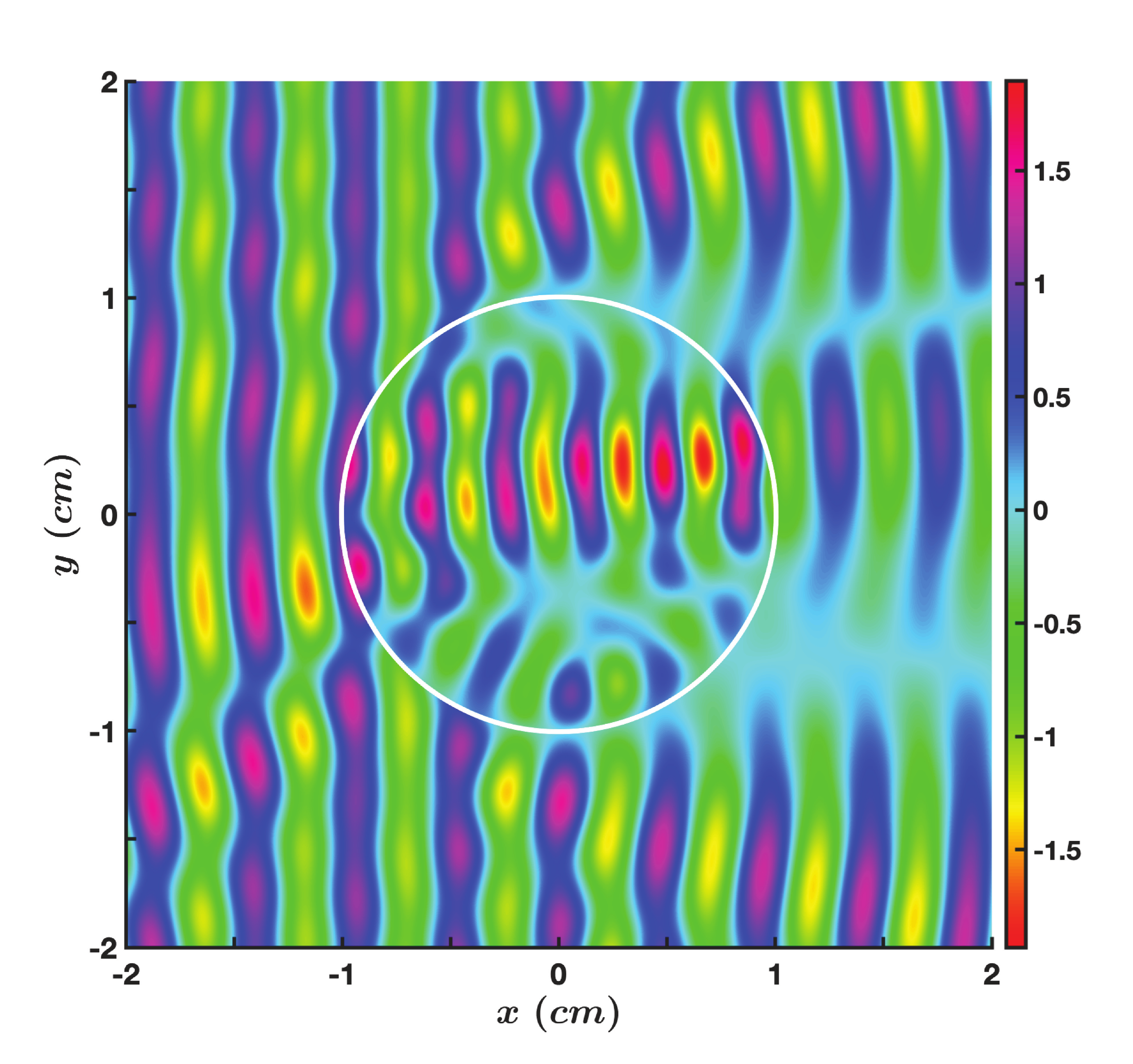}} 
} 
\subfigure[\label{fig:fig14b} \ Contours of ${\rm Re} \left( E_{T y} \right) $ ]{
       {\includegraphics[width=0.455\textwidth]{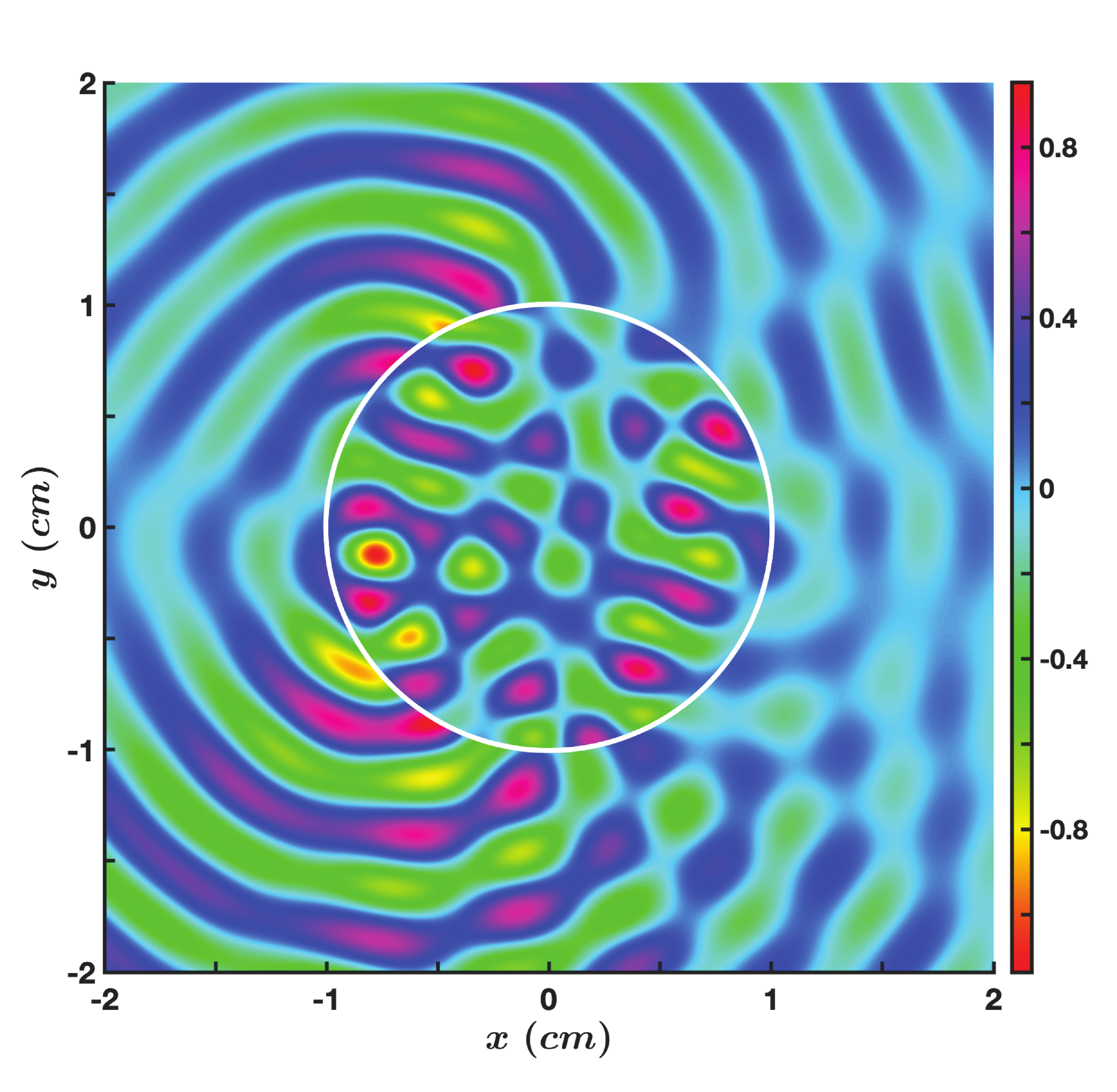}} 
} 
        \captionsetup{justification=raggedright,singlelinecheck=false} %this is for putting the caption justified left 
        \caption{ \label{fig:fig14} Contours of Re$\left( E_{Tx} \right)$ and Re$\left( E_{Ty} \right)$, respectively, when a
propagating slow LH wave is incident from the left. Here, $a = 1$ cm, $n_e^b = 2 \times 10^{19}$ m$^{-3}$, 
$n_e^f = 4 \times 10^{19}$ m$^{-3}$, $B_0 = 4.5$ T, $\nu = 4.6$ GHz, and $n_z = 2$.} 
\end{center}
\end{figure}

%figure15 
\begin{figure}[htp]
\begin{center}
\subfigure[\label{fig:fig15a} \ Contours of ${\rm Re} \left( E_{T x} \right) $ ]{
       {\includegraphics[width=0.48\textwidth]{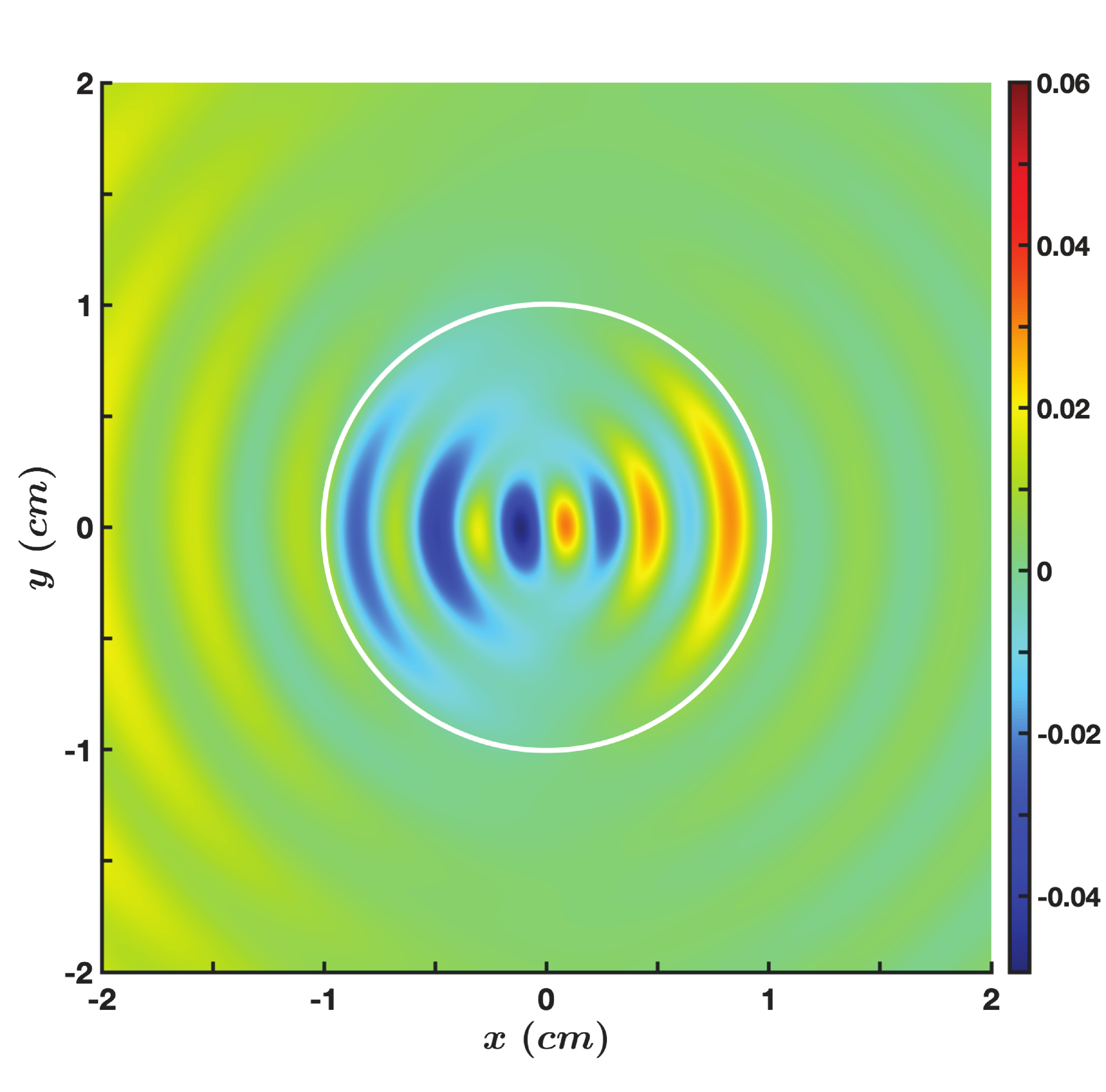}} 
} 
\subfigure[\label{fig:fig15b} \ Contours of ${\rm Re} \left( E_{T y} \right) $ ]{
       {\includegraphics[width=0.48\textwidth]{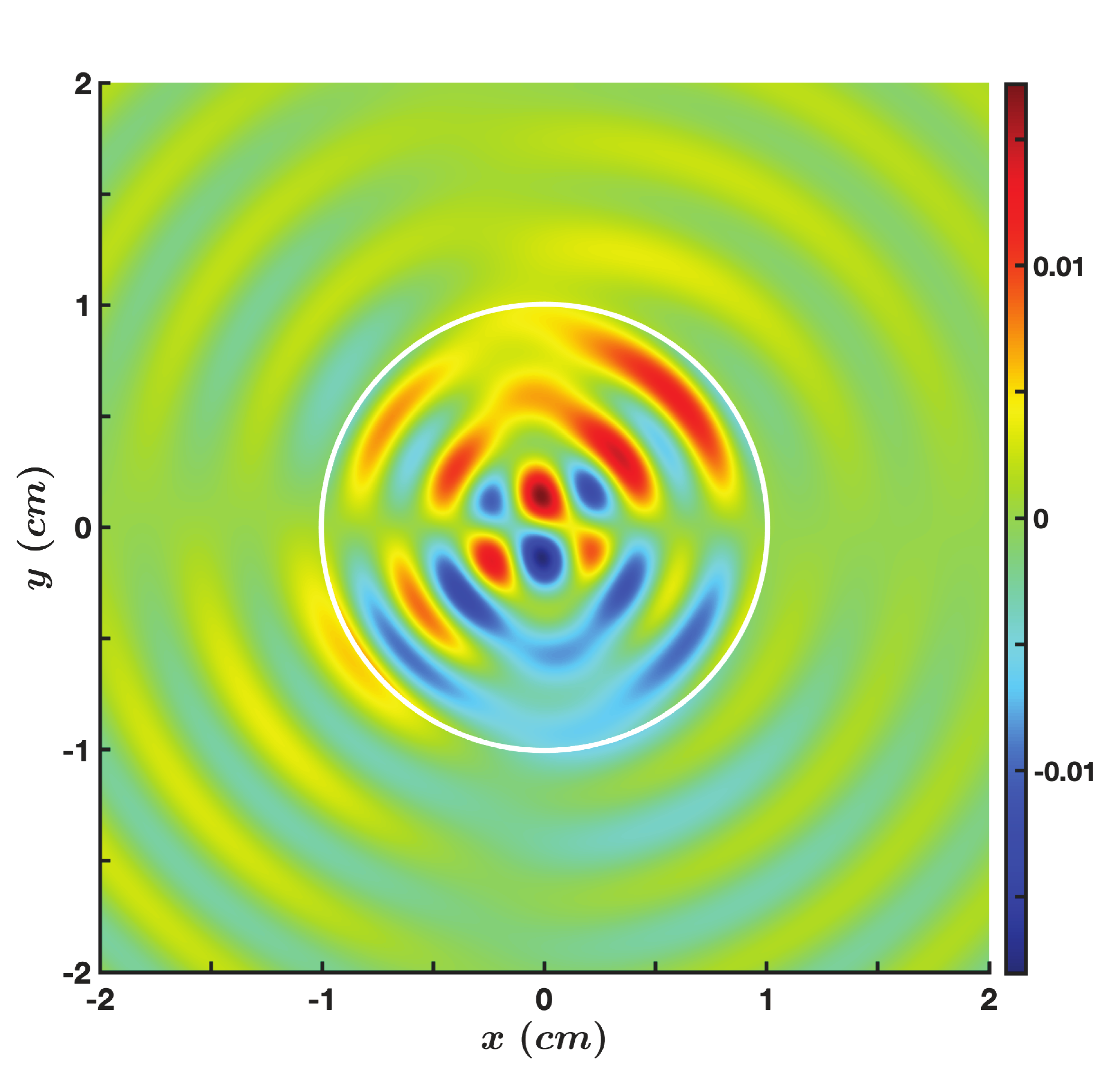}} 
} 
        \captionsetup{justification=raggedright,singlelinecheck=false} %this is for putting the caption justified left 
        \caption{ \label{fig:fig15} Contours of Re$\left( E_{Tx} \right)$ and Re$\left( E_{Ty} \right)$, respectively, when an 
evanescent fast plane LH wave is incident from the left. The plasma and wave parameters are as for Fig. \ref{fig:fig14}. }
\end{center}
\end{figure}

%figure16 
\begin{figure}[ht]
    \begin{center}
       {\scalebox{0.3} {\includegraphics{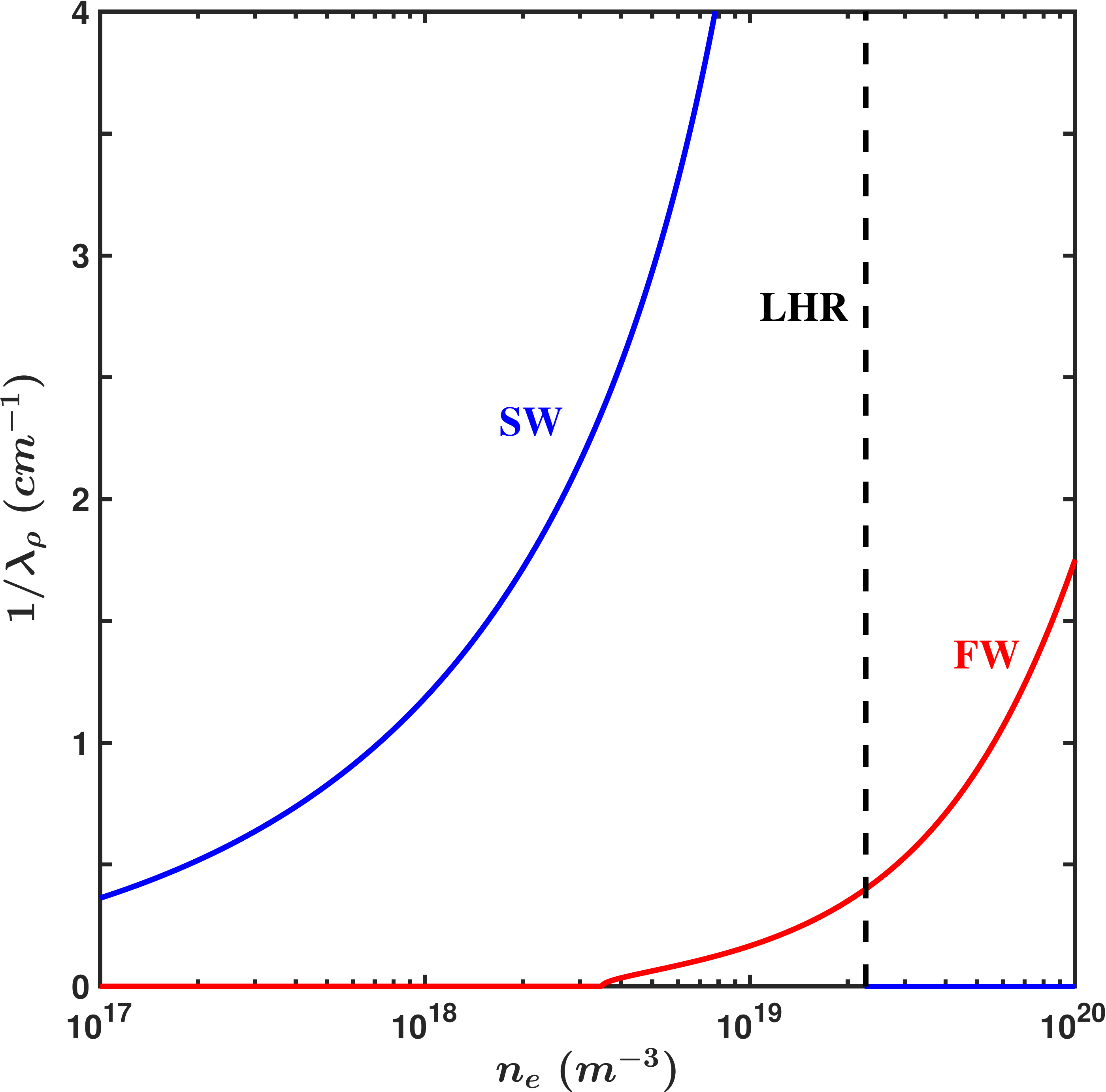}}}
        \captionsetup{justification=raggedright,singlelinecheck=false} %this is for putting the caption justified left 
        \caption{\label{fig:fig16} Variation of the cold plasma dispersion roots for helicon waves, obtained from \eqref{3.10}, 
as a function of the electron density. SW and FW indicate the slow and fast waves, respectively, and LHR marks the density 
location of the lower hybrid resonance. The parameters are, $B_0 = 1.4$ T, $\nu = 476$ MHz, and $n_z = 4$.}  
    \end{center} 
\end{figure}

%figure17 
\begin{figure}[ht]
    \begin{center}
       {\scalebox{0.3} {\includegraphics{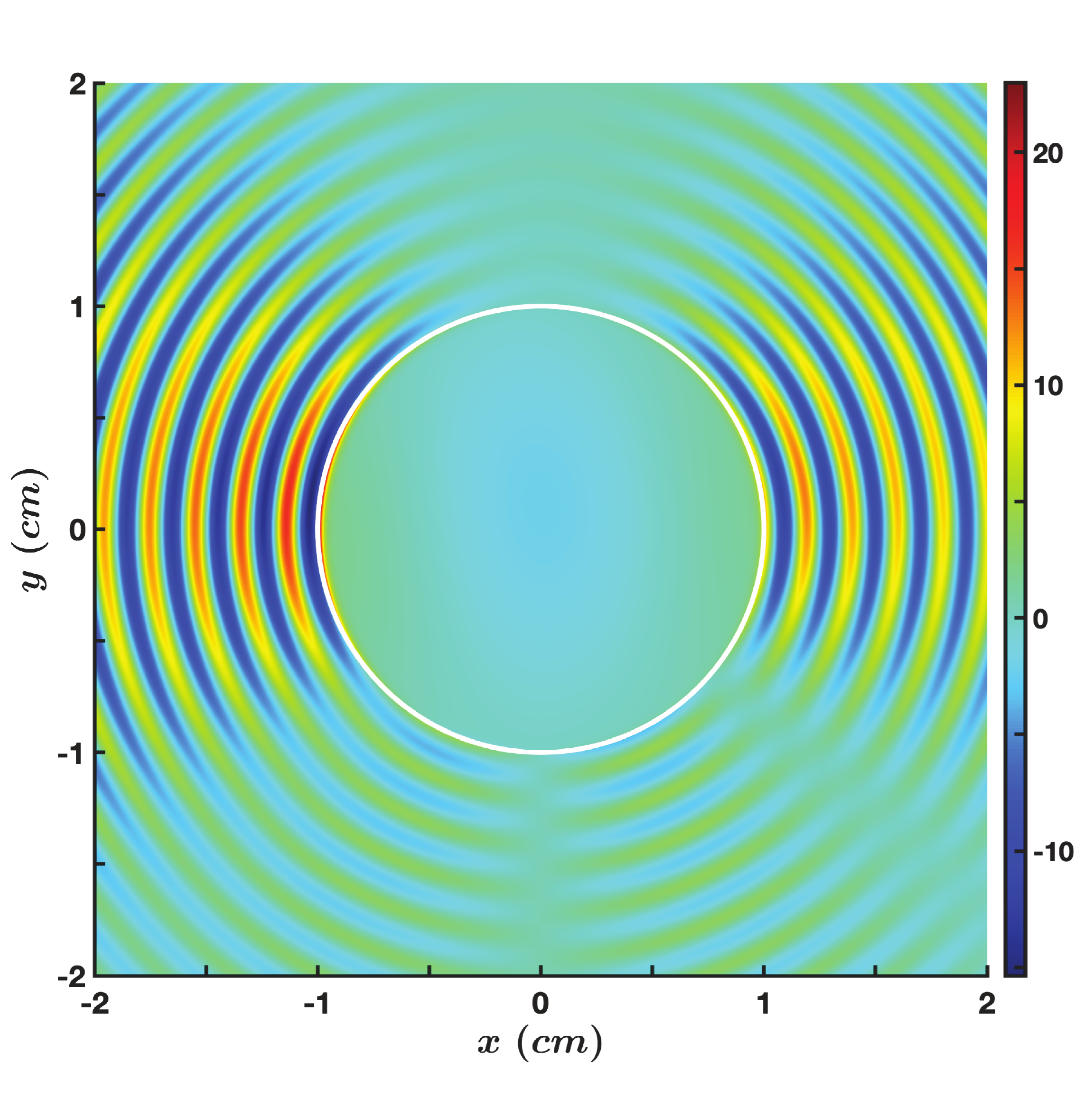}}}
        \captionsetup{justification=raggedright,singlelinecheck=false} %this is for putting the caption justified left 
        \caption{\label{fig:fig17} Contours of Re$\left( E_{Tx} \right)$ in the $x$-$y$ plane for the scattering of a fast helicon wave 
when the plasma density is greater than the cutoff density of the fast wave. The parameters are the same as in Fig. \ref{fig:fig16} along
with $a = 1$ cm, $n_e^b = 10^{19}$ m$^{-3}$, and $n_e^f = 3 \times 10^{19}$ m$^{-3}$.}
    \end{center} 
\end{figure}

%figure18 
\begin{figure}[ht]
    \begin{center}
       {\scalebox{0.3} {\includegraphics{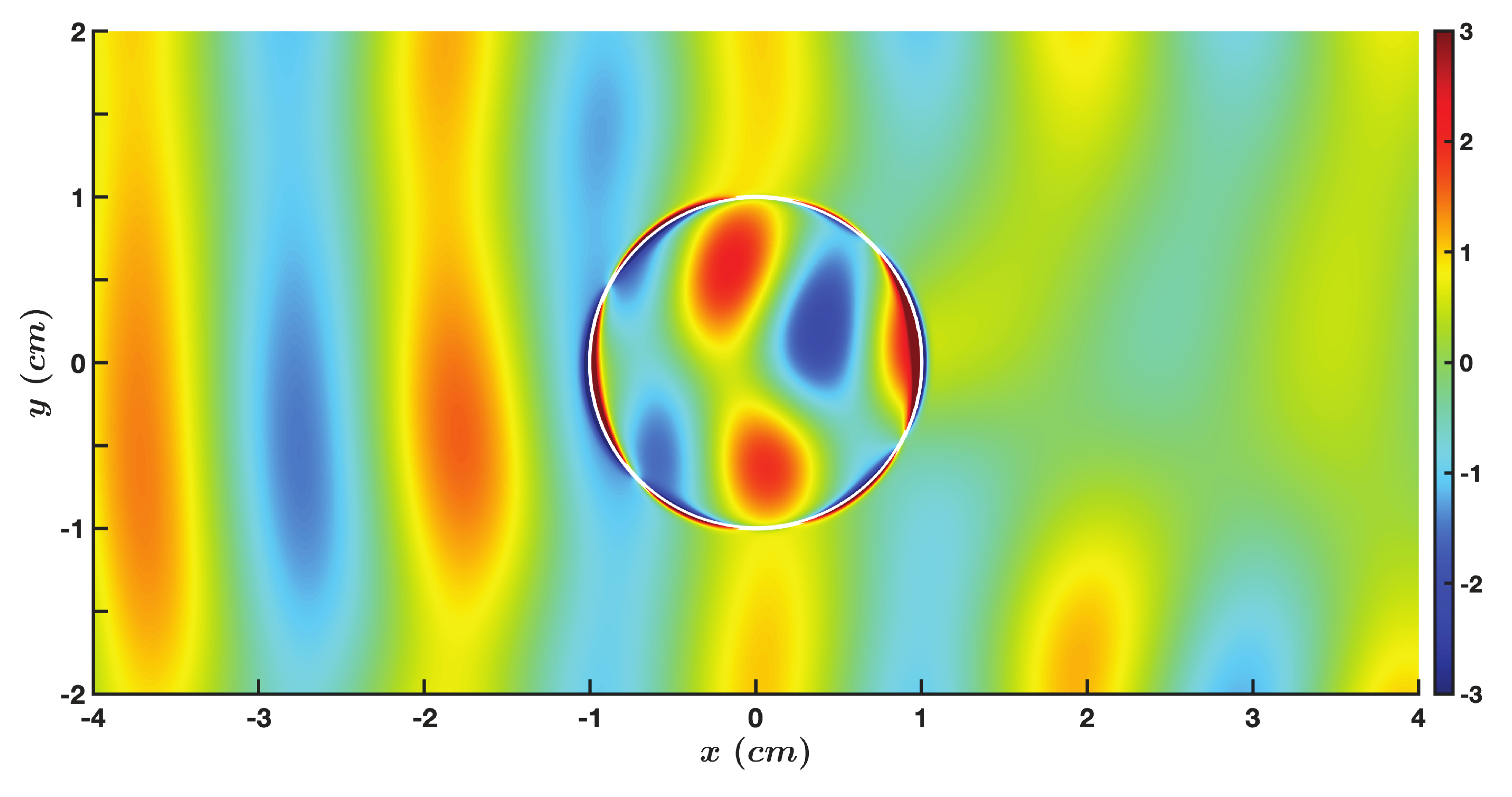}}}
        \captionsetup{justification=raggedright,singlelinecheck=false} %this is for putting the caption justified left 
        \caption{\label{fig:fig18} Contours of Re$\left( E_{Tx} \right)$ in the $x$-$y$ plane for the scattering of a fast helicon wave 
when the plasma density is greater than the density at which the lower hybrid resonance occurs. 
The parameters are the same as in Fig. \ref{fig:fig16} along
with $a = 1$ cm, $n_e^b = 3 \times 10^{19}$ m$^{-3}$, and $n_e^f = 5 \times 10^{19}$ m$^{-3}$. The maximum 
amplitude of the evanescent slow wave field at the filament boundary has been truncated so as to enhance the contrast of the field in other
regions.}
    \end{center} 
\end{figure}

%figure19
\begin{figure}[ht]
    \begin{center}
       {\scalebox{0.3} {\includegraphics{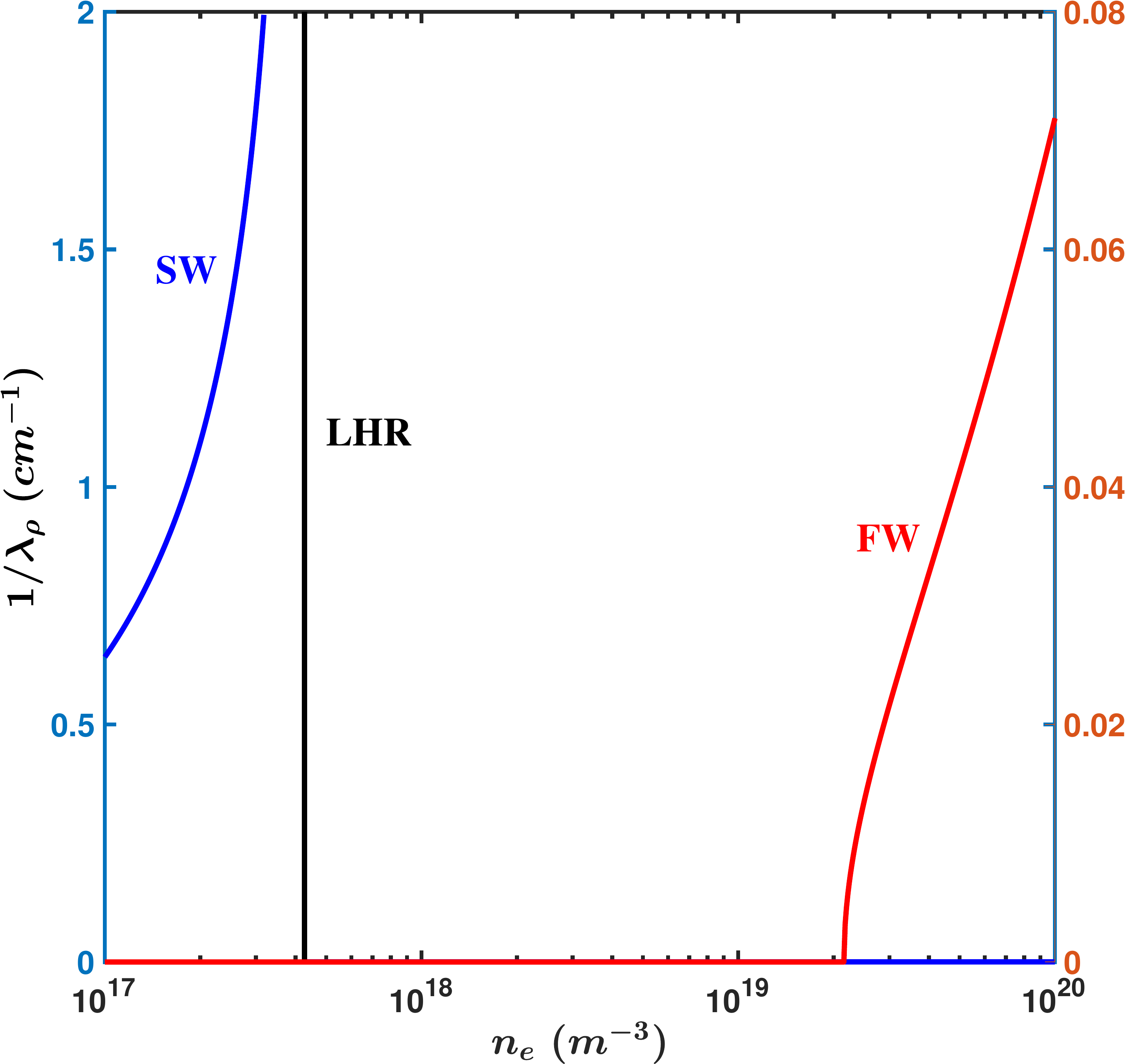}}}
        \captionsetup{justification=raggedright,singlelinecheck=false} %this is for putting the caption justified left 
        \caption{\label{fig:fig19} The cold plasma dispersion roots in the ion cyclotron range of frequencies, obtained from \eqref{3.10}, 
as a function of the electron density. SW and FW indicate the slow and fast Alfv\'en waves, respectively, and LHR marks the density 
location of the lower hybrid resonance. The ordinate scale on the left is for the slow wave and that on the right is for the fast wave.
The parameters are, $B_0 = 9.3$ T, $\nu = 120$ MHz, and $n_z = 6$. }
    \end{center} 
\end{figure}

%figure20 
\begin{figure}[htp]
\begin{center}
\subfigure[\label{fig:fig20a} \ Contours of ${\rm Re} \left( E_{T x} \right) $.]{
       {\includegraphics[width=0.48\textwidth]{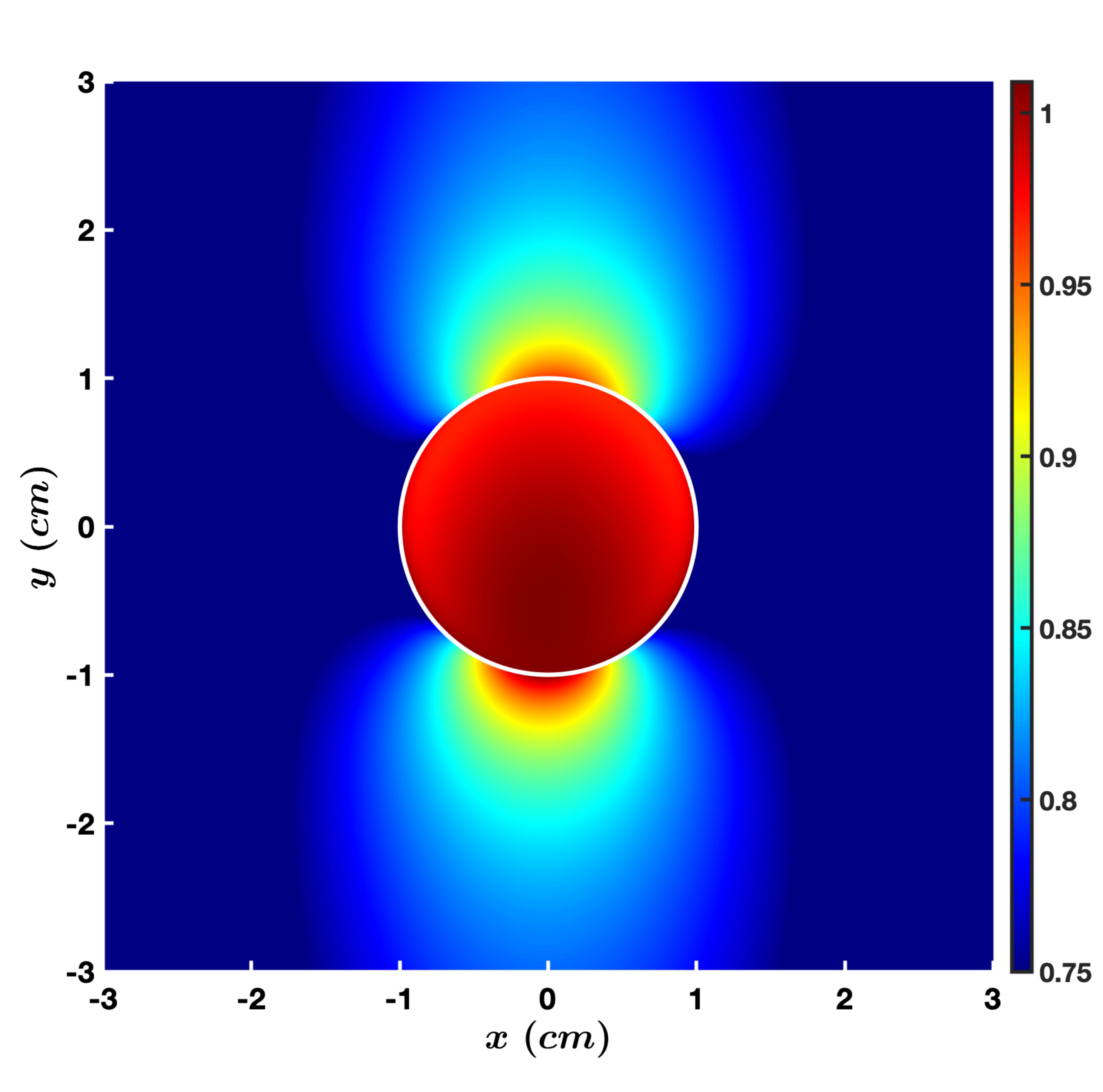}} 
} 
\subfigure[\label{fig:fig20b} \ Contours of $P_z$ ]{
       {\includegraphics[width=0.48\textwidth]{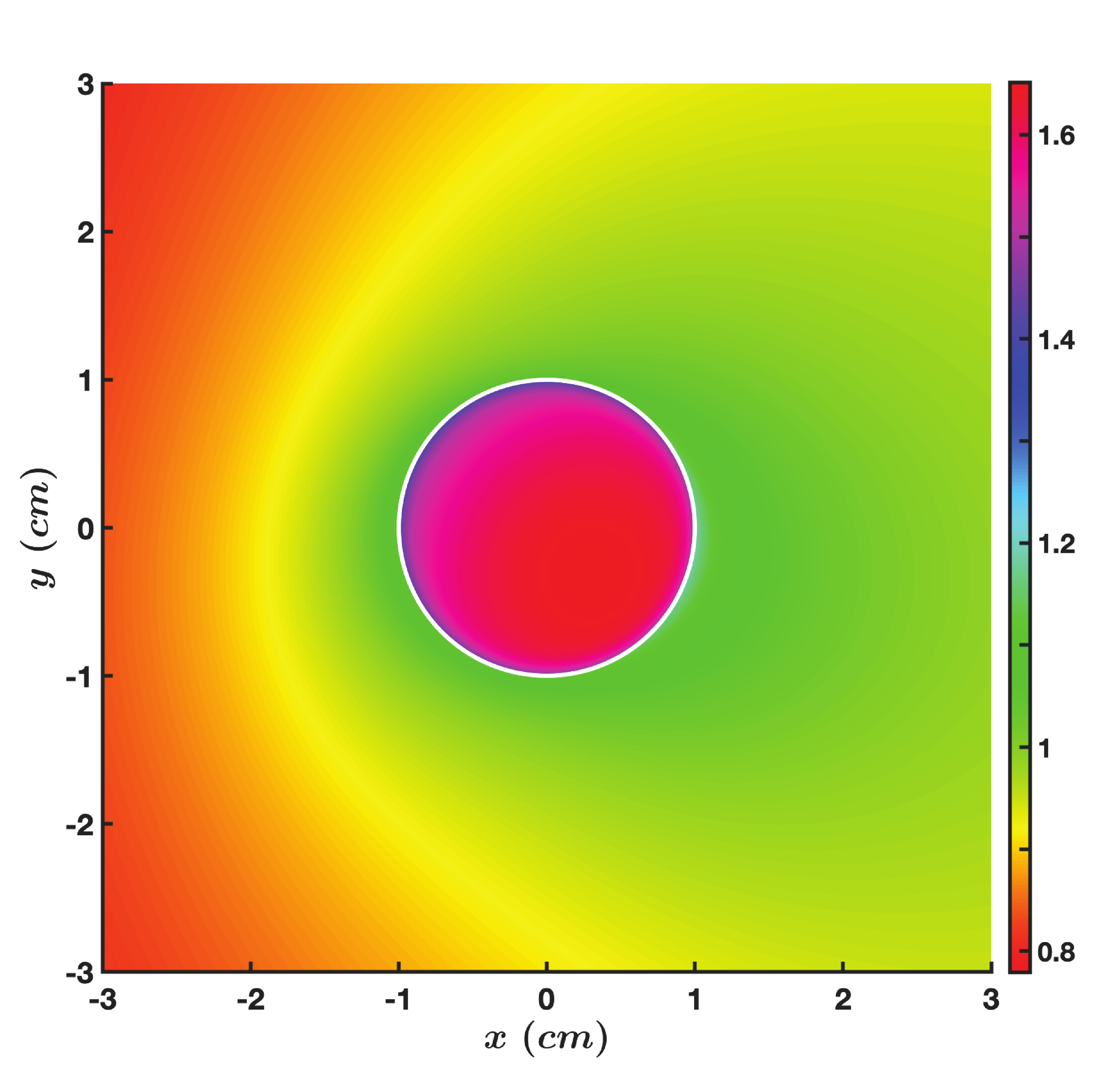}} 
} 
        \captionsetup{justification=raggedright,singlelinecheck=false} %this is for putting the caption justified left 
        \caption{ \label{fig:fig20} Contours of (a) Re$\left( E_{Tx} \right)$ and (b) $P_z$ when a FAW,
incident from the left, scatters off a filament with $a = 1$ cm. The parameters are, $B_0 = 9.3$ T, $\nu = 120$ MHz, $n_z = 6$,
$n_e^b = 4 \times 10^{19}$ m$^{-3}$, and $n_e^f = 7 \times 10^{19}$ m$^{-3}$.}
\end{center}
\end{figure}

%figure21
\begin{figure}[htp]
\begin{center}
\subfigure[\label{fig:fig21a} \ Incident plane wave is the slow LH wave.]{
       {\includegraphics[width=0.48\textwidth]{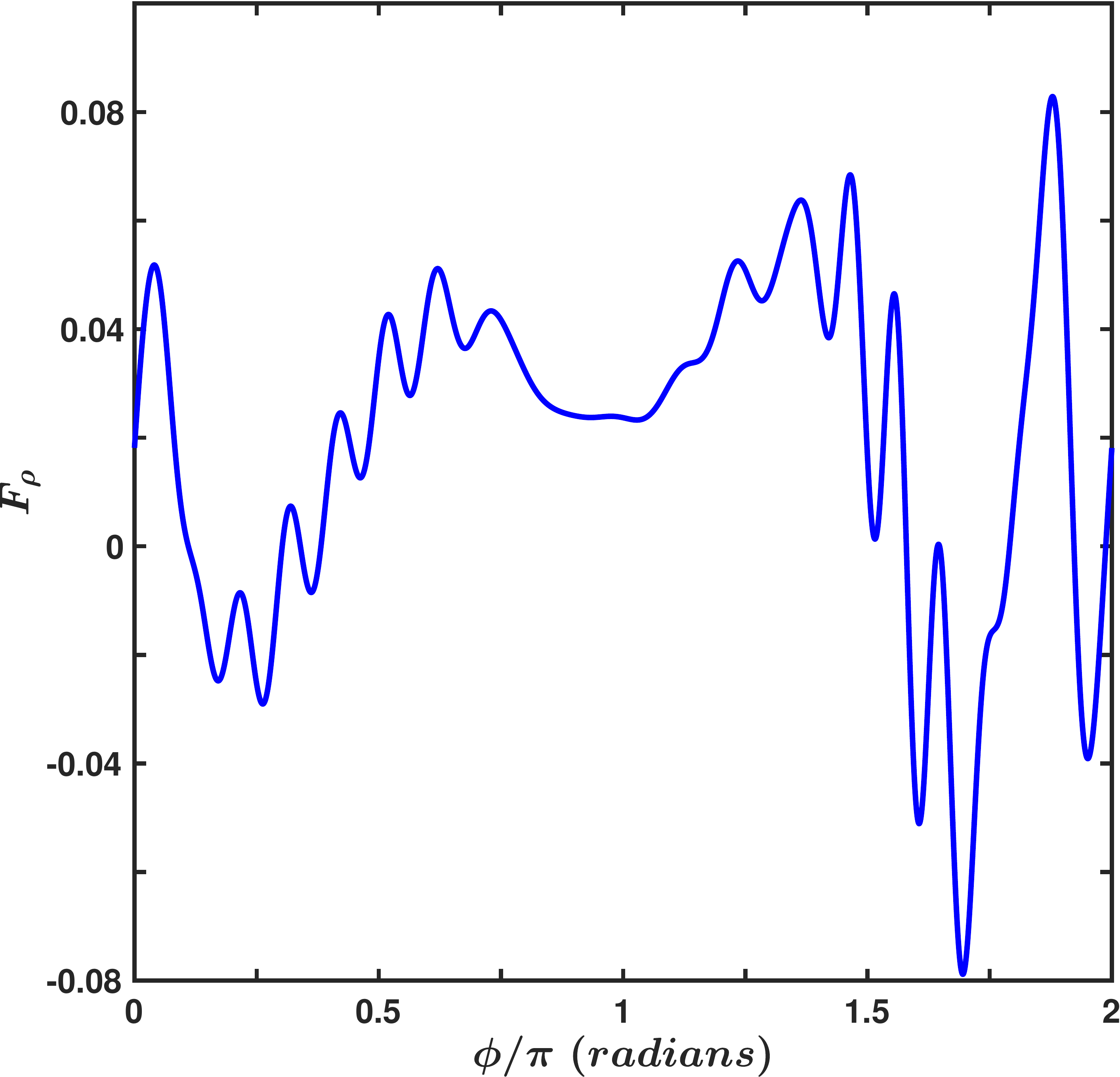}} 
} 
\subfigure[\label{fig:fig21b} \ Incident plane wave is the fast LH wave ]{
       {\includegraphics[width=0.465\textwidth]{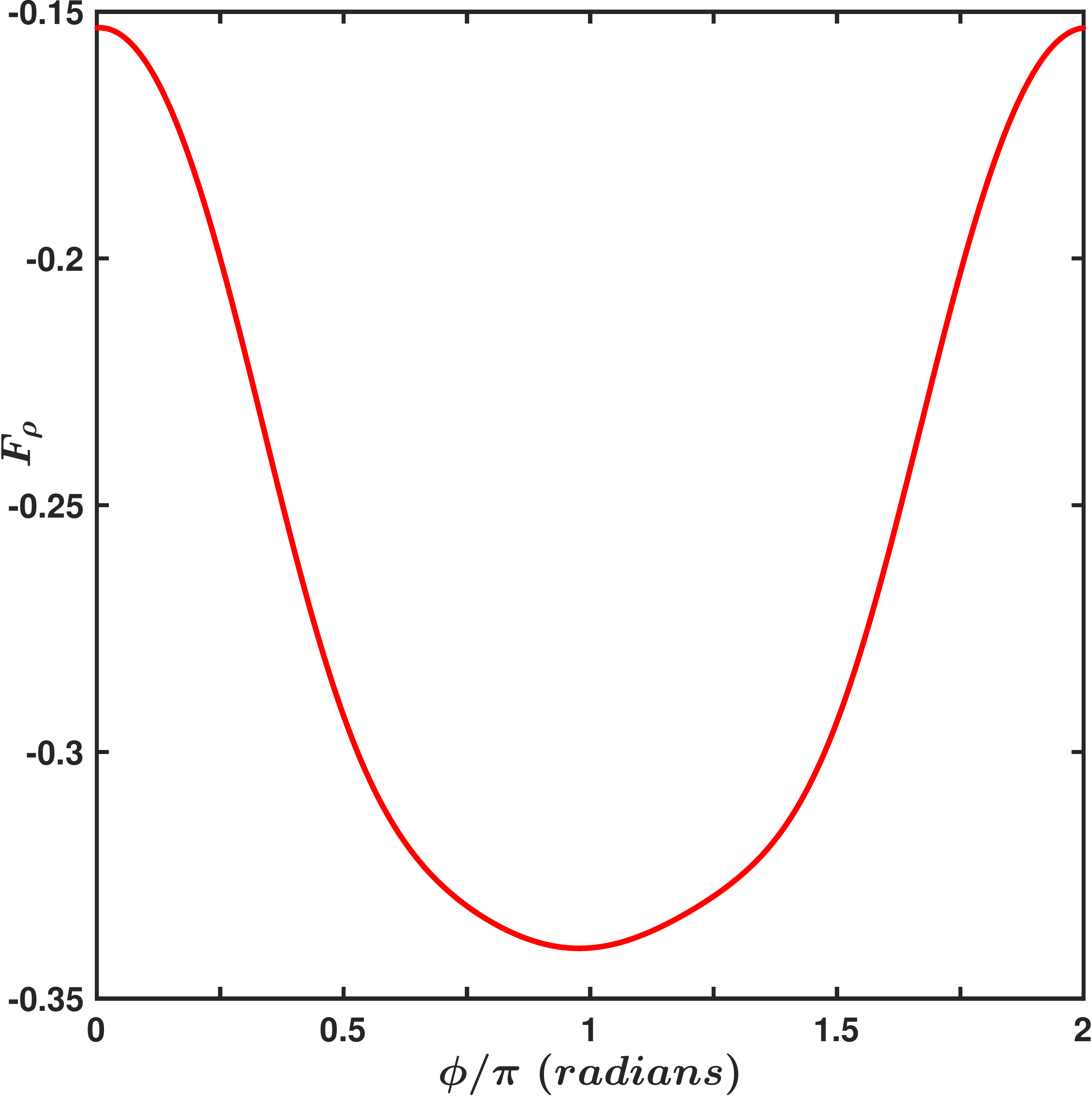}} 
} 
        \captionsetup{justification=raggedright,singlelinecheck=false} %this is for putting the caption justified left 
        \caption{ \label{fig:fig21} The normalized stress force as a function of the azimuthal angle. The parameters for
the slow and fast LH waves are the same
as for Figs. \ref{fig:fig5} and \ref{fig:fig8}, respectively. }
\end{center}
\end{figure}

%\vfill\eject 

\end{document}